\documentclass{bmvc2k}
\usepackage{graphicx}
\usepackage{booktabs}
\usepackage{adjustbox}
\usepackage{amsmath}
\usepackage{amsfonts}
\usepackage{enumitem}
\usepackage{color, soul}

\let\svthefootnote\thefootnote
\newcommand\freefootnote[1]{%
  \let\thefootnote\relax%
  \footnotetext{#1}%
  \let\thefootnote\svthefootnote%
}

\title{A Mixed Quantization Network for Computationally Efficient Mobile Inverse Tone Mapping}

\addauthor{Juan Borrego-Carazo\textsuperscript{1*,}}{juan.borrego@uab.cat}{2}
\addauthor{Mete Ozay}{m.ozay@samsung.com}{1}
\addauthor{Frederik Laboyrie}{f.laboyrie@samsung.com}{1}
\addauthor{Paul Wisbey}{p.wisbey@samsung.com}{1}

\addinstitution{
 Samsung Research UK\\
 Staines-upon-Thames\\
 United Kingdom
}
\addinstitution{
Universitat Autònoma de Barcelona\\
Cerdanyola del Vallès\\
Spain}

\runninghead{MQN}{Computationally Efficient Mobile ITM}

\begin{document}

\maketitle

\begin{abstract}
 
Recovering a high dynamic range (HDR) image from a single low dynamic range (LDR) image, namely inverse tone mapping (ITM), is challenging due to the lack of information in over- and under-exposed regions. Current methods focus exclusively on training high performing but computationally inefficient ITM models, which in turn hinder deployment of the ITM models in resource-constrained environments with limited computing power such as edge and mobile device applications.  

To this end, we propose combining efficient operations of deep neural networks with a novel mixed quantization scheme to construct a well performing but computationally efficient mixed quantization network (MQN) which can perform single image ITM on mobile platforms. In the ablation studies, we explore the effect of using different attention mechanisms, quantization schemes and loss functions on the performance of MQN in ITM tasks. In the comparative  analyses, ITM models trained using MQN perform on par with the state-of-the-art methods on benchmark datasets.  MQN models provide up to  10 times improvement on latency and 25 times improvement on memory consumption. 
\end{abstract}

\section{Introduction}

\freefootnote{* Researched during internship at Samsung Research UK}High dynamic range (HDR) imaging enables capturing, storing, and displaying images and videos that cover the whole range of illuminance values present in natural scenes \cite{reinhard_high_2010}. In contrast, low dynamic range (LDR) imaging technologies only work with a reduced range of values limited by their channel bit depth, thus producing results of inferior perceived quality \cite{akyuz_hdr_2007,hanhart_subjective_2014}. 

In the last two decades, HDR imaging methods have been applied in different fields and sectors, such as digital games \cite{noauthor_hdr_2020, akenine-moller_real-time_2019} and  photography \cite{debevec_recovering_1997, gallo_locally_2015}, among others. Due to the lack of appropriate HDR displays, HDR content was converted to LDR. This operation, called tone mapping (TM) \cite{mantiuk_display_2008, drago_adaptive_2003, reinhard_photographic_2002}, pursues  reproduction of images by reducing the tonal values within the images, which leads to loss of details and inevitable appearance changes in the reproduced images \cite{akyuz_hdr_2007, eilertsen_evaluation_2016}. Recently, there has been a substantial increase in production and commercialization of HDR displays. Sustained by new standards, such as HDR10, the number of HDR TV shipped worldwide has leaped by more than a 10x factor between years 2016-2019 \cite{noauthor_hdr_2020}. In contrast, most content to be reproduced is still LDR, thus not attaining the reproduction capabilities of HDR displays.

\begin{figure}[!t]
	\setlength{\lineskip}{0pt}
	\centering
	\includegraphics[width=0.25\linewidth, height=0.13\textheight]{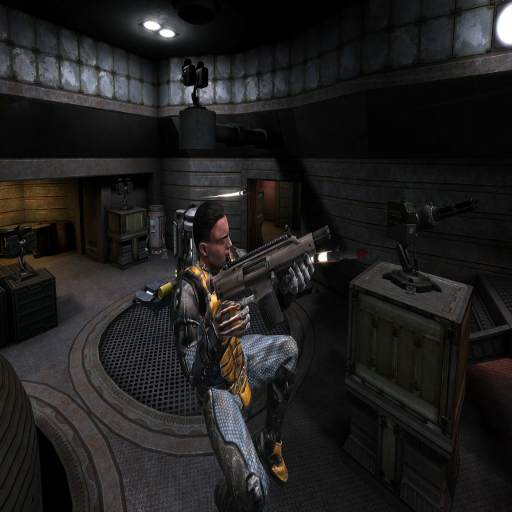}\hspace*{-0.3em}
	\includegraphics[width=0.25\linewidth, height=0.13\textheight]{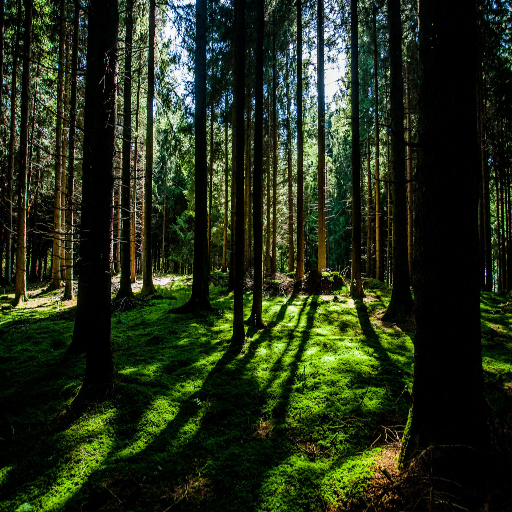}\hspace*{-0.3em}
	\includegraphics[width=0.25\linewidth, height=0.13\textheight]{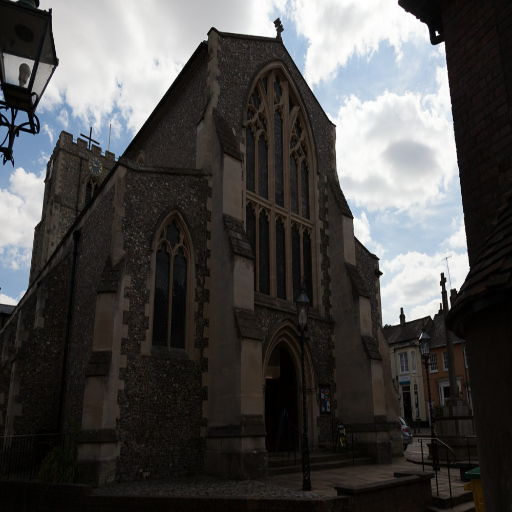}\hspace*{-0.3em}
	\includegraphics[width=0.25\linewidth, height=0.13\textheight]{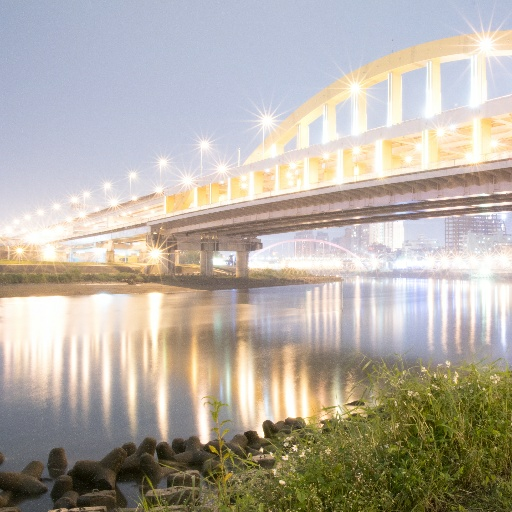}	\\
	\includegraphics[width=0.25\linewidth, height=0.13\textheight]{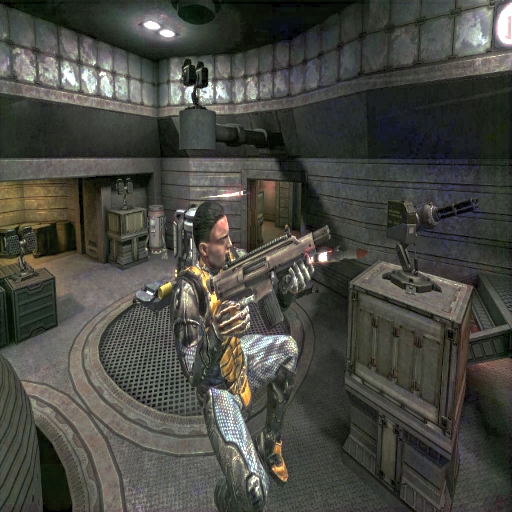}\hspace*{-0.3em}
	\includegraphics[width=0.25\linewidth, height=0.13\textheight]{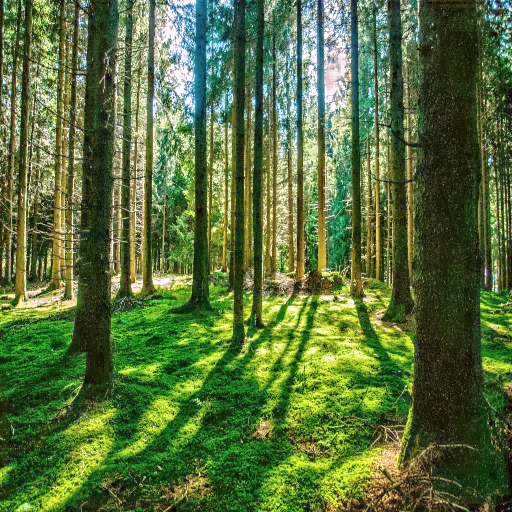}\hspace*{-0.3em}
	\includegraphics[width=0.25\linewidth, height=0.13\textheight]{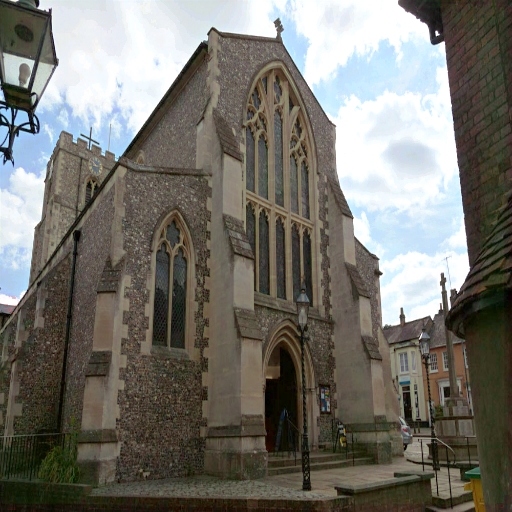}\hspace*{-0.3em}
	\includegraphics[width=0.25\linewidth, height=0.13\textheight]{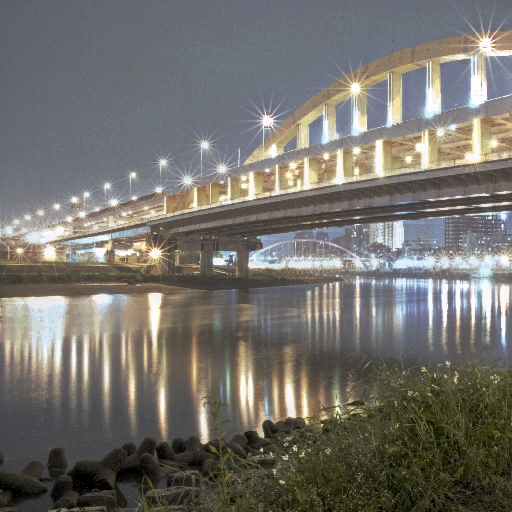}
	\caption{\textbf{Mobile ITM from single LDR images using mixed quantization network (MQN).} Employment of mixed quantization methods and efficient blocks in MQN help reduce  computational complexity of HDR image reconstruction (shown at the bottom row) from single LDR images (shown at the top row), and enable its deployment to mobile platforms, achieving a latency of $\approx$21ms on a Samsung Note 20 Exynos 990.}
	\label{fig:example}
\vspace{-0.5cm}
\end{figure}

This situation shows the need for the conversion of LDR images into HDR content. This  process is referred to as inverse tone mapping  (ITM), and involves recreation of the missing information, and expansion and adaptation of the available information to a higher bit depth. Handcrafted ITM algorithms \cite{banterle_inverse_2006, rempel_ldr2hdr_2007, kuo_content-adaptive_2012, huo_physiological_2014, masia_evaluation_2009}  focused on range expansion to accommodate the LDR content to the new bit depth, as well as the application of tone operators to linearize the content and adjust the missing information. However, such methods did not provide sufficiently appealing results that could match originally produced HDR content \cite{banterle_psychophysical_2009}.

Deep neural networks (DNNs) have been applied to ITM tasks \cite{serrano_convolutional_2016, eilertsen_hdr_2017, kim_end--end_2020} providing state-of-the-art results and products in industrial applications \cite{noauthor_auto_2021}. Nevertheless, current deep-learning-based ITM methods incur high computational costs, both in terms of memory and latency, due to the use of costly operations \cite{endo_deep_2017, marnerides_expandnet_2018},  large models \cite{liu_single-image_2020, lee_deep_2018} or feedback loops \cite{khan_fhdr_2019, kim_end--end_2020}, among other factors. Such burdens impede their deployment and usage in resource-constrained environments, such as edge devices and mobile phones.

In the present work, we propose a novel Mixed Quantization Network (MQN) for computationally efficient mobile ITM by integrating different quantization schemes applied to models, deploying efficient convolutions into  models, and training models using input data. The proposed MQN enables attaining similar accuracy compared to state-of-the-art methods  on a reference ITM dataset and can be deployed to a mobile platform to perform more computationally efficient inference. Further, it can be utilised by different hardware platforms, such as CPU or GPU, due to the flexibility of its components and structure.  More precisely, our models can perform  10x to 100x times faster inference than competing methods (sample results are given in Figure \ref{fig:example}). As far as the authors are aware, this is the first work devoted to constructing computationally efficient DNN-based ITM methods for mobile ITM tasks.

\begin{figure}[t]
	\centering
	\includegraphics[width=\linewidth, height=0.28\textheight]{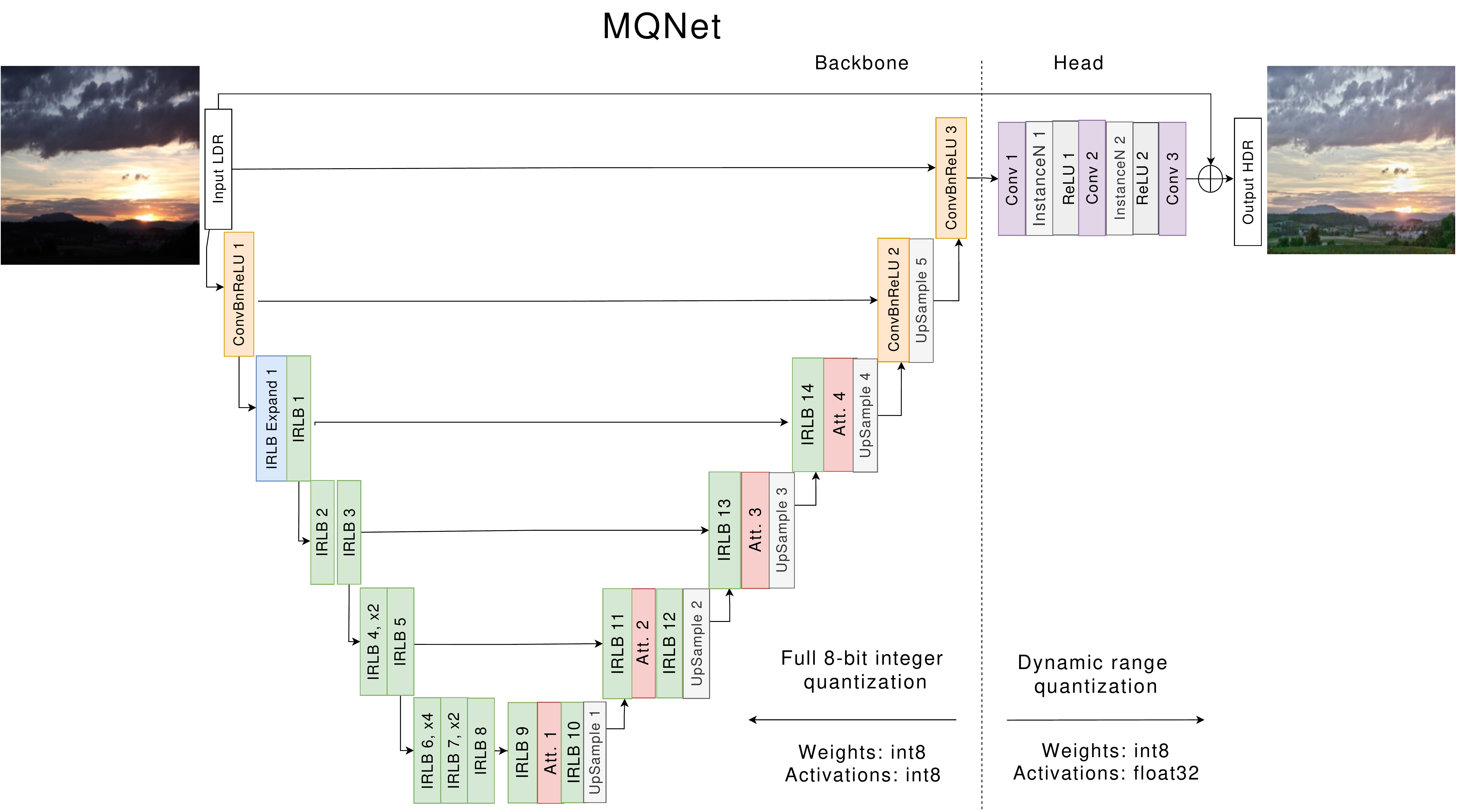}
	\caption{Illustration of the base backbone  and high precision head that comprise the MQN. IRLB is used for fast inference and gated attention mechanisms \cite{zhang_image_2018} for improvement of feature representation learning accuracy. The dotted line indicates the separation between the fully quantized architecture and the dynamically quantized head. Input is added to output of the head to produce the overall output. Hyperparameters are given in the supp. mat.}
	\label{fig:archs}
\vspace{-0.5cm}
\end{figure}

\section{Related Work}\label{RW}

\textbf{Vanilla ITM Methods}. Vanilla ITM methods have been implemented  using tone operators \cite{banterle_inverse_2006}, expansion algorithms \cite{banterle_inverse_2006, masia_dynamic_2017, masia_evaluation_2009}, such as gamma curve expansion and over-exposed region enhancement \cite{rempel_ldr2hdr_2007}, among other techniques \cite{akyuz_hdr_2007, huo_physiological_2014}. A great benefit, in general, of such methods is their low computing power requirement. However, they struggle in generation of high quality image content on over- and under-exposed image regions \cite{endo_deep_2017}.

\textbf{DNN-based ITM methods}. Recently, DNNs have been employed on ITM tasks. Although most of these methods blatantly differ regarding network architectures and model training components, they can be categorized into three main groups.

The methods in the first group \cite{zhou_unmodnet_2020, liu_single-image_2020, khan_fhdr_2019, marnerides_expandnet_2018, ning_learning_2018, yang_image_2018, zhang_learning_2017, eilertsen_hdr_2017, vien_single-shot_2021, ye_single_2021, sharif_two-stage_2021, perez-pellitero_ntire_2021, akhil_single_2021, chen_single-image_2021} suffer from the problem of not accurately processing  over- and under-exposed regions, but provide  more compact systems. Methods in the second group use a group of bracketed over- and under-exposed images as input to directly learn to generate  HDR output \cite{pan_metahdr_2021, yan_attention-guided_2019, niu_hdr-gan_2020, yan_deep_2020, wu_deep_2018, kalantari_deep_2017, liu_adnet_2021, pan_metahdr_2021, prabhakar_labeled_2021}, as similarly utilized in photographic HDR generation. Methods belonging to these two groups differ mainly according to  their network components, such as non-local blocks \cite{yan_deep_2020} and attention mechanisms \cite{yan_attention-guided_2019}.
Methods in the third group train models to generate over- and under-exposed images from an LDR input, and  merge them to obtain an HDR image \cite{kim_end--end_2020, lee_deep_2018, lee_deep_2018-1, endo_deep_2017}. Their main distinction from the other groups lies in the methods used to generate different exposures, such as deconvolution \cite{endo_deep_2017}, sequential generation \cite{lee_deep_2018, lee_deep_2018-1}, or recurrency \cite{kim_end--end_2020}. 

\textbf{Single Image HDR Reconstruction}. In this work, we focus on single image HDR reconstruction. The task is substantially more challenging than  multi-input HDR imaging. Among the first  works on this task,  \cite{eilertsen_hdr_2017} propose using a U-Net type DNN to learn only representations of the over-exposed regions, while the rest of the image is only linearized through a default function.  \cite{marnerides_expandnet_2018} use a three-branch network with different dilation ratios and sizes. Recently, \cite{liu_single-image_2020} achieved the state-of-the-art by developing a model that reverses and unravels the camera pipeline to reproduce the final HDR, using a different DNN for each step, thus resulting in a computationally heavy system.  

\textbf{Computationally Efficient ITM Methods}. Most DNN-based methods use heuristics which make them unsuitable for deployment in resource-constrained platforms. A common trick to enhance training that burdens inference is using feedback loops \cite{khan_fhdr_2019, kim_end--end_2020, lee_deep_2018-1}, which are commonly used to generate  differently exposed bracketed images. Another case is when different DNNs  are stacked sequentially and/or in parallel \cite{lee_deep_2018, liu_single-image_2020, wu_deep_2018}. There are also studies that make use of  custom network blocks, such as non-local blocks \cite{yan_deep_2020} or 3D convolutions \cite{endo_deep_2017}, which would hinder their deployment to mobile platforms \cite{Sud20}. 

\section{A Mixed Quantization Network for Mobile ITM}\label{ProposedSystem}

The purpose of this work is to develop a learning-based ITM system with fast inference, especially devoted to the deployment to platforms with limited computational power. To this end, we propose a Mixed Quantization Network (MQN) by designing its architecture and components with state-of-the-art fast inference techniques such as quantization and efficient convolutions. To be able to accelerate inference while maintaining
the required high precision output needed for ITM, we implement a mixed quantization (MQ)
scheme as depicted in  Figure \ref{fig:archs}. Moreover, we train models to learn multi-scale feature representations of HDR content over the input using a backbone network.
The proposed MQN has two components: (1) a feature learning backbone network, endowed with full integer post-training quantization, and (2) a smaller network equipped with a head utilizing dynamic quantization to obtain a target precision for ITM.

\subsection{Backbone and High Precision Head of MQN}\label{BA}

In order to learn multi-scale feature representations, we implement the backbone using a U-Net architecture with skip connections which is also utilized in other single image ITM methods \cite{liu_single-image_2020, eilertsen_hdr_2017, endo_deep_2017} due to its strong accuracy to speed trade-off compared to single-scale (resolution) networks \cite{marnerides_expandnet_2018}. As the encoder of the backbone, we use a MobileNetV2 (MBV2) \cite{sandler_mobilenetv2_2018} to obtain fast inference. We extensively use IRLB  blocks \cite{DBLP:journals/corr/HowardZCKWWAA17}, which have a reduced computation cost\footnote{The computational cost reduction from a vanilla convolution to a depthwise separable is of $\frac{1}{N} + \frac{1}{D_{k}^{2}}$, where $N$ is the number of output channels and $D_{k}$ is the size of the convolution kernel.} compared to vanilla convolutions, and can be used in different hardware platforms (GPU \cite{tan_efficientnet_2020, noauthor_depth-wise_nodate}, TPU \cite{noauthor_depth-wise_nodate}, CPU \cite{DBLP:journals/corr/HowardZCKWWAA17, sandler_mobilenetv2_2018}, NPU \cite{lee_s3nas_2020, choi_prema_2020}) with improvements in latency. To implement the decoder of the backbone, we favor upsampling in contrast to transposed convolutions, since the former is faster and does not produce artifacts \cite{liu_single-image_2020}. To construct the decoder, after each upsampling we concatenate the upsampled feature maps and selected intermediate outputs of the encoder with skip connections. Following the concatenation, we add several IRLB blocks operating on the same resolution. Instead of IRLB blocks, the last two blocks of the decoder are composed of pointwise convolutions followed by batch normalization and ReLU activations. We design the whole  network to have a reduced number of filters and convolutions compared to other U-Net structures (U-Net \cite{ronneberger_u-net_2015} has 7.76M parameters, U-Net++ \cite{zhou_unet_2018} has 9.04M parameters and our model has about 1M parameters), thus reducing latency and memory consumption.

We employ a gated attention mechanism (depicted by Att. in Figure \ref{fig:archs}) after IRLB blocks to improve performance. In the analyses, we explore three methods to implement attention. First, we implement Spatial Attention (SA) \cite{woo_cbam_2018} gated blocks. Second,  we add channel information through a depthwise convolution in parallel to the SA mechanism, which define the channel spatial attention (CSA) mechanism at a given layer by
\begin{equation}
	O_{f} = ((\sigma\circ D_{f})(I_{f}))\odot (((\sigma\circ C_{1})(I_{f}))\odot I_{f})
\end{equation} 
where $I_{f}$ and $O_{f}$  are the input and output of the layer with $f$ channels respectively. $C_{1}$ denotes a convolution with a $1 \times 1$ kernel, $\sigma$ is a sigmoid activation, $D_{f}$ is a depthwise convolution, and $\circ$ indicates function composition. Finally, we also examine channel attention (CA) blocks \cite{zhang_image_2018}, although these have a higher computational cost due to pooling mechanisms. In the  analyses (Section~\ref{ab}), CA provides higher accuracy compared to SA and CSA. All such attention mechanisms can be seen as reduced one-head attention models without dense connections, thus being faster but also less powerful than those employed in, for example, transformer networks \cite{vaswani_attention_2017}. More details about the structure of the attention mechanisms  can be found in the supplementary material.

The head is in charge of recovering both the required detail and style of the input LDR image in the output HDR image. For this reason, the head is composed of three layers (convolution, instance normalization (IN) \cite{huang_arbitrary_2017} and ReLU) and a residual connection with the original input of the model. Then, the head produces the final HDR prediction by ${\hat{H}} = \sigma({I} + \phi({O}))$, where $\phi$ denotes a hyperbolic tangent activation function, $I$ is the input LDR image  and $O$ is the output HDR image of the system. We use $\phi$ to learn the nonlinear transformation between pixel values of the LDR and HDR images, and the purpose of using $\sigma$ is to map to relative illuminance values, i.e. [0, 1] interval \cite{marnerides_expandnet_2018}. The entire MQN architecture is illustrated in Figure \ref{fig:archs}.

\subsection{Mixed Quantization and Fusion}
\label{F}

Full \textit{8-bit} integer quantization \cite{jacob_quantization_2018} cannot be used directly in ITM, since a higher precision output (HDR image) is required. Hence, direct ITM methods \cite{zhou_unmodnet_2020, liu_single-image_2020, khan_fhdr_2019, marnerides_expandnet_2018, ning_learning_2018, yang_image_2018, zhang_learning_2017, eilertsen_hdr_2017} cannot use full \textit{8-bit} quantization and benefit from size and latency reduction on devices supporting only integer valued operations. To address this problem, we define a mixed quantization scheme. The backbone of MQN is quantized to full \textit{8-bit} integer quantization of both weights and activations, to obtain  the most acceleration at inference time,  thus opening the door to deployment in integer only hardware accelerators, such as NPUs, but without restricting the application in other platforms, such as FPGAs \cite{baskin_streaming_2018} or GPUs \cite{gysel_ristretto_2018}. Meanwhile, the remaining part of the network, the head, which produces output with equivalent resolution to that of the input, is quantized by dynamic range post-training quantization \cite{jacob_quantization_2018}. That is,  \textit{8-bit} integer quantization is applied to the weights and 32-bit float activations are used in order to obtain the required high precision output for the mobile ITM tasks.

\subsection{Loss functions}\label{losses}

We train MQN models using the following loss functions of ground-truth and predicted HDR images ${H}$ and $\hat{H}$:

\begin{itemize}[noitemsep,topsep=0pt,leftmargin=*]
    \item $\ell_1$ loss function $\mathcal{L}_1(H, \hat{H}) = ||{\hat{H}} - {H}||_{1}$, where $\|\cdot \|_1$ is the $\ell_1$ norm.
    \item $\ell_2$ loss function $\mathcal{L}_2(H, \hat{H}) = ||{\hat{H}} - {H}||_{2}$, where $\|\cdot \|_2$ is the $\ell_2$ norm.
    \item Cosine loss function
$\mathcal{L}_{CS}(H, \hat{H}) = 1 - \frac{1}{K}\sum_{k=1}^{K}\frac{\langle { \hat{\mathbf{h}}}_{k}, \mathbf{h}_{k}\rangle }{||{\hat{\mathbf{h}}}_{k}||_{2}||{\mathbf{h}}_{k}||_{2}},$
where $\langle\cdot,\cdot\rangle$ denotes the inner product, and ${\hat{\mathbf{h}}}_{k} \in \mathbb{R}^3$ and ${\mathbf{h}}_{k} \in \mathbb{R}^3$ is the $k$-th pixel vector of the image ${\hat{H}}$ and $H$.

\item As part of the problem is content generation, we include a perceptual loss in the form of a variant of the feature reconstruction (FR) loss function as in \cite{johnson_perceptual_2016} to force the network to match the feature traits of the original HDR images. In this case, we use a VGG16 to produce the necessary feature maps to compute the FR loss by
	$\mathcal{L}_{FR} (\hat{H}, {H}) = \sum_{i=1}^{3} \frac{1}{K}\sum_{k=0}^{K} | F_{i}(h_k) - F_{i}({\hat{h}_k})|$,
where $F_{i}$ denotes the output feature map obtained from the $i^{th}$ pooling block of the VGG16 and $|\cdot|$ is the absolute value function.
\end{itemize}
 Thus, the overall loss function used for training the MQN model is defined by
\begin{equation}
	\mathcal{L_{MQN}} (\hat{H}, {H}) = \lambda_{1}\mathcal{L}_{1} (\hat{H}, {H})+ \lambda_{2}\mathcal{L}_{2} (\hat{H}, {H})+ \lambda_{3}\mathcal{L}_{CS}(\hat{H}, {H}) + \lambda_{4}\mathcal{L}_{FR}(\hat{H}, {H}),
\end{equation}
where $\lambda_i >0, i=1,2,3,4$ are parameters used to balance range of loss functions.

\section{Experimental Analyses}

We first describe the experimental setup and evaluation methodologies, Next, we present the results from the architecture alternatives defined in Section \ref{ProposedSystem}. Finally, we compare our model with state-of-the-art methods. In the supplemental material, we provide implementation details such as further information on datasets, training, and evaluation procedures. We also provide a more extensive comparison with state-of-the-art models, extended and additional analyses, and a video that shows prediction results on a video game. The code is available at \href{https://github.com/BCJuan/ITMMQNet}{https://github.com/BCJuan/ITMMQNet}.

\subsection{Experimental Setup}\label{ExperimentalSetup}

\textbf{Datasets}. We build our training data from a collection of HDR image datasets \cite{funt_effect_2010, ward_high_2006, noauthor_pfstools_2015, hasinoff_burst_2016}, consisting of 3768 HDR images, split into a training set of 3580 images and a validation set of 188 images. Most of these datasets  do not contain unprocessed LDR images which can be used as input. We opt then for creating the LDR images through TM \cite{marnerides_expandnet_2018}, that is, we apply a tone mapping operator (TMO) \cite{drago_adaptive_2003, mantiuk_display_2008, reinhard_photographic_2002} to the original HDR images to produce LDR images.  For testing and comparing with state-of-the-art methods, we use publicly available datasets, HDR-Eye \cite{nemoto_visual_2015}, HDR-Real \cite{liu_single-image_2020}, and RAISE-1K \cite{dang-nguyen_raise_2015}. These datasets contain LDR and HDR images, enabling a fair evaluation between methods.

\textbf{Accuracy measures}. We measure the accuracy of methods using Peak Signal Noise Ratio (PSNR), Structural Similarity (SSIM) and HDRVDP-2 \cite{mantiuk_hdr-vdp-2_2011}. To measure latency, we perform inference on the CPU on a Samsung Note 20 Exynos 990 (SN20E990). 

\textbf{Training parameters}. We use the Adam optimizer with an initial learning rate of ${5\times10^{-5}}$, a decreasing learning schedule with a decay factor of 0.99 applied at every 4 epochs and a batch size of 4. In the analyses,  the best results are obtained using $\lambda_{1} = 1$, $\lambda_{2} = 1$, $\lambda_{3} = 0.1$ and $\lambda_{4} = 0.05$ for integrating loss functions.

\begin{figure}[!t]
	\setlength{\lineskip}{0pt}
	\includegraphics[width=0.11\linewidth, height=0.08\linewidth]{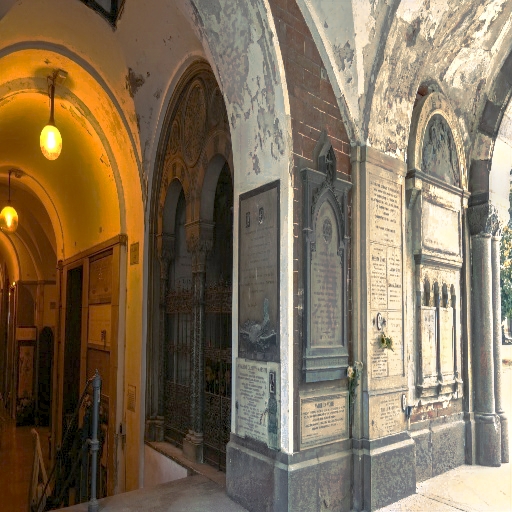}\hspace*{-0.1em}
	\includegraphics[width=0.11\linewidth, height=0.08\linewidth]{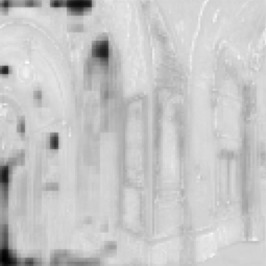}\hspace*{-0.1em}
	\includegraphics[width=0.11\linewidth, height=0.08\linewidth]{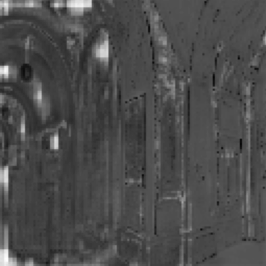}\hspace*{-0.1em}
	\includegraphics[width=0.11\linewidth, height=0.08\linewidth]{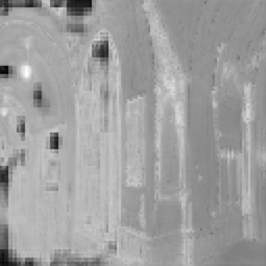}\hspace*{-0.1em}
	\includegraphics[width=0.11\linewidth, height=0.08\linewidth]{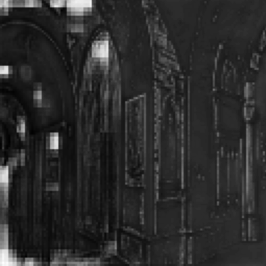}\hspace*{-0.1em}
	\includegraphics[width=0.11\linewidth, height=0.08\linewidth]{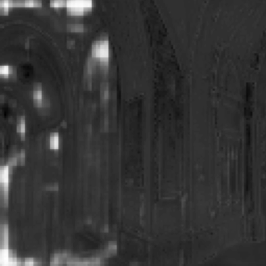}\hspace*{-0.1em}
	\includegraphics[width=0.11\linewidth, height=0.08\linewidth]{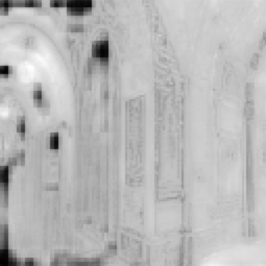}\hspace*{-0.1em}
	\includegraphics[width=0.11\linewidth, height=0.08\linewidth]{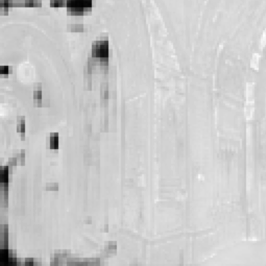}\hspace*{-0.1em}
	\includegraphics[width=0.11\linewidth, height=0.08\linewidth]{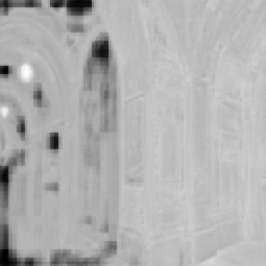}\hspace*{-0.1em}
	\setlength{\lineskip}{0pt}\\
	\includegraphics[width=0.11\linewidth, height=0.08\linewidth]{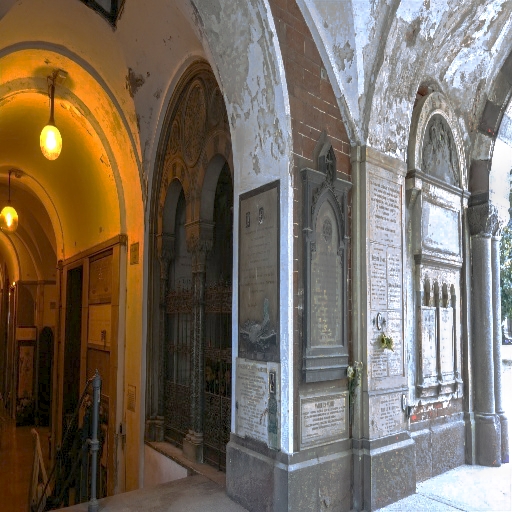}\hspace*{-0.1em}
	\includegraphics[width=0.11\linewidth, height=0.08\linewidth]{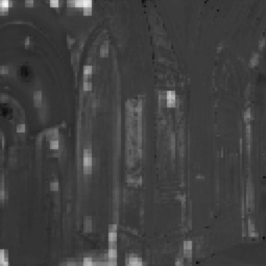}\hspace*{-0.1em}
	\includegraphics[width=0.11\linewidth, height=0.08\linewidth]{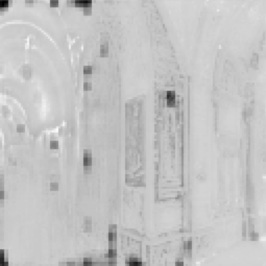}\hspace*{-0.1em}
	\includegraphics[width=0.11\linewidth, height=0.08\linewidth]{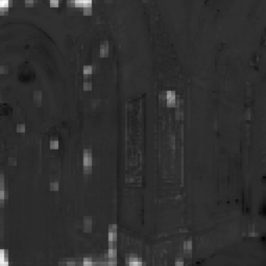}\hspace*{-0.1em}
	\includegraphics[width=0.11\linewidth, height=0.08\linewidth]{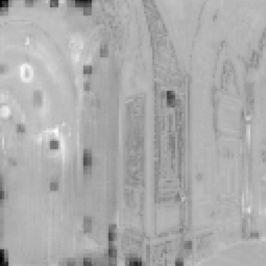}\hspace*{-0.1em}
	\includegraphics[width=0.11\linewidth, height=0.08\linewidth]{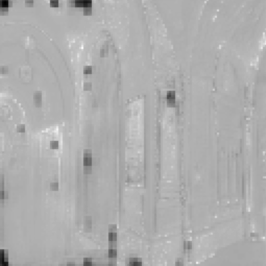}\hspace*{-0.1em}
	\includegraphics[width=0.11\linewidth, height=0.08\linewidth]{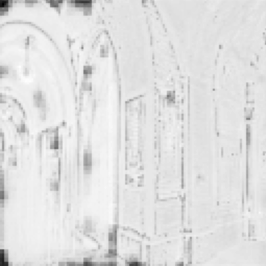}\hspace*{-0.1em}
	\includegraphics[width=0.11\linewidth, height=0.08\linewidth]{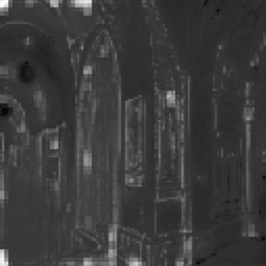}\hspace*{-0.1em}
	\includegraphics[width=0.11\linewidth, height=0.08\linewidth]{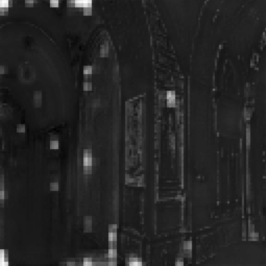}\hspace*{-0.1em}
	\setlength{\lineskip}{0pt}\\
	\includegraphics[width=0.11\linewidth, height=0.08\linewidth]{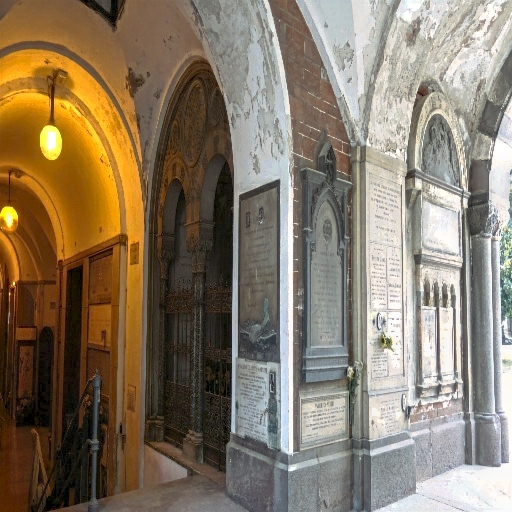}\hspace*{-0.1em}
	\includegraphics[width=0.11\linewidth, height=0.08\linewidth]{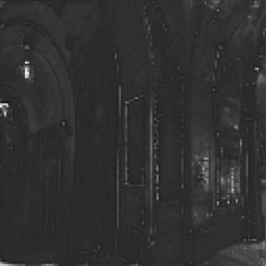}\hspace*{-0.1em}
	\includegraphics[width=0.11\linewidth, height=0.08\linewidth]{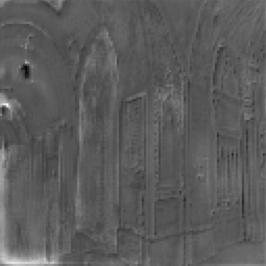}\hspace*{-0.1em}
	\includegraphics[width=0.11\linewidth, height=0.08\linewidth]{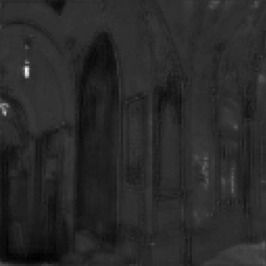}\hspace*{-0.1em}
	\includegraphics[width=0.11\linewidth, height=0.08\linewidth]{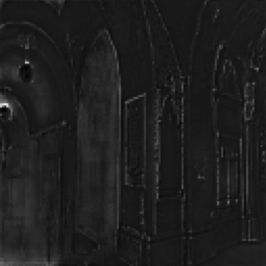}\hspace*{-0.1em}
	\includegraphics[width=0.11\linewidth, height=0.08\linewidth]{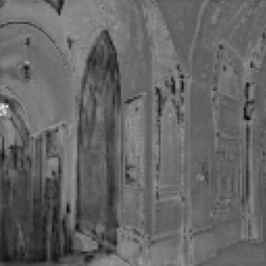}\hspace*{-0.1em}
	\includegraphics[width=0.11\linewidth, height=0.08\linewidth]{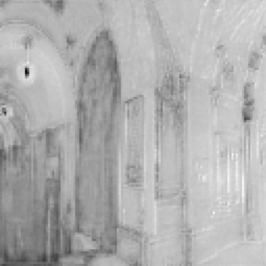}\hspace*{-0.1em}
	\includegraphics[width=0.11\linewidth, height=0.08\linewidth]{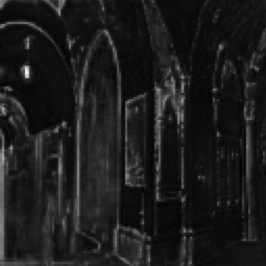}\hspace*{-0.1em}
	\includegraphics[width=0.11\linewidth, height=0.08\linewidth]{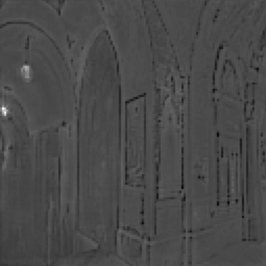}\hspace*{-0.1em}
	\setlength{\lineskip}{0pt}\\
	\includegraphics[width=0.11\linewidth, height=0.08\linewidth]{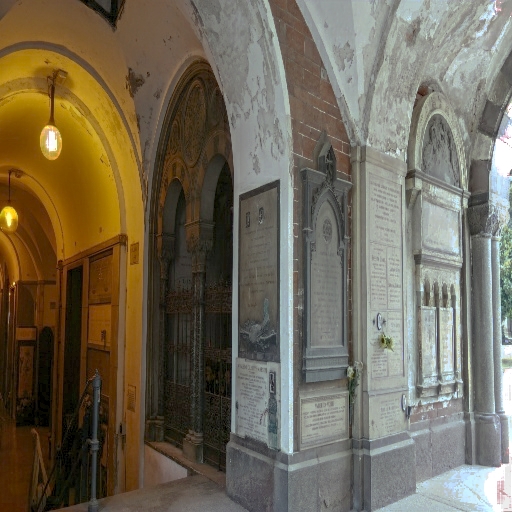}\hspace*{-0.1em}
	\includegraphics[width=0.11\linewidth, height=0.08\linewidth]{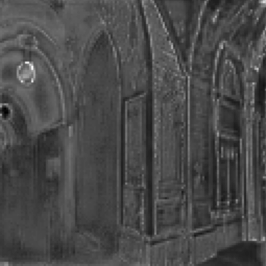}\hspace*{-0.1em}
	\includegraphics[width=0.11\linewidth, height=0.08\linewidth]{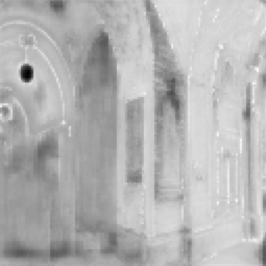}\hspace*{-0.1em}
	\includegraphics[width=0.11\linewidth, height=0.08\linewidth]{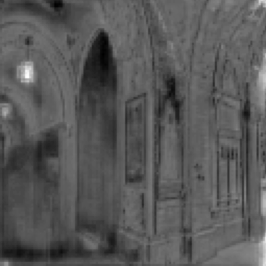}\hspace*{-0.1em}
	\includegraphics[width=0.11\linewidth, height=0.08\linewidth]{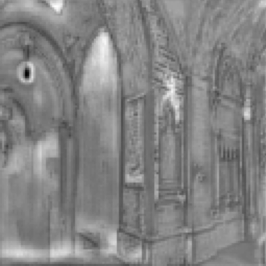}\hspace*{-0.1em}
	\includegraphics[width=0.11\linewidth, height=0.08\linewidth]{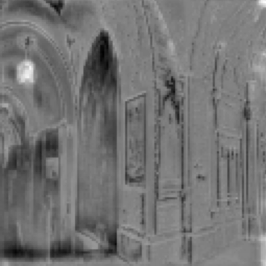}\hspace*{-0.1em}
	\includegraphics[width=0.11\linewidth, height=0.08\linewidth]{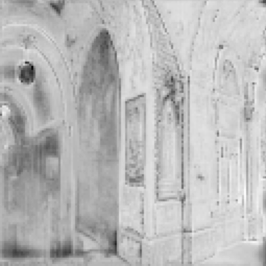}\hspace*{-0.1em}
	\includegraphics[width=0.11\linewidth, height=0.08\linewidth]{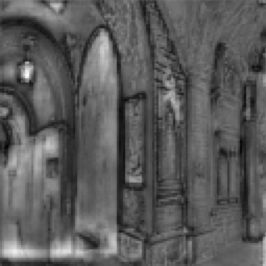}\hspace*{-0.1em}
	\includegraphics[width=0.11\linewidth, height=0.08\linewidth]{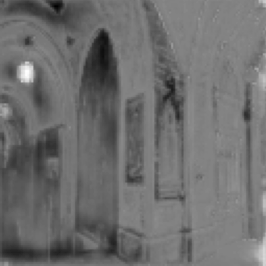}\hspace*{-0.1em}

	\caption{Illustration of feature maps learned at the Att.~4 layer of the MQN depicted in Figure \ref{fig:archs}. The rows show in order results obtained without using an attention mechanism, followed by using SA, CSA and CA. The first column shows the predicted HDR image $\hat{H}$ and the rest show the feature maps learned at different channels of the Att.~4 layer. \label{fig:attention}}
\vspace{-0.2cm}
\end{figure}

\begin{table}[!t]
	
	\centering
	\caption{(Left) Analyses of accuracy and latency measurements (on the SN20E990) for different attention mechanisms. None indicates that no attention mechanism is used and B indicates backbone. (Right) Analyses of accuracy for different loss combinations.\label{tab:attention}. Blue and red indicate the best and the second best accuracy, respectively.}

	\begin{adjustbox}{max width=0.45\textwidth}
	\begin{tabular}{|l|c|c|c|}
		\hline
		\textbf{Attention} &  \textbf{Latency B (ms}) & \textbf{\textit{PSNR}-$TM$} & \textbf{\textit{SSIM}-$TM$} \\
		\specialrule{.2em}{.2em}{.2em}
		None & \textcolor{blue}{11.55 $\pm$ 0.27} & 20.76 $\pm$ 3.21 & 0.8440 $\pm$ 0.0741 \\
		\hline
		SA & \textcolor{red}{11.68 $\pm$ 0.49} & \textcolor{red}{21.12 $\pm$ 3.03} & 0.8333 $\pm$ 0.0800 \\
		\hline
		CSA & 12.17 $\pm$ 0.25 & 20.75 $\pm$ 3.06  & 0.8559 $\pm$ 0.0597\\
		\hline
		CAB & 12.40 $\pm$ 0.32 & \textcolor{blue}{21.25 $\pm$ 3.11} & \textcolor{blue}{0.8782 $\pm$ 0.0520} \\
		\hline			
	\end{tabular}
	\end{adjustbox}
	\hspace{0.025\textwidth}
	\begin{adjustbox}{max width=0.45\textwidth}
	\begin{tabular}{|l|c|c|}
		\hline
		\textbf{Loss} &  \textbf{\textit{PSNR}-$TM$} & \textbf{\textit{SSIM}-$TM$} \\
		\specialrule{.2em}{.2em}{.2em}
		(i) $\mathcal{L}1$ & 20.00 $\pm$ 3.10 & 0.8295 $\pm$ 0.0646\\
		\hline
		(ii) $\mathcal{L}_1$, $\mathcal{L}_2$ & 19.96 $\pm$ 3.05 & 0.8281 $\pm$  0.0656\\
		\hline
		(iii) $\mathcal{L}_1$, $\mathcal{L}_2$, $\mathcal{L}_{CS}$  &  20.22 $\pm$ 3.02&  0.8268 $\pm$ 0.0625\\
		\hline
		(iv) $\mathcal{L}_1$,  $\mathcal{L}_2$, $\mathcal{L}_{CS}$, $\mathcal{L}_{FR}$ & 21.25 $\pm$ 3.11 & \textcolor{blue}{0.8782 $\pm$ 0.0520} \\
		\hline
		(v) $\mathcal{L}_1$, $\mathcal{L}_{FR}$ & 21.19 $\pm$ 2.85 & 0.8261 $\pm$ 0.0832 \\
		\hline
		(vi) $\mathcal{L}_2$, $\mathcal{L}_{FR}$ & \textcolor{red}{21.53 $\pm$ 2.92} & 0.8343 $\pm$ 0.0675 \\
		\hline
		(vii) $\mathcal{L}_1$, $\mathcal{L}_2$, $\mathcal{L}_{FR}$ & \textcolor{blue}{21.54 $\pm$ 2.85} & \textcolor{red}{0.8538 $\pm$ 0.0597} \\
		\hline
	\end{tabular}
	
	\end{adjustbox}
\vspace{-0.2cm}
\end{table}

\subsection{Ablation Studies}
\label{ab}
In this section, we study ablations with regard to various attention mechanisms, the quantization schemes, loss functions and deployment of MQN to hardware platforms for efficient mobile ITM.

\begin{figure}[!t]

    \setlength{\lineskip}{0pt}
	\includegraphics[width=0.22\linewidth, height=0.12\linewidth]{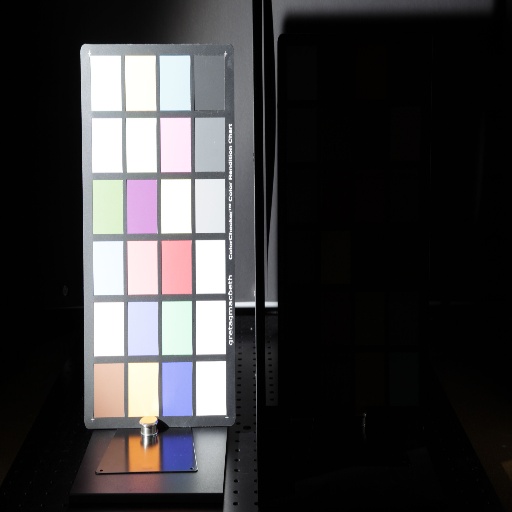}\hspace*{-0.1em}	
    \includegraphics[width=0.22\linewidth, height=0.12\linewidth]{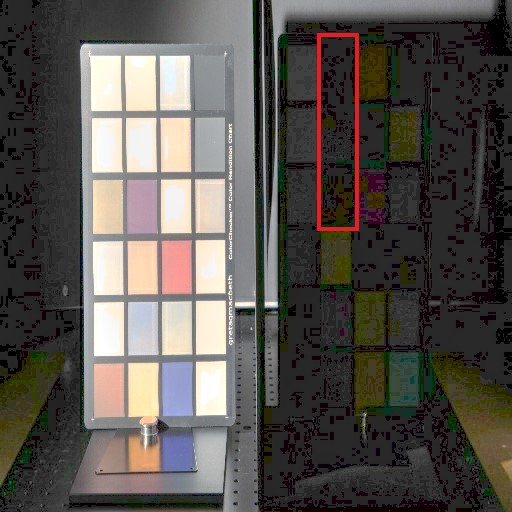}\hspace*{-0.1em}	
    \includegraphics[width=0.06\linewidth, height=0.12\linewidth]{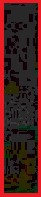}\hspace*{-0.1em}	
	\includegraphics[width=0.22\linewidth, height=0.12\linewidth]{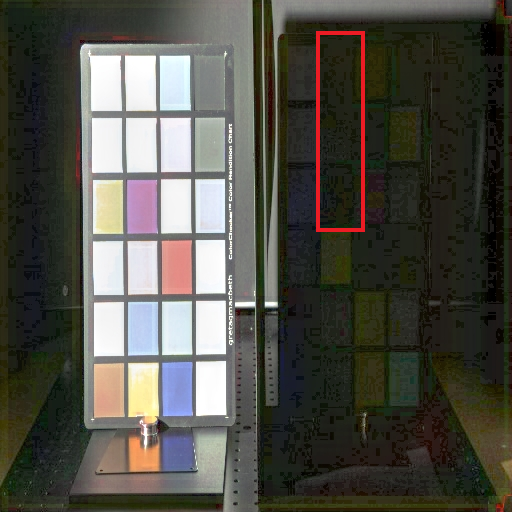}\hspace*{-0.1em}
	\includegraphics[width=0.06\linewidth, height=0.12\linewidth]{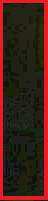}\hspace*{-0.1em}
	\includegraphics[width=0.22\linewidth, height=0.12\linewidth]{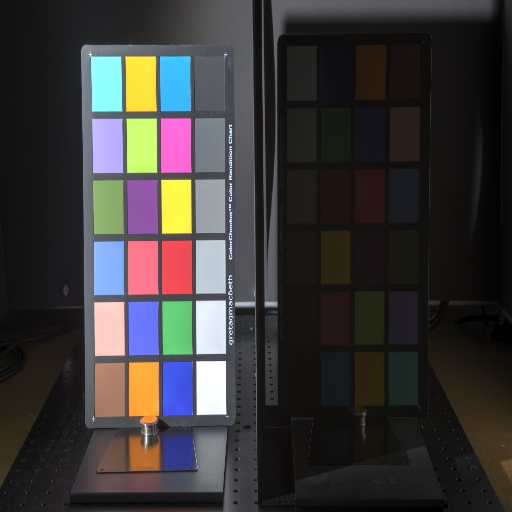}\hspace*{-0.1em}\\
   \setlength{\lineskip}{0pt}
	\includegraphics[width=0.22\linewidth, height=0.12\linewidth]{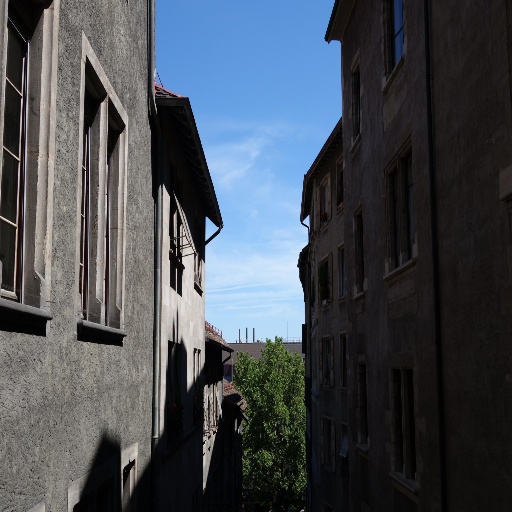}\hspace*{-0.1em}	
    \includegraphics[width=0.22\linewidth, height=0.12\linewidth]{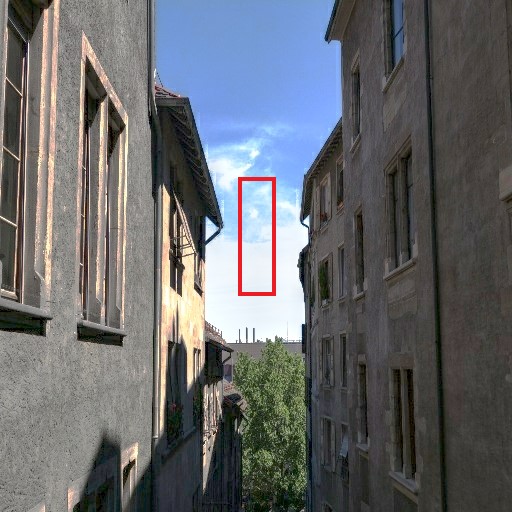}\hspace*{-0.1em}	
    \includegraphics[width=0.06\linewidth, height=0.12\linewidth]{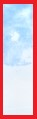}\hspace*{-0.1em}	
	\includegraphics[width=0.22\linewidth, height=0.12\linewidth]{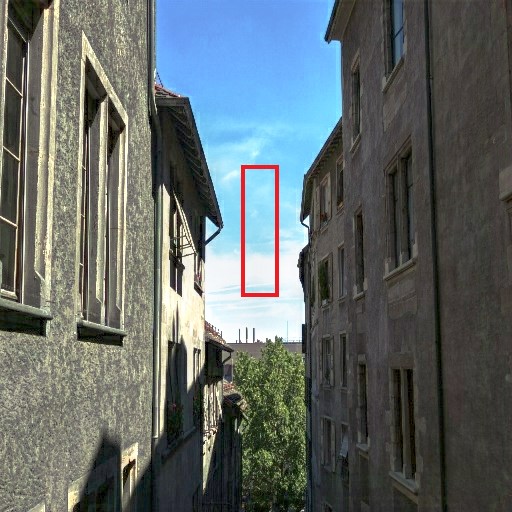}\hspace*{-0.1em}
	\includegraphics[width=0.06\linewidth, height=0.12\linewidth]{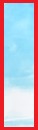}\hspace*{-0.1em}
	\includegraphics[width=0.22\linewidth, height=0.12\linewidth]{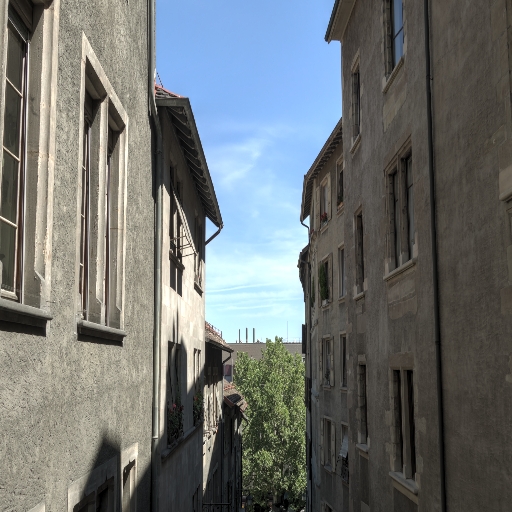}\hspace*{-0.1em}
	
	    \hspace{0.06\linewidth} Input $I$. \hspace{0.13\linewidth} $\hat{H}$ wo/$\mathcal{L}_{FR}$. \hspace{0.13\linewidth} $\hat{H}$ w/$\mathcal{L}_{FR}$. \hspace{0.13\linewidth} Ground-truth ${H}$.
	\caption{Visual analyses of the effect of training models with and without using $\mathcal{L}_{FR}$ on predictions $\hat{H}$. It helps models gain structural  coherency and improve color details.}
	\label{fig:losses}
\vspace{-0.2cm}
\end{figure}

\begin{table}[!t]
	
	\centering
	\caption{ Results obained using different quantization schemes. Latency (L.) is measured on the deployment platform (SN20E990).  \label{tab:sr}}
    
    \begin{adjustbox}{max width=0.8\textwidth}
	\begin{tabular}{|l|l|c|c|c|c|}
		\hline
		\textbf{Backbone} & \textbf{Head} & \textbf{L. Backbone} (ms) & \textbf{L. Head} (ms) &  \textbf{\textit{PSNR}-$T$} & \textbf{\textit{SSIM}-$T$} \\
		\specialrule{.2em}{.2em}{.2em}
		Quant. & Dynamic & 11.52 $\pm$ 0.82 & 9.63 $\pm$ 0.12 & 21.25 $\pm$ 3.11 & 0.8782 $\pm$ 0.05\\
		\hline
		Quant. & Float32 & 11.52 $\pm$ 0.82 & 20.95 $\pm$ 0.99 & 21.34 $\pm$ 3.08 & 0.8793 $\pm$ 0.05\\
		\hline
		Float32 & Float32 & 21.13 $\pm$ 0.42 & 20.95 $\pm$ 0.99 & 21.58 $\pm$ 3.14 & 0.8727 $\pm$ 0.05\\
	    \hline
	\end{tabular}
	\end{adjustbox}
\vspace{-0.2cm}
\end{table}

\textbf{Attention Mechanisms}. We explore different attention mechanisms specified in Section~\ref{BA}. Results given in Table~\ref{tab:attention} show that accuracy increases when moving from SA to CAB, which provides the best accuracy with an increase of $3\%$ on SSIM, albeit with an increase in latency of  $\approx 1$ ms. We examine the features learned using different attention mechanisms in  Figure~\ref{fig:attention}. We observe that better feature representations of edges and surfaces are learned  when models are trained using CSA and CA. For instance, in both cases, the lamp is well captured with large feature activation values, enabling the dimming effect in the prediction.

\begin{figure}[!t]
	\setlength{\lineskip}{0pt}
	\includegraphics[width=0.16\linewidth, height=0.1\linewidth]{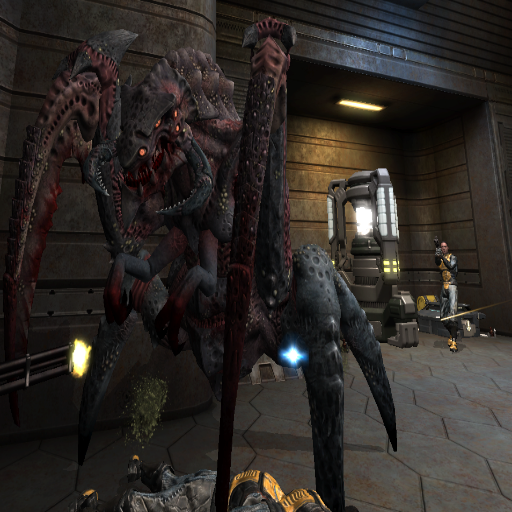}\hspace*{-0.1em}
	\includegraphics[width=0.16\linewidth, height=0.1\linewidth]{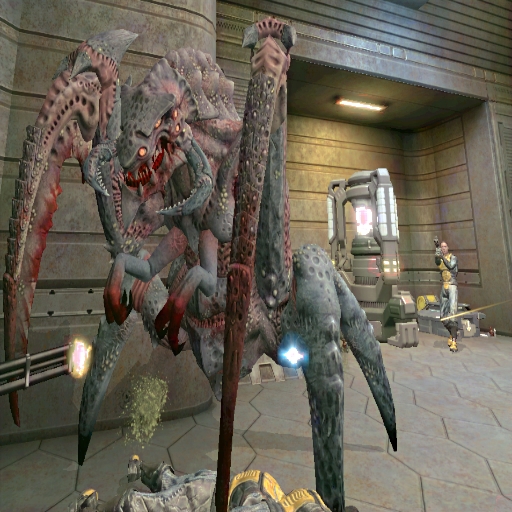}\hspace*{-0.1em}
	\includegraphics[width=0.16\linewidth, height=0.1\linewidth]{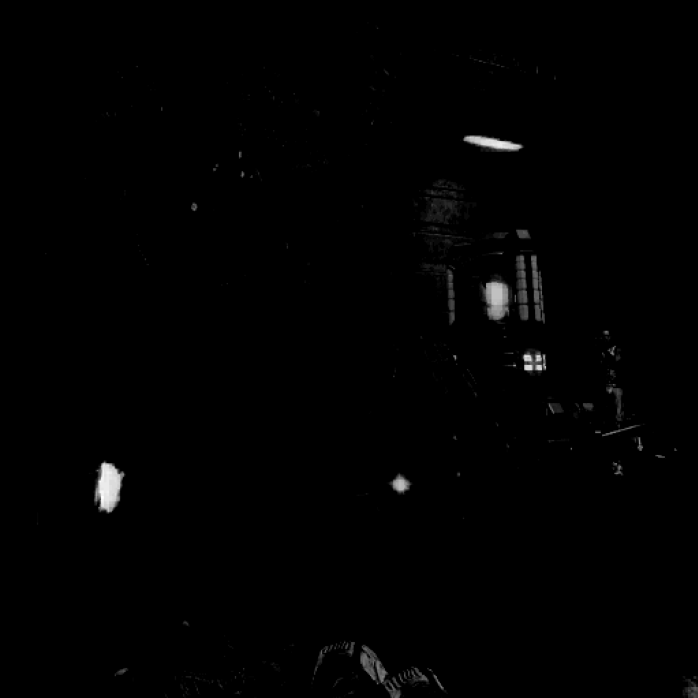}\hspace*{-0.1em}
	\includegraphics[width=0.16\linewidth, height=0.1\linewidth]{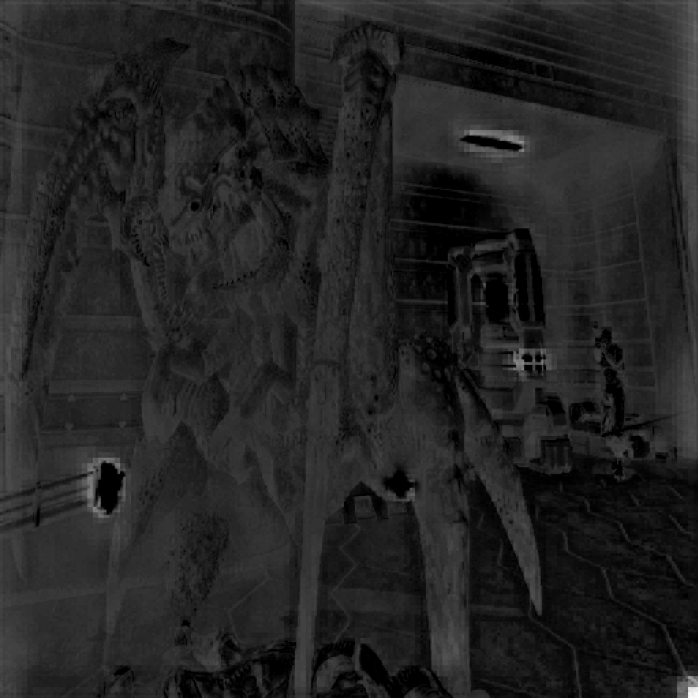}\hspace*{-0.1em}
	\includegraphics[width=0.16\linewidth, height=0.1\linewidth]{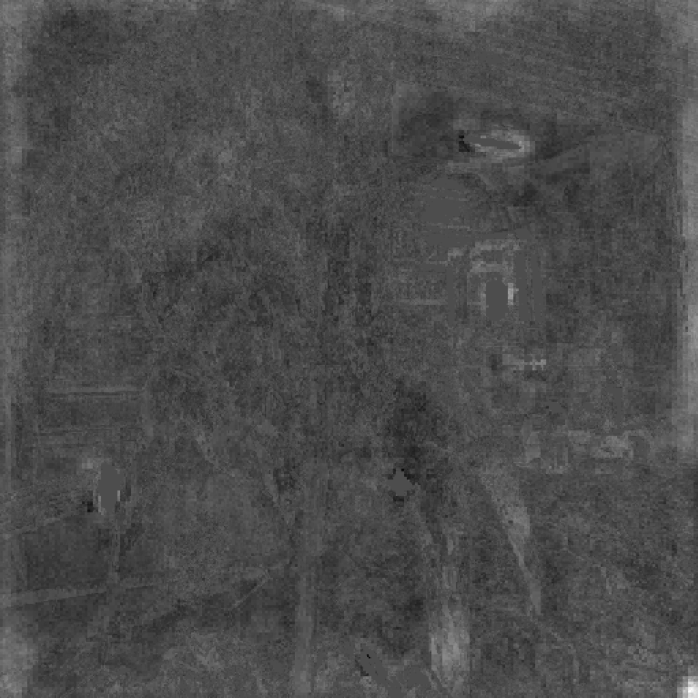}\hspace*{-0.1em}
	\includegraphics[width=0.16\linewidth, height=0.1\linewidth]{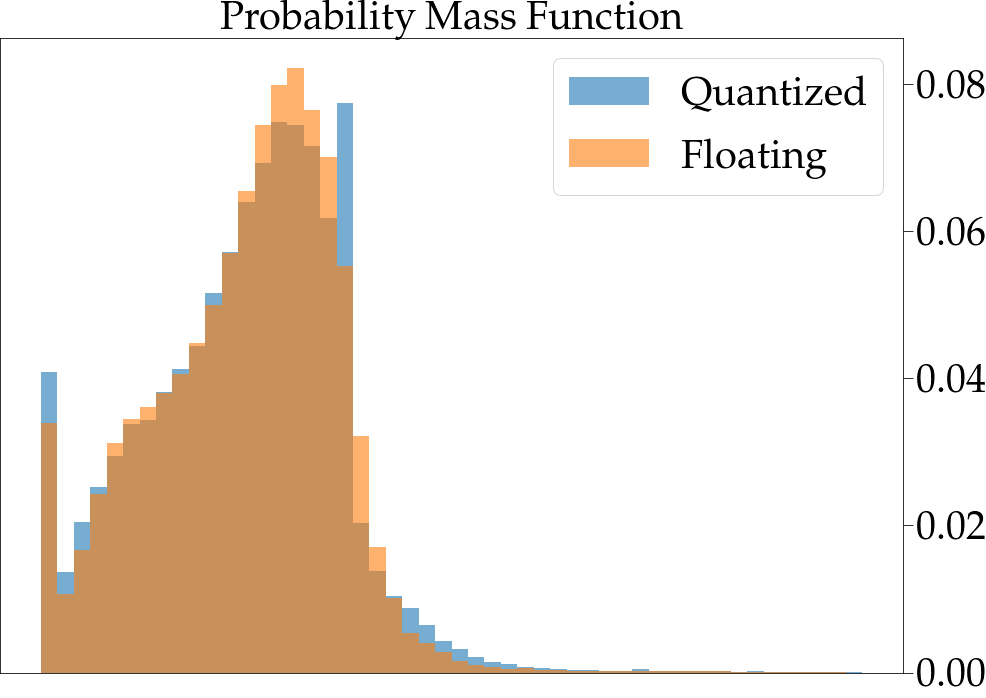}\hspace*{-0.1em}\\
	\includegraphics[width=0.16\linewidth, height=0.1\linewidth]{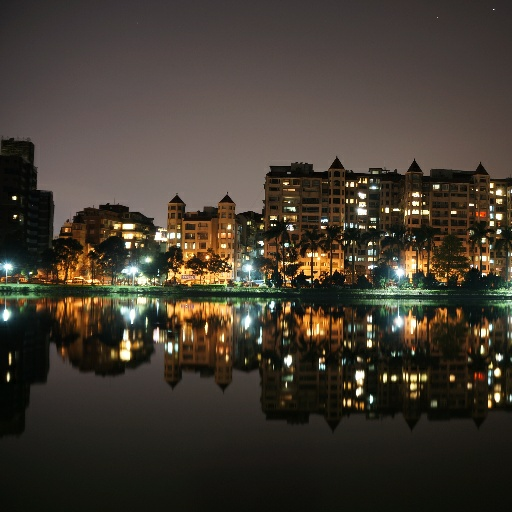}\hspace*{-0.1em}
	\includegraphics[width=0.16\linewidth, height=0.1\linewidth]{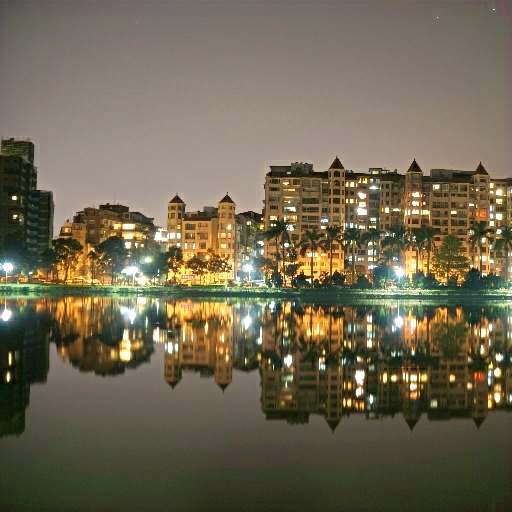}\hspace*{-0.1em}
	\includegraphics[width=0.16\linewidth, height=0.1\linewidth]{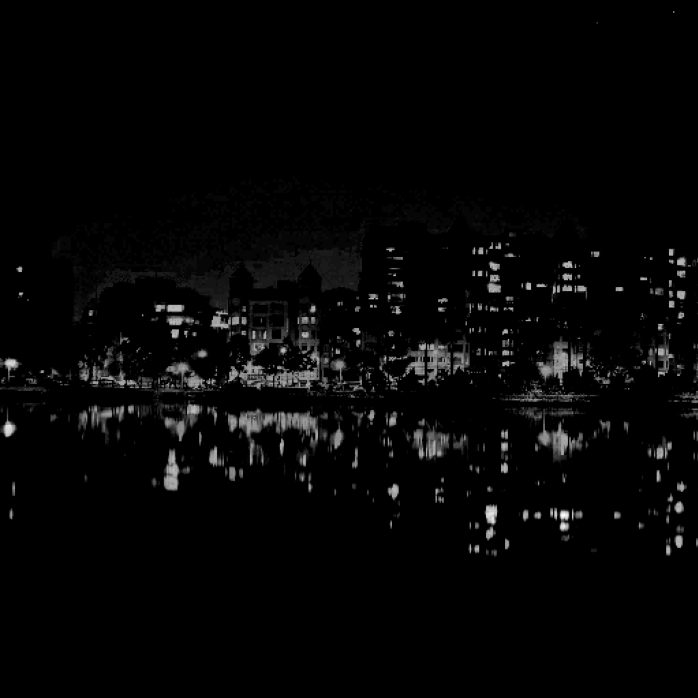}\hspace*{-0.1em}
	\includegraphics[width=0.16\linewidth, height=0.1\linewidth]{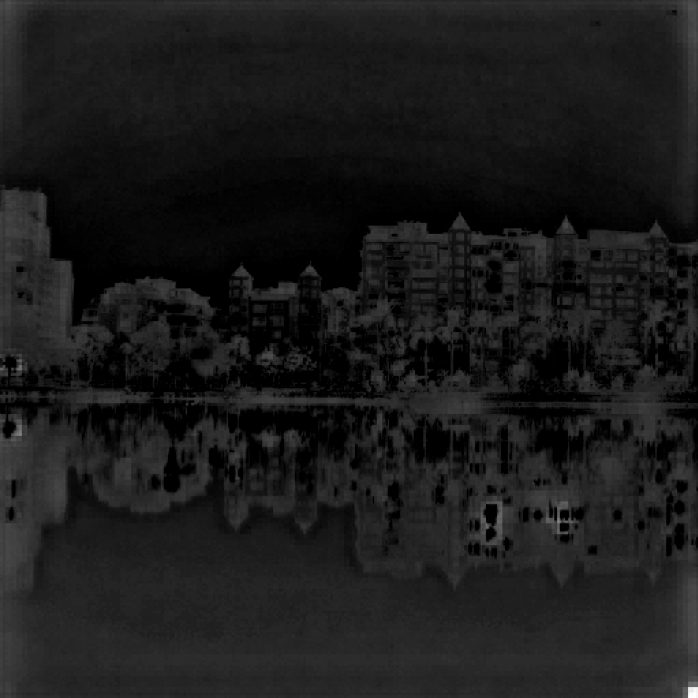}\hspace*{-0.1em}
	\includegraphics[width=0.16\linewidth, height=0.1\linewidth]{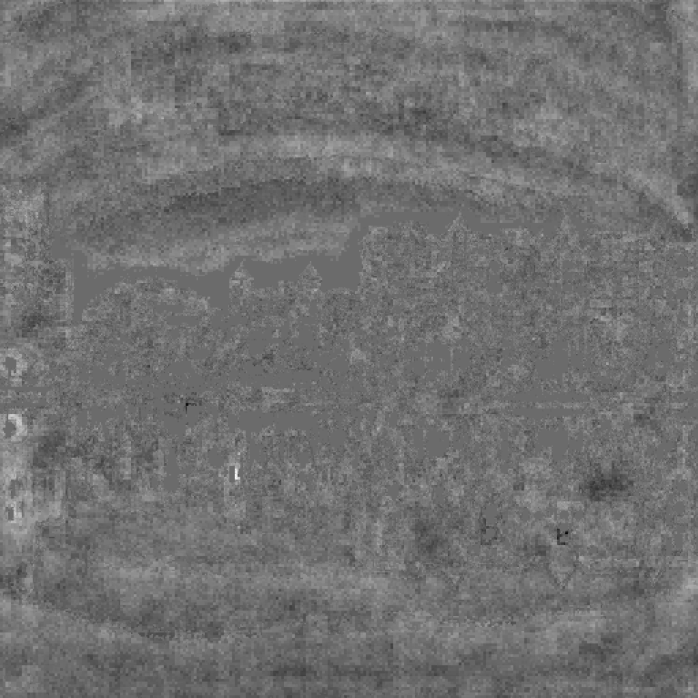}\hspace*{-0.1em}
	\includegraphics[width=0.16\linewidth, height=0.1\linewidth]{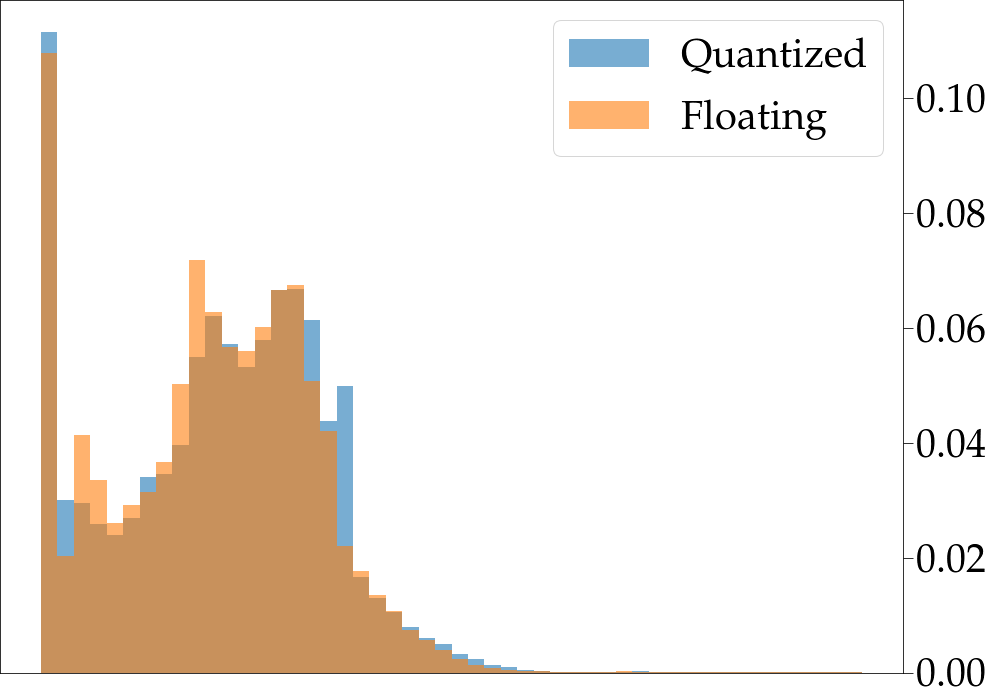}\hspace*{-0.1em}\\
	\includegraphics[width=0.16\linewidth, height=0.1\linewidth]{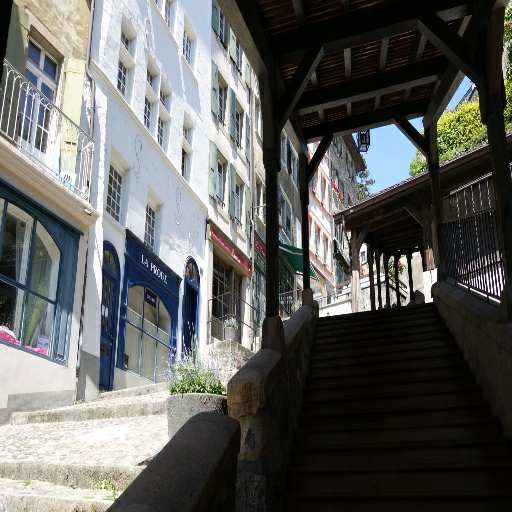}\hspace*{-0.1em}
	\includegraphics[width=0.16\linewidth, height=0.1\linewidth]{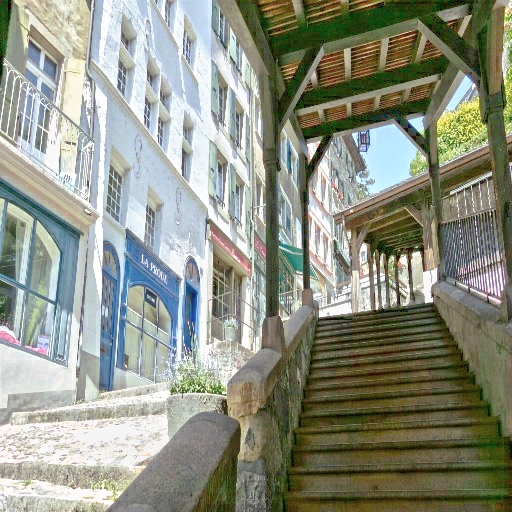}\hspace*{-0.1em}
	\includegraphics[width=0.16\linewidth, height=0.1\linewidth]{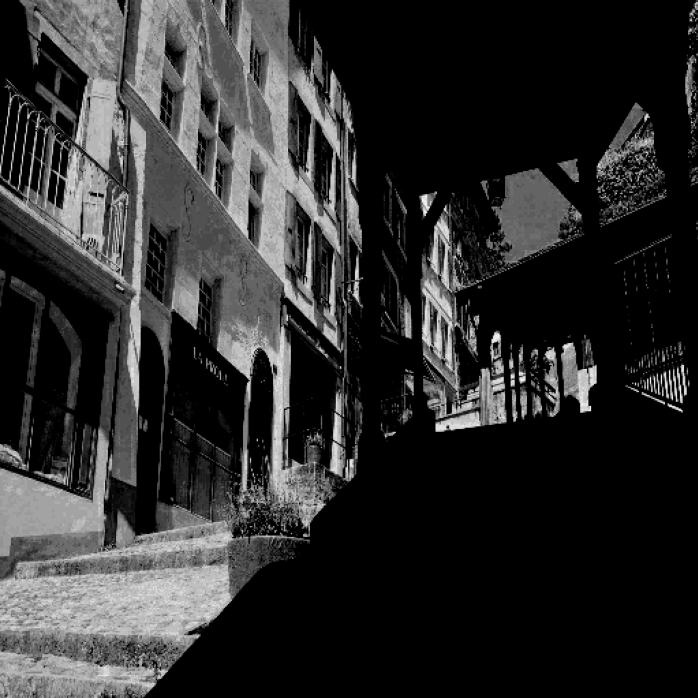}\hspace*{-0.1em}
	\includegraphics[width=0.16\linewidth, height=0.1\linewidth]{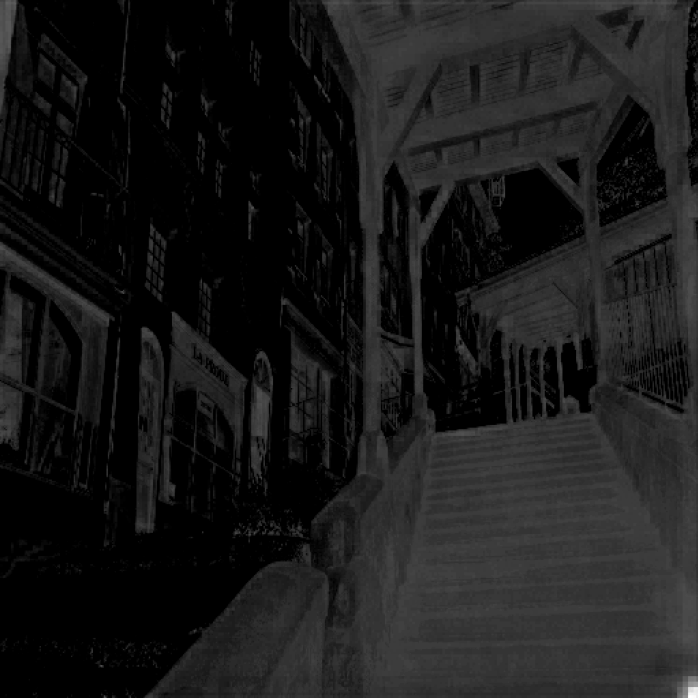}\hspace*{-0.1em}
	\includegraphics[width=0.16\linewidth, height=0.1\linewidth]{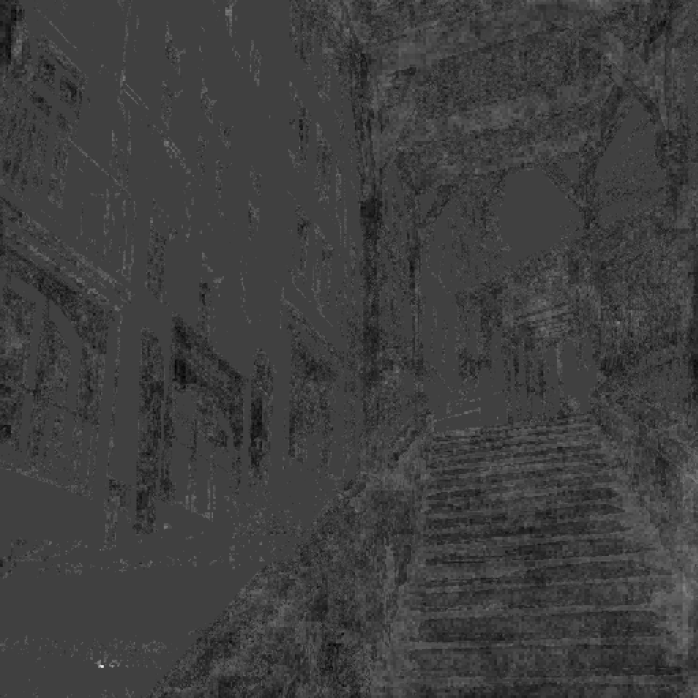}\hspace*{-0.1em}
	\includegraphics[width=0.16\linewidth, height=0.1\linewidth]{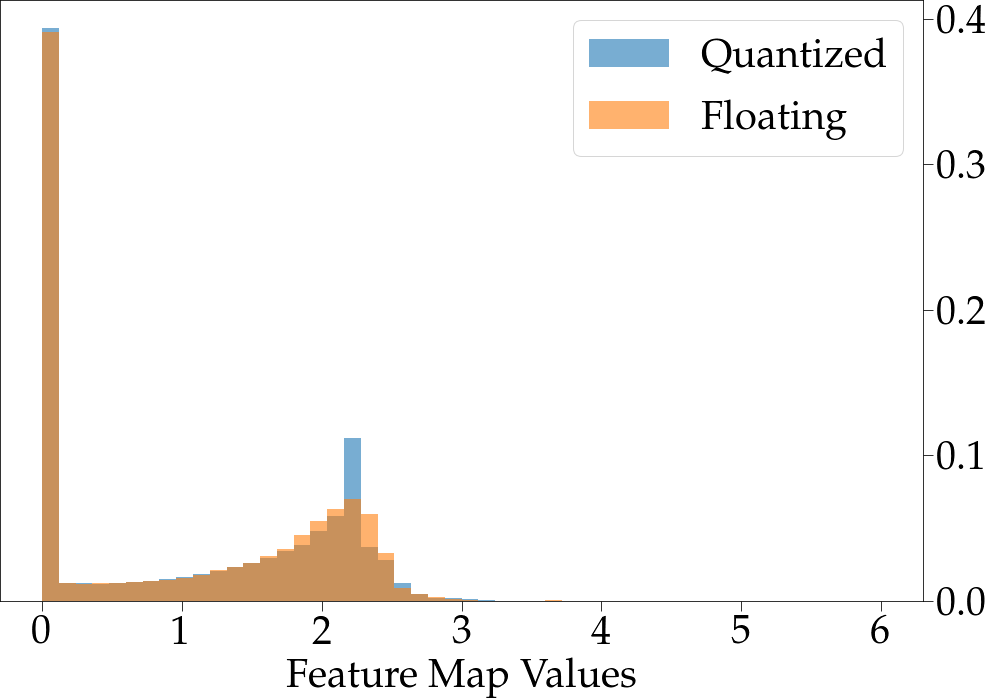}\hspace*{-0.1em}

	\hspace{0.05\linewidth} a) $I$. \hspace{0.10\linewidth}  b) $\hat{H}$. \hspace{0.09\linewidth}   c) $f_1$. \hspace{0.09\linewidth}  d) $f_2$. \hspace{0.08\linewidth}  e) $\Delta f_2$. \hspace{0.06\linewidth} f) PMFs.
	\caption{Analyses of the quantized $Q(\mathbf{f})$ and floating point $F(\mathbf{f})$ features $\mathbf{f}$ learned by the MQN at the ConvBnReLU3 layer at Figure \ref{fig:archs} using a) three sample input images with b) predictions $\hat{H}$. Visualization of (c) the first channel $f_1$ and (d) the second channel $f_2$ of $\mathbf{f}$, (e) the difference map $\Delta f_2 = \| Q(f_2) - F(f_2)\|_1$. We show the probability mass function (PMF) of $Q(f_2)$ and $F(f_2)$ in (f). \label{fig:interface}}
\vspace{-0.5cm}
\end{figure}

\textbf{Quantization Schemes}. Next, we study the behavior of features $\mathbf{f}$ learned at the interface between the backbone and the head (i.e., at the ConvBnReLU 3 layer depicted in Figure \ref{fig:archs}), as well as how quantization affects the interface and the head. In Table~\ref{tab:sr}, we analyze how the performance and latency of models change for different  quantization methods. The results show that the proposed quantization scheme enables to obtain similar PSNR/SSIM accuracy whilst showing improvements in latency.

In order to analyze the effect of quantization on statistical properties of features, we compute histograms approximating probability mass functions (PMFs) of features and their quantized versions.  The results given in Figure \ref{fig:interface}.e show that distributions of quantized features $Q(\mathbf{f})$ and features with floating point values $F(\mathbf{f})$ have similar distributions.  This result suggests that the quantization scheme preserves statistical information of features. 

We also study what is learned in the interface as well as the effect of the network on a computer graphics image. In Figure \ref{fig:interface} (c and d), we present two feature maps $f_1$ and $f_2$ corresponding to two channels of $\mathbf{f}$. The maps show that two very different representations are learned in these channels: in (c) light sources are identified, striking its relevancy for the ITM task, while in (d) general edge structures are learned. Moreover, we can see in the first row, as well as in Figure \ref{fig:example}, that the network can be applied to computer graphics images without having special artifacts or distortions, opening the door for employment of MQN in computer graphics.

\textbf{Loss functions}. We analyze the effect of using different loss functions defined in Section~\ref{losses} in Table~\ref{tab:attention}. The results show that employing $\mathcal{L}_{FR}$ with $\mathcal{L}_{CS}$ and $\mathcal{L}_{2}$ increases accuracy substantially.  Meanwhile, $\mathcal{L}_{CS}$ and $\mathcal{L}_{2}$ seem to provide a slightly negative effect on their own. Moreover, as observed in Figure~\ref{fig:losses}, adding the FR loss to $\mathcal{L}_{1}$, $\mathcal{L}_{2}$, and $\mathcal{L}_{CS}$ helps improve color details and structural coherency.

\begin{table}[!t]
	\caption{Comparison with other state-of-the-art single image HDR reconstruction methods.  Performance metric and latency values reproduced with the same evaluation criteria and original codes. Blue and red indicates the best and second best accuracy. P. indicates number of parameters, L. M. indicates latency for mobile, M. RAM the maximum RAM memory consumed by the model, and O. the numer of operations in multiply-accumulate units. Performance values are given in HDRVDP-Q score. *FHDR\cite{khan_fhdr_2019} uses recurrence: the present value is computed taking into account two iterations.}\label{tab:final}
	\centering
	\small
	\resizebox{\textwidth}{!}{%
		\begin{tabular}{|c|c|c|c|c|c|c|c|c|}
			\hline
			\textbf{Model}  &  \textbf{P. (M)} & \textbf{L. GPU (ms)}  &\textbf{L. M. (ms)} & \textbf{O. (GMAC)} &\textbf{M. RAM (MB)} & \textbf{HDR-Eye} &  \textbf{Raise-1K} &\textbf{HDR-Real}\\
			\hline 

			HDRCNN \cite{eilertsen_hdr_2017} & 29.44 &  247 & - & 30.35 & - & 51.16 $\pm$ 4.43 &   51.89 $\pm$ 2.77 & 45.56 $\pm$	8.18\\
			\hline
			SingleHDR \cite{liu_single-image_2020} & 29.01 & 976 & - & 112.75 & -& \textcolor{blue}{53.05 $\pm$ 5.08} &  51.69 $\pm$ 2.56 & \textcolor{blue}{48.72 $\pm$ 4.03} \\
			\hline
			FHDR \cite{khan_fhdr_2019} & 0.571 & 54 & 4970 $\pm$ 434 & 72.34* & 832.40 & 51.41 $\pm$ 6.72 & \textcolor{blue}{53.13 $\pm$ 1.71}  & \textcolor{red}{45.82 $\pm$ 8.67}  \\
			\hline
			 HDRUnet \cite{chen_hdrunet_2021} & 1.651  & \textcolor{red}{17} & 808 $\pm$ 22 & 23.42 & 353.46 &  50.32 $\pm$ 4.07
			 &     51.42 $\pm$ 3.35  & 44.60 $\pm$ 7.30   \\
			\hline
			ExpandNet \cite{marnerides_expandnet_2018} & \textcolor{blue}{0.45} & 21 & 474 $\pm$ 7 & 13.66 & 262.77 & 50.52 $\pm$ 3.94 &  51.83 $\pm$ 1.68 & 44.86 $\pm$ 8.21\\

			\hline
			 DeepHDR \cite{santos_single_2020} & 51.545 & \textcolor{red}{17} & \textcolor{red}{238 $\pm$ 4} & 18.94 & \textcolor{red}{251.17} & 51.11 $\pm$ 4.45 &    51.66 $\pm$ 2.79 &  45.81 $\pm$ 8.34\\
			\hline
			 TwoStage \cite{sharif_two-stage_2021} & 1.088 & 32 & 3338 $\pm$ 456  & 54.91 & 397.41 & 49.68 $\pm$ 3.7 & \textcolor{red}{52.95 $\pm$ 2.36} &  43.46 $\pm$    7.55 \\
			\specialrule{.15em}{.15em}{.15em}
			Ours (best) & \textcolor{red}{0.928} & \textcolor{blue}{11} & \textcolor{blue}{21 $\pm$ 1} & 0.5 & \textcolor{blue}{9.45} & \textcolor{red}{51.59 $\pm$ 4.61} & 51.81 $\pm$ 1.56 & 45.15 $\pm$ 8.15 \\
			\hline
	\end{tabular}}
\vspace{-0.4cm}
\end{table}

\textbf{Deployment Platforms}. In the experimental analyses, we used CPU as our main deployment hardware platform. However, as our objective has been to develop a well performing but efficient model employing mixed quantization, attention, and efficient operations, our model can also be extended to other hardware platforms. For this reason, we also deploy our model to the GPU (Arm $Mali^{TM}$-G77 MP11) of SN20E990, elucidating the flexibility of our model with regards to deployment of MQN models on platforms with different hardware configurations. We obtained a latency of 10.5 ms for our backbone and  9.3 ms for our high precision head, improving our results even further and showcasing the hardware flexibility for implementation of our MQN.

\begin{figure}[!t]		

	\setlength{\lineskip}{0pt}
	\includegraphics[width=0.2\linewidth, height=0.14\linewidth]{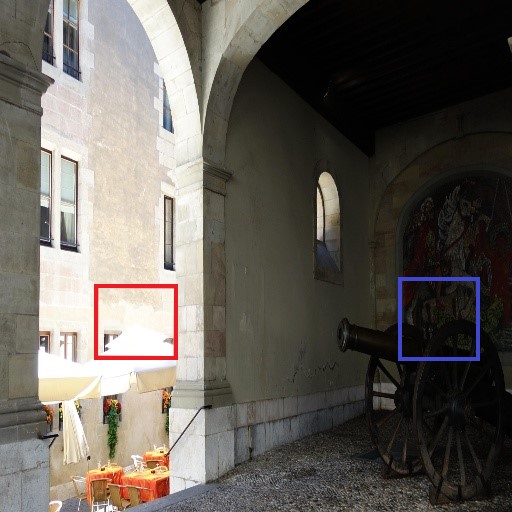}\hspace*{-0.1em}
	\includegraphics[width=0.2\linewidth, height=0.14\linewidth]{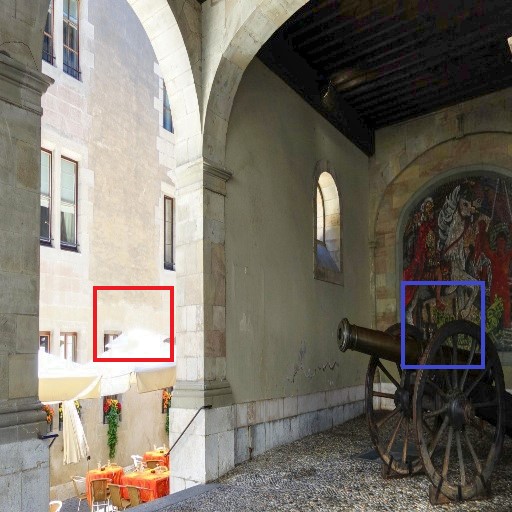}\hspace*{-0.1em}
	\includegraphics[width=0.2\linewidth, height=0.14\linewidth]{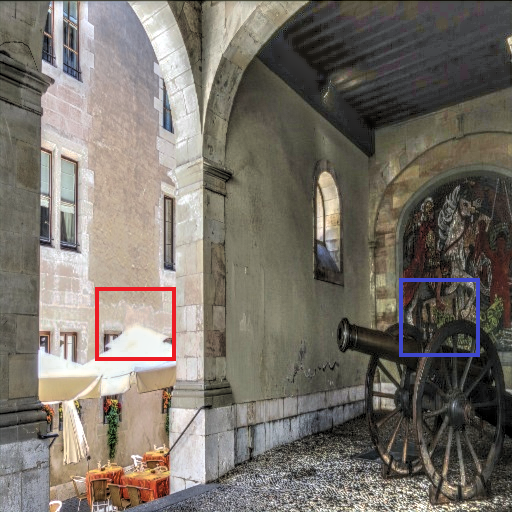}\hspace*{-0.1em}
	\includegraphics[width=0.2\linewidth, height=0.14\linewidth]{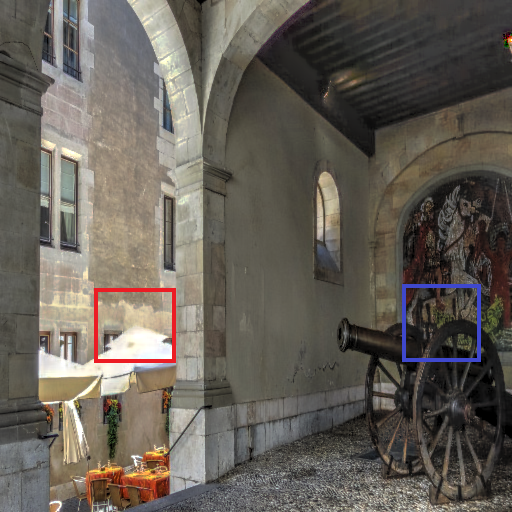}\hspace*{-0.1em}
	\includegraphics[width=0.2\linewidth, height=0.14\linewidth]{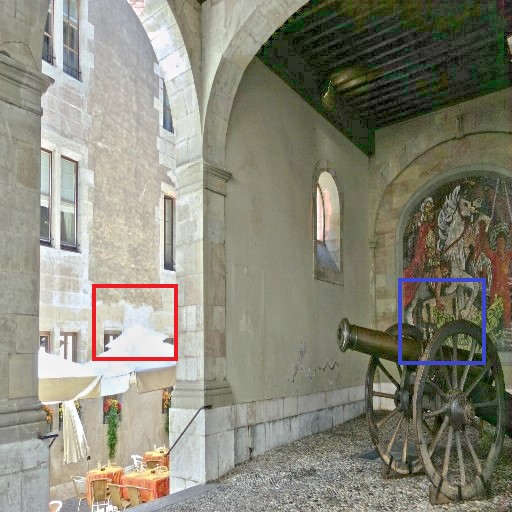}\hspace*{-0.1em}
\\
\setlength{\lineskip}{1pt}
	\includegraphics[width=0.1\linewidth, height=0.06\linewidth]{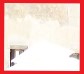}\hspace*{-0.2em}
	\includegraphics[width=0.1\linewidth, height=0.06\linewidth]{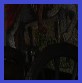}\hspace*{-0.1em}
	\includegraphics[width=0.1\linewidth, height=0.06\linewidth]{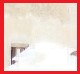}\hspace*{-0.2em}
	\includegraphics[width=0.1\linewidth, height=0.06\linewidth]{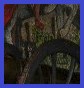}\hspace*{-0.1em}
	\includegraphics[width=0.1\linewidth, height=0.06\linewidth]{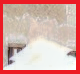}\hspace*{-0.2em}
	\includegraphics[width=0.1\linewidth, height=0.06\linewidth]{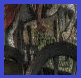}\hspace*{-0.1em}
	\includegraphics[width=0.1\linewidth, height=0.06\linewidth]{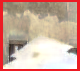}\hspace*{-0.2em}
	\includegraphics[width=0.1\linewidth, height=0.06\linewidth]{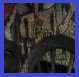}\hspace*{-0.1em}
	\includegraphics[width=0.1\linewidth, height=0.06\linewidth]{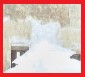}\hspace*{-0.2em}
	\includegraphics[width=0.1\linewidth, height=0.06\linewidth]{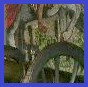}\hspace*{-0.1em}\\
	\setlength{\lineskip}{0pt}
	\includegraphics[width=0.2\linewidth, height=0.14\linewidth]{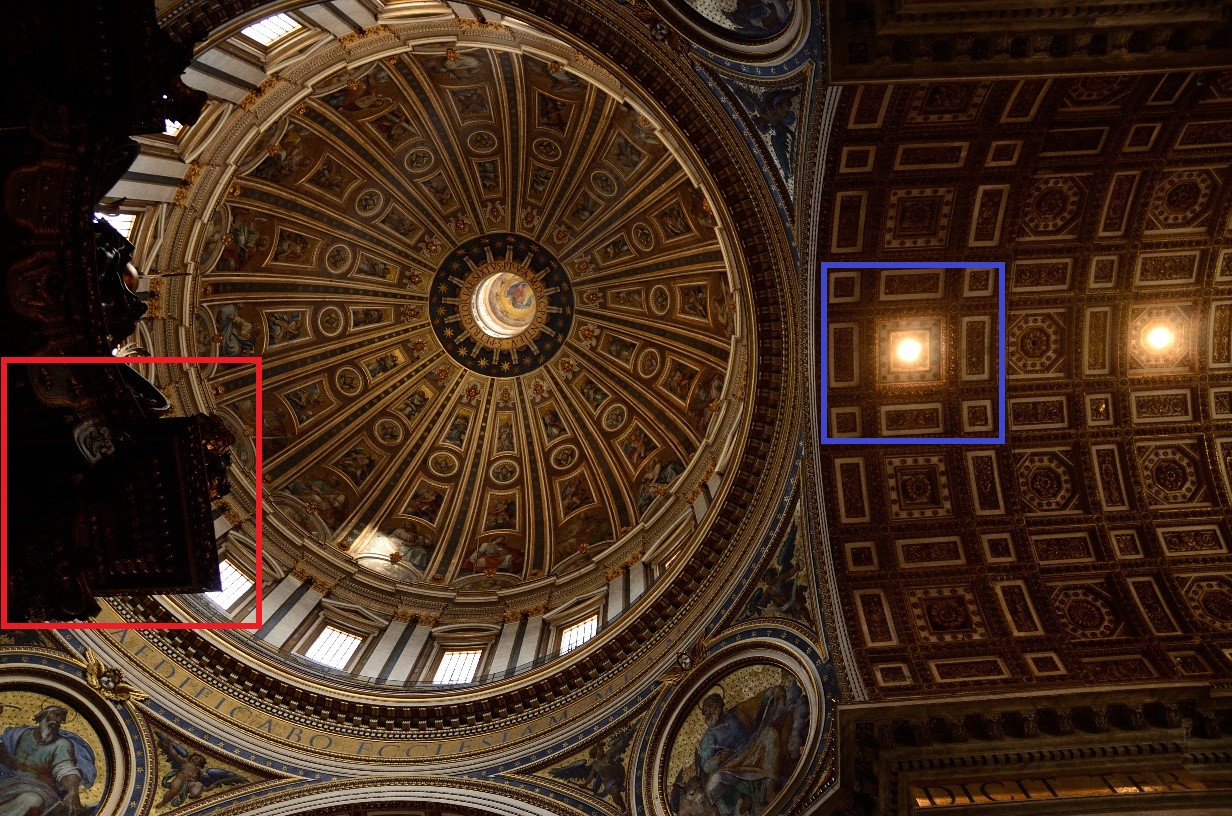}\hspace*{-0.1em}
	\includegraphics[width=0.2\linewidth, height=0.14\linewidth]{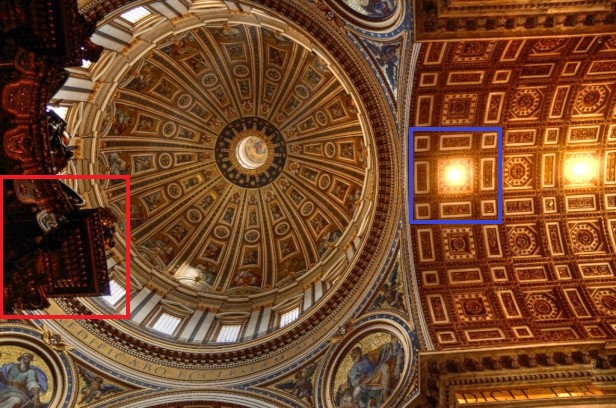}\hspace*{-0.1em}
	\includegraphics[width=0.2\linewidth, height=0.14\linewidth]{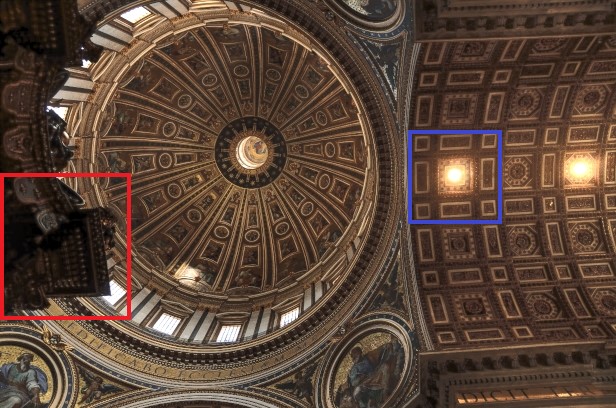}\hspace*{-0.1em}
	\includegraphics[width=0.2\linewidth, height=0.14\linewidth]{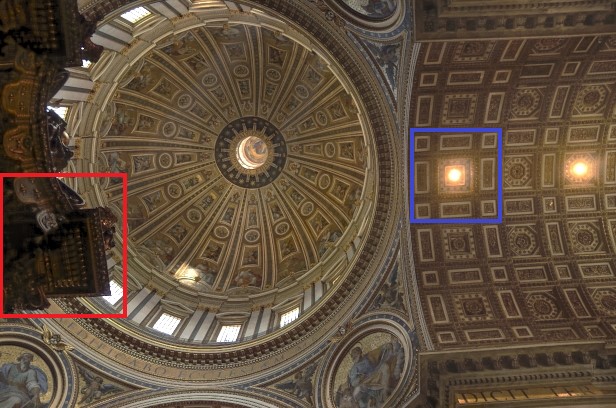}\hspace*{-0.1em}
	\includegraphics[width=0.2\linewidth, height=0.14\linewidth]{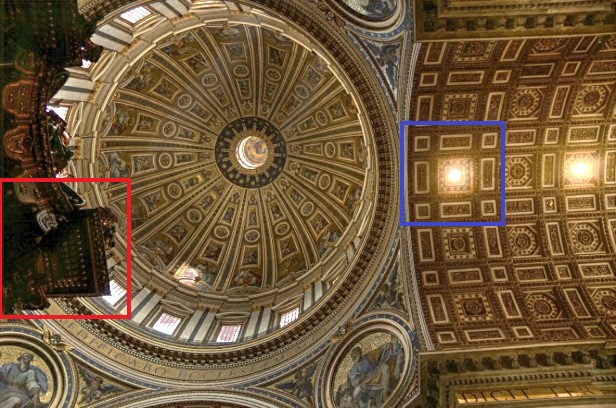}\hspace*{-0.1em}\\
    \setlength{\lineskip}{1pt}
	\includegraphics[width=0.1\linewidth, height=0.06\linewidth]{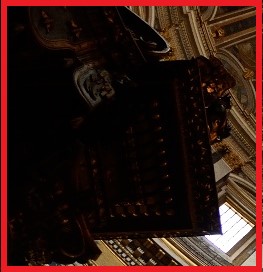}\hspace*{-0.2em}
	\includegraphics[width=0.1\linewidth, height=0.06\linewidth]{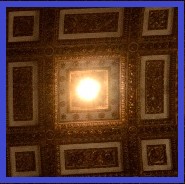}\hspace*{-0.1em}
	\includegraphics[width=0.1\linewidth, height=0.06\linewidth]{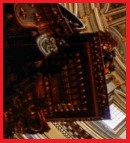}\hspace*{-0.2em}
	\includegraphics[width=0.1\linewidth, height=0.06\linewidth]{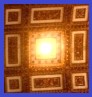}\hspace*{-0.1em}
	\includegraphics[width=0.1\linewidth, height=0.06\linewidth]{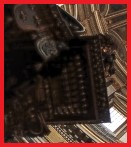}\hspace*{-0.2em}
	\includegraphics[width=0.1\linewidth, height=0.06\linewidth]{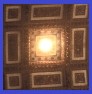}\hspace*{-0.1em}
	\includegraphics[width=0.1\linewidth, height=0.06\linewidth]{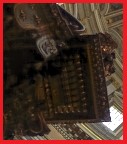}\hspace*{-0.2em}
	\includegraphics[width=0.1\linewidth, height=0.06\linewidth]{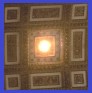}\hspace*{-0.1em}
	\includegraphics[width=0.1\linewidth, height=0.06\linewidth]{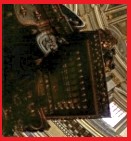}\hspace*{-0.2em}
	\includegraphics[width=0.1\linewidth, height=0.06\linewidth]{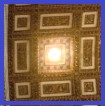}\hspace*{-0.1em}\\
	\setlength{\lineskip}{1pt}
	\includegraphics[width=0.2\linewidth, height=0.14\linewidth]{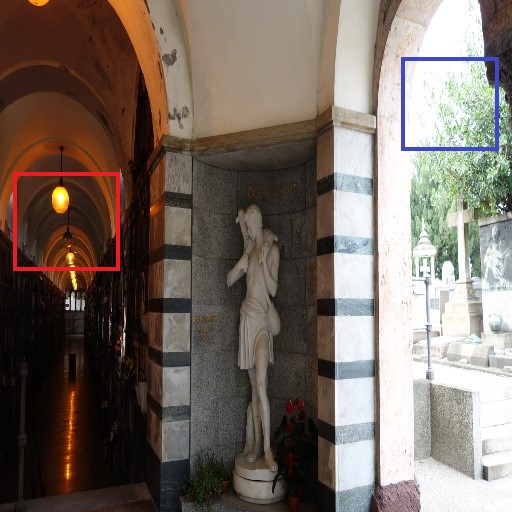}\hspace*{-0.1em}
	\includegraphics[width=0.2\linewidth, height=0.14\linewidth]{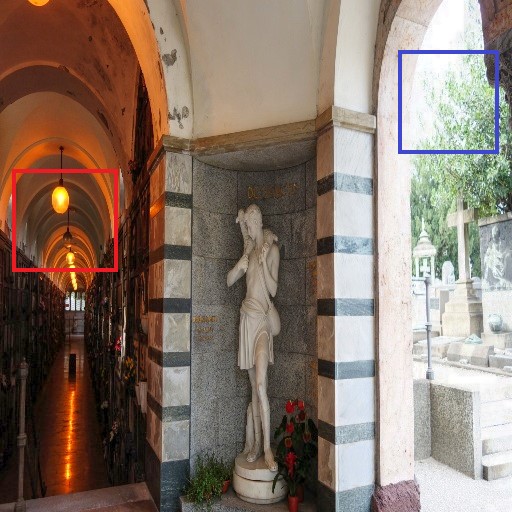}\hspace*{-0.1em}
	\includegraphics[width=0.2\linewidth, height=0.14\linewidth]{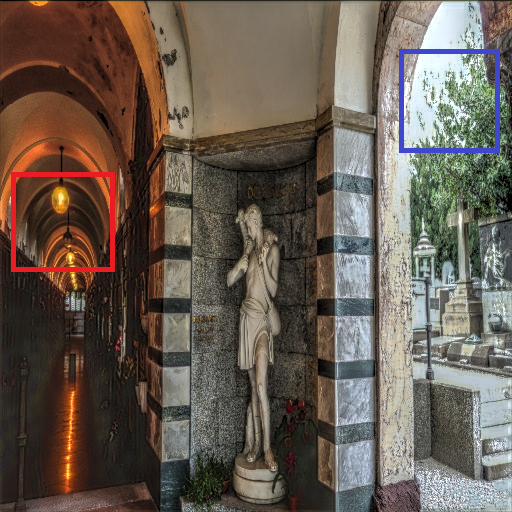}\hspace*{-0.1em}
	\includegraphics[width=0.2\linewidth, height=0.14\linewidth]{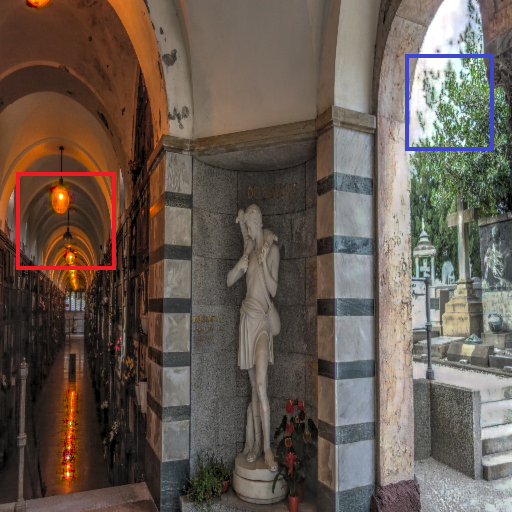}\hspace*{-0.1em}
	\includegraphics[width=0.2\linewidth, height=0.14\linewidth]{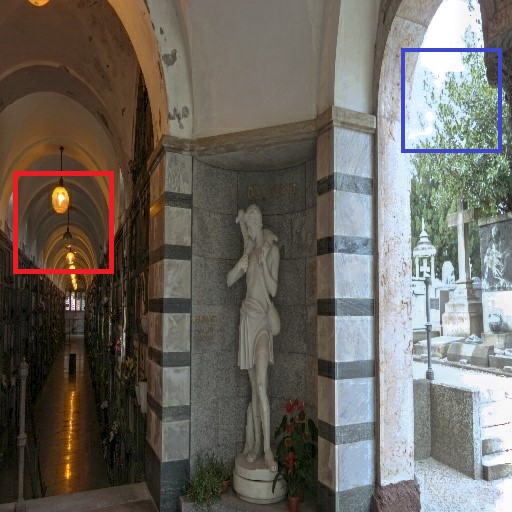}\hspace*{-0.1em}\\
	\includegraphics[width=0.1\linewidth, height=0.06\linewidth]{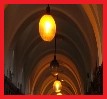}\hspace*{-0.2em}
	\includegraphics[width=0.1\linewidth, height=0.06\linewidth]{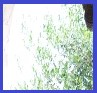}\hspace*{-0.1em}
	\includegraphics[width=0.1\linewidth, height=0.06\linewidth]{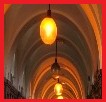}\hspace*{-0.2em}
	\includegraphics[width=0.1\linewidth, height=0.06\linewidth]{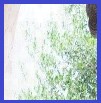}\hspace*{-0.1em}
	\includegraphics[width=0.1\linewidth, height=0.06\linewidth]{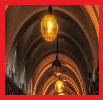}\hspace*{-0.2em}
	\includegraphics[width=0.1\linewidth, height=0.06\linewidth]{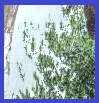}\hspace*{-0.1em}
	\includegraphics[width=0.1\linewidth, height=0.06\linewidth]{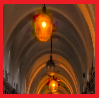}\hspace*{-0.2em}
	\includegraphics[width=0.1\linewidth, height=0.06\linewidth]{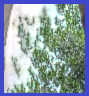}\hspace*{-0.1em}
	\includegraphics[width=0.1\linewidth, height=0.06\linewidth]{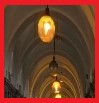}\hspace*{-0.2em}
	\includegraphics[width=0.1\linewidth, height=0.06\linewidth]{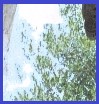}\hspace*{-0.1em}
	
	\caption{Visual comparison of results obtained using, from left to right; input LDR images, HDRCNN \cite{eilertsen_hdr_2017}, ExpandNet \cite{marnerides_expandnet_2018}, SingleHDR \cite{liu_single-image_2020} and our proposed MQN.  All images are produced with Balanced TMO from the suite Photomatix \cite{noauthor_photo_2017}, similarly to \cite{liu_single-image_2020}.} 
	
	\label{fig:comparison}
\vspace{-0.6cm}
\end{figure}

\subsection{Comparison with State-of-the-art Methods}

In Table~\ref{tab:final}, we compare accuracy, latency, and size of models trained using MQN and state-of-the-art methods. We selected competing methods according to two criteria: (1) We considered methods that promised a reasonable accuracy-latency trade-off \cite{chen_hdrunet_2021, santos_single_2020} due to their simplicity, to evaluate the efficiency gain introduced with our method. (2) To estimate the cost of adding efficiency as a factor in network design, we compared MQN to state-of-the-art or baseline methods \cite{liu_single-image_2020, eilertsen_hdr_2017} that did not take accuracy-efficiency trade-off into account. The results show that MQN models provide accuracy (HDRVDP-Q score) on par with the larger state-of-the-art models and with a notable reduction in latency and RAM consumption.

In terms of latency, our MQN provides the fastest model, both in experiments conducted on GPU and mobile deployment platforms. Comparison of the latency of the models on the GPU platform is not fair, since our MQN model uses \textit{8-bit} integer quantization and employs depthwise convolutions which are not suited for GPU platforms \cite{sandler_mobilenetv2_2018}.  To obtain a fair comparison on the mobile deployment platform, we adapt the four methods HDRUNet \cite{chen_hdrunet_2021}, ExpandNet \cite{marnerides_expandnet_2018}, DeepHDR \cite{santos_single_2020} and TwoStage \cite{sharif_two-stage_2021} that perform the fastest inference on the GPU. The results show that the difference between the latency of our MQN models and these four models increases further on the mobile platform. More precisely, our MQN model is running in real-time ({\color{blue} 21ms}) while others run from a quarter of a second ({\color{red} DeepHDR with 238ms}) to more than 3 seconds (TwoStage with 3338ms). The same trend is observed with memory consumption (maximum RAM MB) where our model stands as the most efficient with a reduction of a factor of x26 or more. The main factor for such differences is employment of our MQ scheme with computationally efficient components, such as IRLB blocks. These facts, along with learning representations of HDR content over the input, enables us to obtain a  faster model with similar accuracy. However, other models use methods that hinder efficient deployment, such as by utilising images with same resolution \cite{marnerides_expandnet_2018, sharif_two-stage_2021}, inefficient network composition \cite{liu_single-image_2020}, convolution operations and special blocks \cite{chen_hdrunet_2021, santos_single_2020}. 

In Figure~\ref{fig:comparison}, we compare the ITM models visually. In the analyses, our MQN model performs well with good quality in under-exposed regions. For instance, our MQN model recovers details better than others as seen in the detail of the image in row one. In the case of over-exposed regions, our model can recover details better than HDRCNN \cite{eilertsen_hdr_2017} and similarly to ExpandNet \cite{marnerides_expandnet_2018} as seen in the lamp, tree and tent details. Although SingleHDR \cite{liu_single-image_2020}  performs slightly better on over-exposed regions, the SingleHDR model has 29 M parameters and takes almost a second to perform inference on a desktop GPU, while our model is 29x times smaller and almost 100x faster.
\vspace{-0.3cm}
\section{Conclusion}

In this work, we proposed a novel DNN-based method called Mixed Quantization Network (MQN), for computationally efficient ITM.  The proposed MQN has  produced competitive accuracy with better computational efficiency compared to the state-of-the-art, being the first DNN-based single image ITM method targeting computationally efficient mobile ITM. Moreover, we have proven the flexibility of our framework by deploying it to both CPU and GPU platforms. Future work could consider addressing optimization of latency and accuracy trade-off for mobile ITM using additional network design and search methods, such as structured pruning and neural architecture search. 
\vspace{-0.5cm}
\section*{Acknowledgements}

Authors thank the staff of Samsung R\&D UK for the support to this project. Special gratitude to Cristian Szabo for his support in model deployment and also to Albert Saa-Garriga, Karthikeyan Saravanan, and Daniel Ansorregui for their support and orientation.

\bibliography{toCite}
\end{document}


\maketitle

\section{Overview}

In the present document, we provide more insights with regards to data, evaluation procedure, ablation studies, and inference examples.
 
\section{Datasets, Architecture and Hyperparameters of the MQN}

\subsection{Training Datasets}

We employ 4 different datasets for training: Ward, Funt, PFSTools, and HDRPlus. Most of them only contain high bit depth images, and thus input LDR images have to be recreated as detailed in Section 4.1, Experimental Setup of the main paper. The four tone mapping operators (TMOs) used are: Drago \cite{drago_adaptive_2003}, Mantiuk \cite{mantiuk_display_2008}, Reinhard \cite{reinhard_photographic_2002} and Exposure obtained from the OpenCV  library \cite{noauthor_opencv_2021}. For data augmentation, we use a batch of ground-truth HDR images. For each image, a TMO is chosen randomly and the TMO is applied on the image with random parameters.

\subsubsection{Ward and PFSTools}

The Ward dataset \cite{ward_high_2006} is a collection of 33 HDR images originally intended to compare different HDR formats (OpenEXR, Radiance RGBE and XYZE, 24-bit and 32-bit LogLuv TIFF, and others).  The PFSTools \cite{noauthor_pfstools_2015} is a collection of 8 HDR images of both outside and interior scenes.

\subsubsection{Funt}

The Funt collection \cite{funt_effect_2010}  is a set of 105 HDR images built by bracketing 9 differently exposed LDR images. The LDR images have a difference of 1 EV between them, a rate of capture of 5 seconds and the f-stop is not fixed. The Final HDR images are created from raw LDR bursts by alignment and filtering.

\subsubsection{HDRPlus}

HDR+ \cite{hasinoff_burst_2016} is a content enhancement pipeline dataset consisting on 3640 bursts (made up of 28461 images in total) resulting from the Google HDR+ system. The dataset also contains an intermediate DNG burst merge image and the 8-bit image resulting from the pipeline. We use the merged burst image in DNG format as our HDR output and do not use the 8-bit images defined as a result of the pipeline or the raw input images.

\subsection{Test Datasets}

For testing our method, we choose three datasets that, contrarily to our training data, have LDR-HDR paired images, thus enabling a fair evaluation.

\subsubsection{HDR Eye}

HDR Eye \cite{nemoto_visual_2015} consists of 46 LDR-HDR pairs taken with the Sony DSC-RX100 II, Sony NEX-5N, and Sony 6000 cameras. The HDR images are generated by combining LDR images with exposures (-2.7, -2, -1.3, -0,7, 0, 0,7, 1,3, 2, 2.7). Evaluation is performed using images with size 256x256 as suggested in \cite{lee_deep_2018-1}.

\subsubsection{HDR Real}

HDR Real \cite{liu_single-image_2020} is a photographic dataset specially designed for extreme HDR contexts. It consists of 1838 LDR-HDR pairs taken by amateur photographers, employing 42 different cameras, using different exposures and covering the whole range of lighting conditions: from near pitch-black to extremely saturated images. We perform evaluation using 256x256 images, following \cite{lee_deep_2018-1}. Instead of preprocessing the the datasets as employed in \cite{liu_single-image_2020}, we employ the dataset directly without preprocessing.

\subsubsection{RAISE 1K}

The RAISE-1K \cite{dang-nguyen_raise_2015} consists of a subset of 1000 RAW-TIFF images pairs selected from the original RAISE dataset. Originally intended for digital forensics, it contains high-resolution images captured in diverse scenarios: indoor, outdoor, man-made and natural. Following \cite{liu_single-image_2020}, we consider using unprocessed RAW images (with 12- or 14-bit depth) as ground truth HDR images, and the TIFF images as 8-bit LDR input images. We convert the RAW images to \textit{.hdr} format and the TIFF images to JPEG. For evaluation we downsize images to a quarter of their original size respecting proportion ratios, resulting in images which are approximately 720p.

\section{Evaluation Procedure}

For evaluation of methods using Peak Signal Noise Ration (PSNR) and Structural Similarity (SSIM), we use tone mapped images and predictions. Inspired by \cite{liu_single-image_2020}, and unlike our training procedure where we use OpenCV tone mapping operators, we tone map images using four tone mapping operators from the Photomatix suite \cite{noauthor_photo_2017} for evaluation: (1) detailed, (2) balanced, (3) realistic and (4) photographic.

Regarding HDR-VDP metric evaluation, we use version 2.2.2 for evaluation. Our predictions take values from [0, 1] (relative luminance). We reescale them to a display range of  $1000\, cd/m^{2}$ and align the 0.01 and 0.99 percentiles of both prediction and ground truth. For a fair comparison, we use the same parameters as utilized in \cite{lee_deep_2018, lee_deep_2018-1} to obtain the \textit{pixel-per-degree} parameter, which are 24-inch display, 0.5 distance and 1080p display resolution.

With regards to latency measurements, we test models on both a desktop GPU, NVIDIA GTX 1080 Ti, and a mobile platform, Samsung Note 20 Exynos 990. The latency calculations are performed on the desktop platform taking into account the process between the reading of the LDR image and the output of the final HDR image, that is to say that the reading and writing computational costs are not considered. So, for those methods, such as \cite{eilertsen_hdr_2017} or ours, that use the input mixed with the output to create the final HDR image, the combination procedure is also included in the latency computation. In the mobile platform, latency is tested through the native benchmark application for the arch64 architecture offered by Tensorflow \cite{noauthor_performance_2021}, always running on a CPU with 4 threads for 300 runs to average the results using images of size 256x256.

\subsection{Training and Network Details}

The entire training takes approximately 5 days on a machine with an Intel Core i7-6850K and an Nvidia GTX 1080 Ti.

\begin{figure}[!hbtp]
	\centering
    \includegraphics[width=0.7\linewidth, height=0.8\textheight]{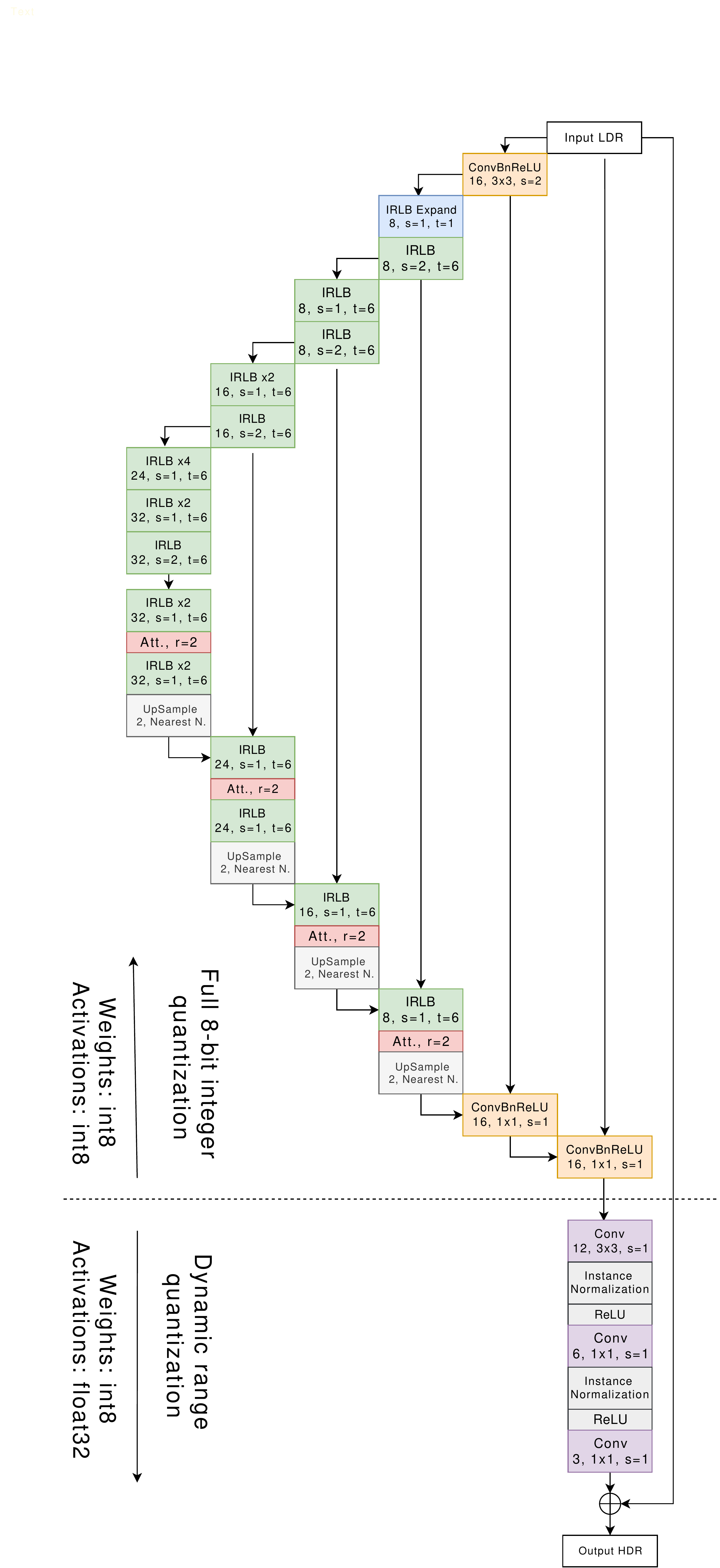}   
	\caption{Illustration of the base encoder-decoder architecture and high precision head that conform the MQN. The first number indicates filters, \textit{s} stands for stride, \textit{t} denotes the expansion ratio in the Inverted Residual Linear Bottleneck (IRLB) blocks, \textit{r} is the ratio in the attention blocks. Kernel size is indicated when appropriate by \textit{kxk}. The dotted line indicates the separation between the fully quantized architecture and the dynamically quantized head.}
	\label{fig:arch_detailed}
\end{figure}

For the encoder of the backbone, we use a MobileNetV2 (MBV2) with width factor $\alpha=0.35$ and skip connections applied at the activations of layers (1, 3, 6, 13) and output of batch normalization at layer 16. Further details of each of the layers are given in Figure \ref{fig:arch_detailed}.

\section{Additional Quantitative Evaluation and Analyses}

\subsection{Extended State-of-the-art Comparison}

\subsubsection{Histograms of HDRVDP Q Values}

\begin{figure}[!t]
\includegraphics[width=0.48\textwidth]{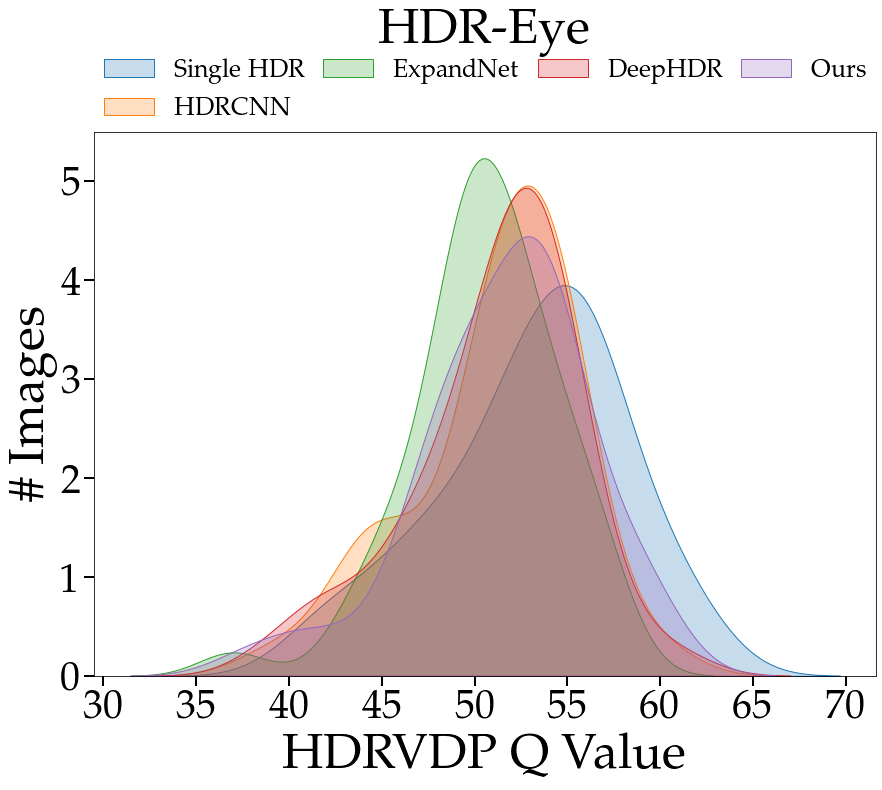}
\includegraphics[width=0.48\textwidth]{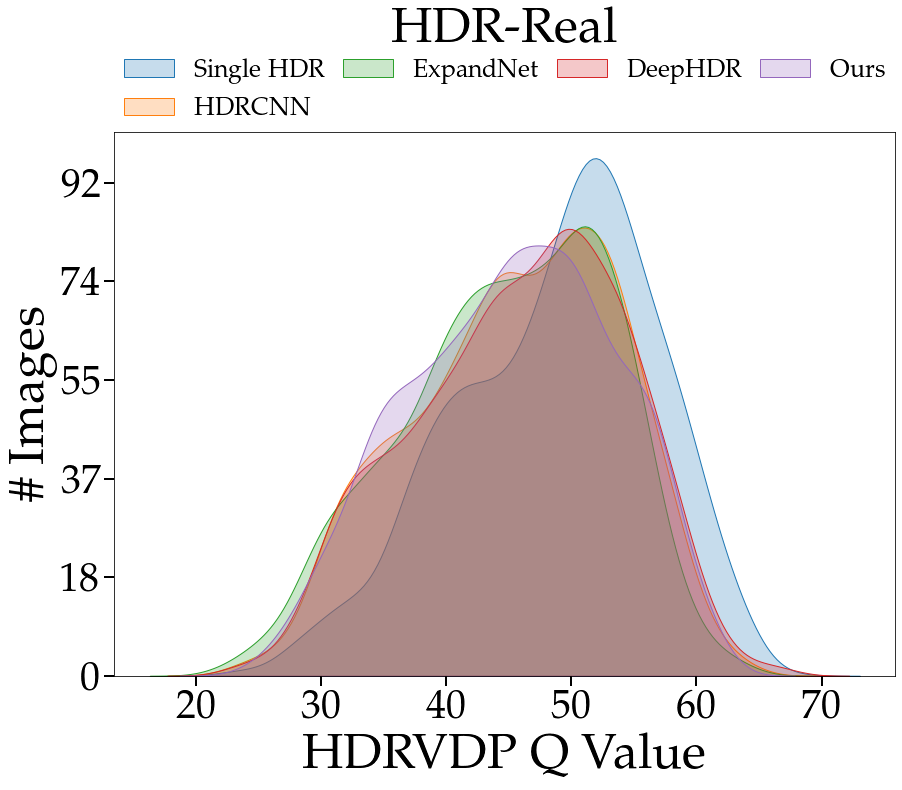}\\
\begin{center}
\includegraphics[width=0.5\textwidth]{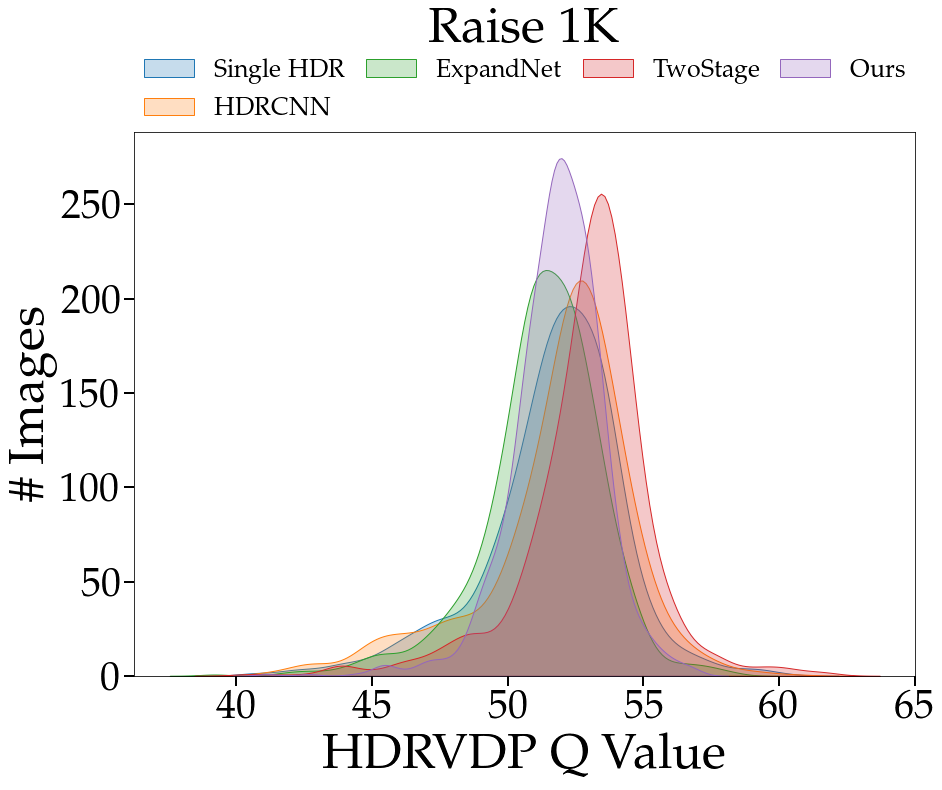}
\end{center}
\caption{Distribution of HDRVDP Q score values for three test datasets and 5 best competing methods in each case.}
\label{fig:histogramshdrvdp}
\end{figure}

In Figure \ref{fig:histogramshdrvdp}, we plot the distribution of the Q value scores obtained from HDRVDP for HDRCNN \cite{eilertsen_hdr_2017}, SingleHDR \cite{liu_single-image_2020}, ExpandNet \cite{marnerides_expandnet_2018}, and our method for all three test sets: HDR-Eye, HDR-Real and Raise 1K. As the results show, our method performs similarly or better than HDRCNN and ExpandNet, and slightly worse than SingleHDR, even though our method is purposed for fast inference and not for performance quality, as other competing methods are.

\subsubsection{Quantitative Evaluation on Tone Mapped Images}

\begin{figure}		

	\setlength{\lineskip}{0pt}
	\includegraphics[width=0.2\linewidth, height=0.15\textheight]{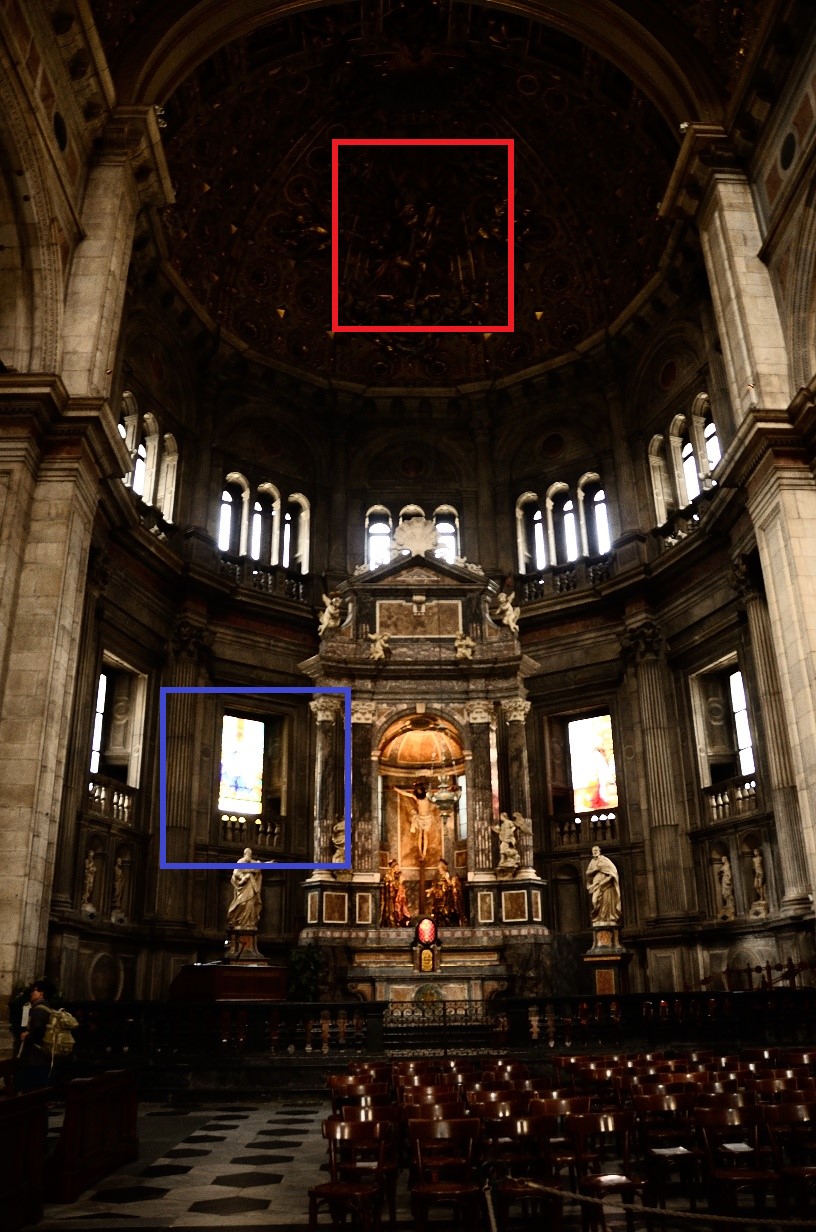}\hspace*{-0.1em}
	\includegraphics[width=0.2\linewidth, height=0.15\textheight]{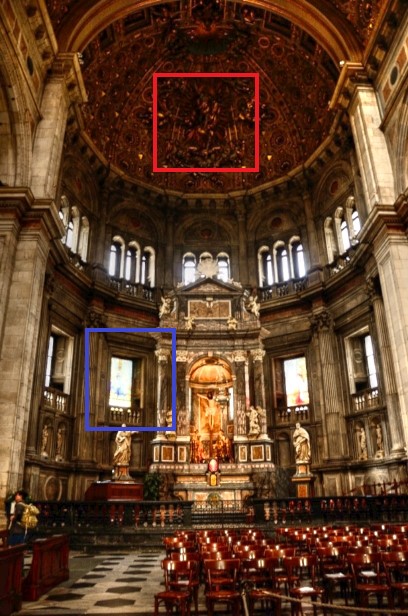}\hspace*{-0.1em}
	\includegraphics[width=0.2\linewidth, height=0.15\textheight]{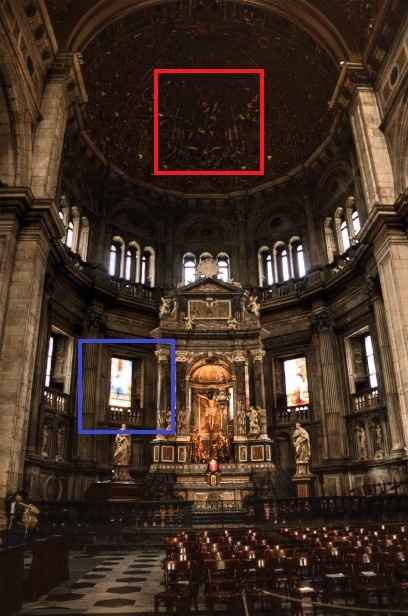}\hspace*{-0.1em}
	\includegraphics[width=0.2\linewidth, height=0.15\textheight]{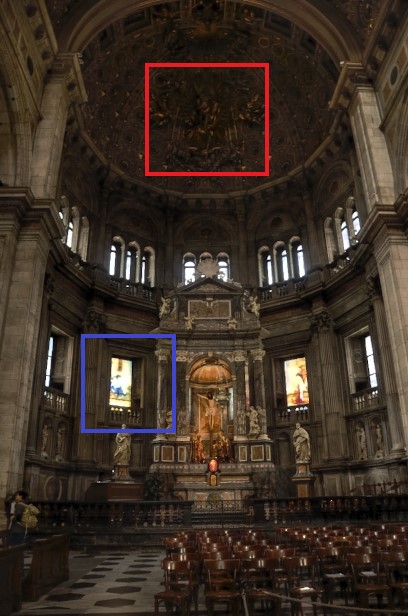}\hspace*{-0.1em}
	\includegraphics[width=0.2\linewidth, height=0.15\textheight]{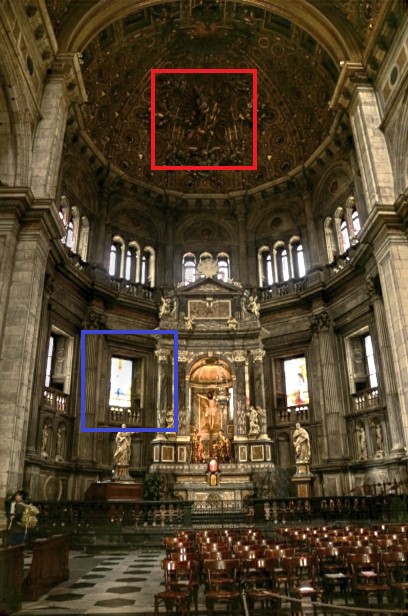}\hspace*{-0.1em}\\
\setlength{\lineskip}{1pt}
	\includegraphics[width=0.1\linewidth, height=0.08\textheight]{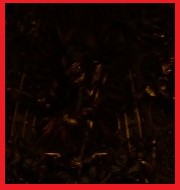}\hspace*{-0.2em}
	\includegraphics[width=0.1\linewidth, height=0.08\textheight]{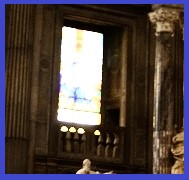}\hspace*{-0.1em}
	\includegraphics[width=0.1\linewidth, height=0.08\textheight]{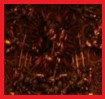}\hspace*{-0.2em}
	\includegraphics[width=0.1\linewidth, height=0.08\textheight]{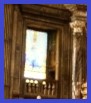}\hspace*{-0.1em}
	\includegraphics[width=0.1\linewidth, height=0.08\textheight]{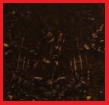}\hspace*{-0.2em}
	\includegraphics[width=0.1\linewidth, height=0.08\textheight]{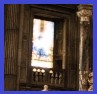}\hspace*{-0.1em}
	\includegraphics[width=0.1\linewidth, height=0.08\textheight]{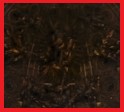}\hspace*{-0.2em}
	\includegraphics[width=0.1\linewidth, height=0.08\textheight]{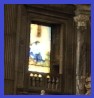}\hspace*{-0.1em}
	\includegraphics[width=0.1\linewidth, height=0.08\textheight]{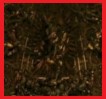}\hspace*{-0.2em}
	\includegraphics[width=0.1\linewidth, height=0.08\textheight]{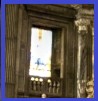}\hspace*{-0.1em}\\
	\setlength{\lineskip}{1pt}
	\includegraphics[width=0.2\linewidth, height=0.13\textheight]{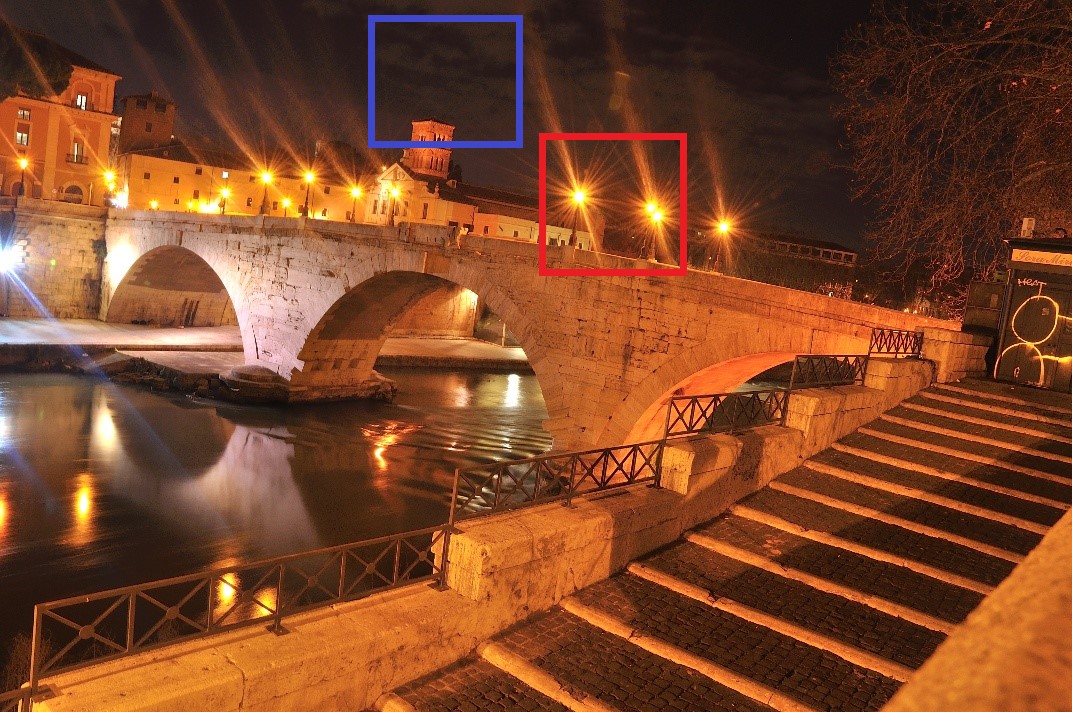}\hspace*{-0.1em}
	\includegraphics[width=0.2\linewidth, height=0.13\textheight]{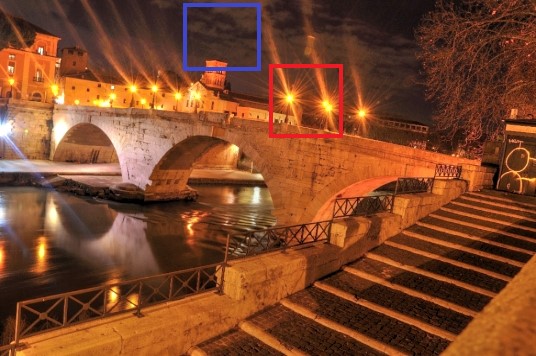}\hspace*{-0.1em}
	\includegraphics[width=0.2\linewidth, height=0.13\textheight]{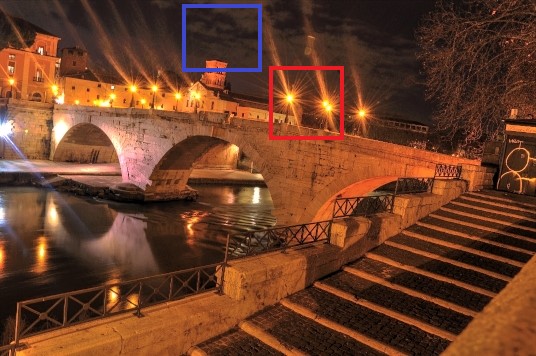}\hspace*{-0.1em}
	\includegraphics[width=0.2\linewidth, height=0.13\textheight]{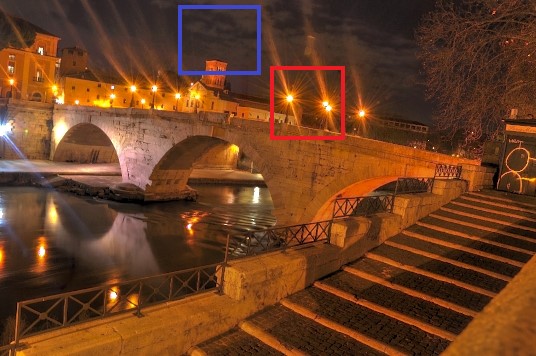}\hspace*{-0.1em}
	\includegraphics[width=0.2\linewidth, height=0.13\textheight]{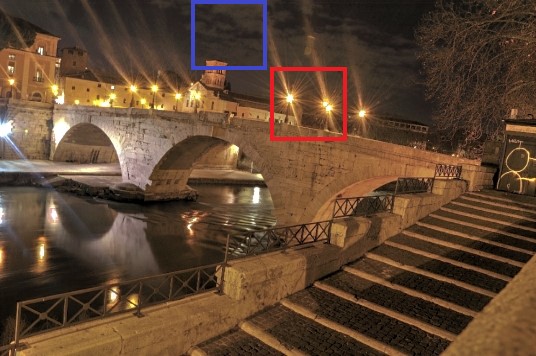}\hspace*{-0.1em}\\
	\includegraphics[width=0.1\linewidth, height=0.07\textheight]{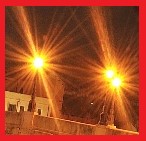}\hspace*{-0.2em}
	\includegraphics[width=0.1\linewidth, height=0.07\textheight]{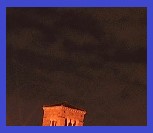}\hspace*{-0.1em}
	\includegraphics[width=0.1\linewidth, height=0.07\textheight]{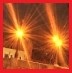}\hspace*{-0.2em}
	\includegraphics[width=0.1\linewidth, height=0.07\textheight]{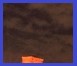}\hspace*{-0.1em}
	\includegraphics[width=0.1\linewidth, height=0.07\textheight]{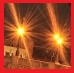}\hspace*{-0.2em}
	\includegraphics[width=0.1\linewidth, height=0.07\textheight]{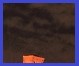}\hspace*{-0.1em}
	\includegraphics[width=0.1\linewidth, height=0.07\textheight]{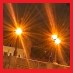}\hspace*{-0.2em}
	\includegraphics[width=0.1\linewidth, height=0.07\textheight]{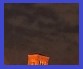}\hspace*{-0.1em}
	\includegraphics[width=0.1\linewidth, height=0.07\textheight]{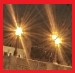}\hspace*{-0.2em}
	\includegraphics[width=0.1\linewidth, height=0.07\textheight]{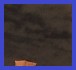}\hspace*{-0.1em}\\
	\setlength{\lineskip}{1pt}
	\includegraphics[width=0.2\linewidth, height=0.14\textheight]{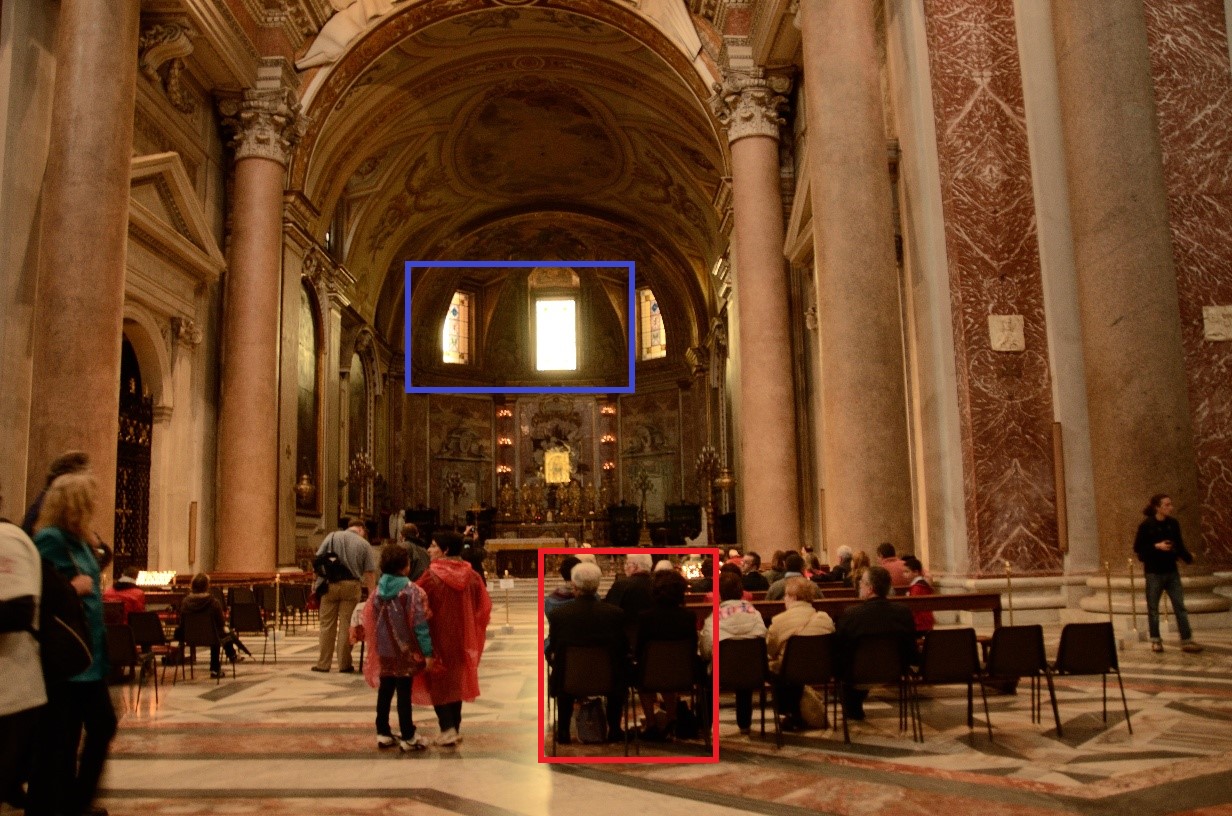}\hspace*{-0.1em}
	\includegraphics[width=0.2\linewidth, height=0.14\textheight]{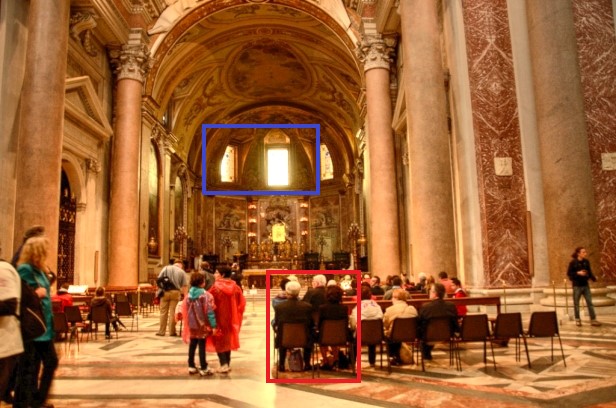}\hspace*{-0.1em}
	\includegraphics[width=0.2\linewidth, height=0.14\textheight]{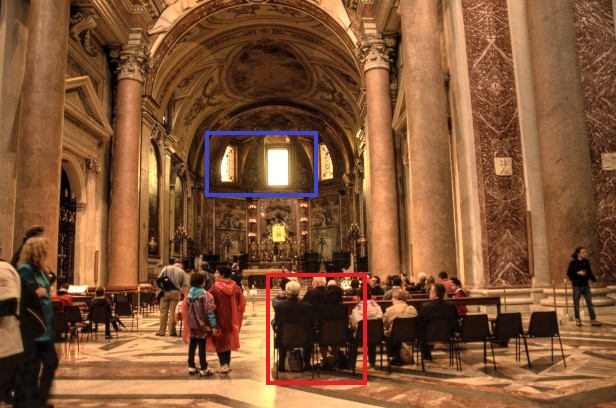}\hspace*{-0.1em}
	\includegraphics[width=0.2\linewidth, height=0.14\textheight]{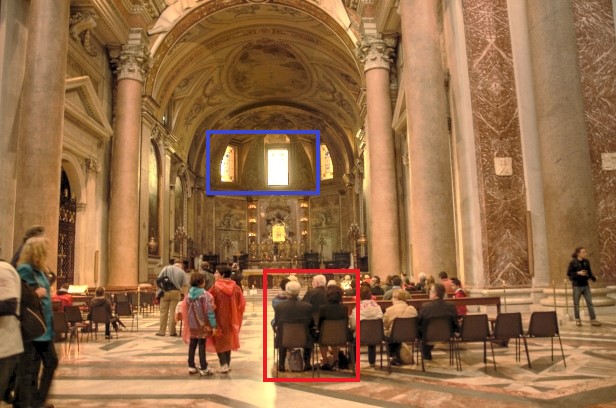}\hspace*{-0.1em}
	\includegraphics[width=0.2\linewidth, height=0.14\textheight]{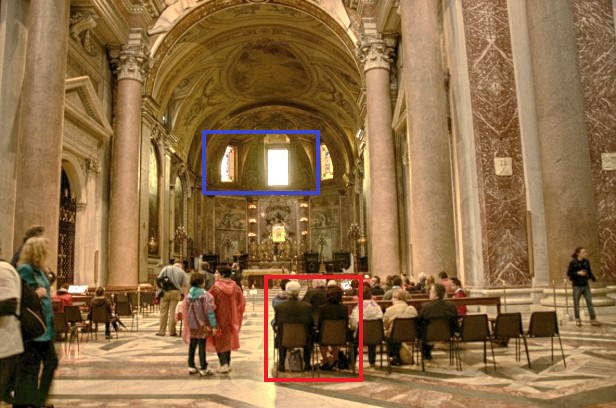}\hspace*{-0.1em}\\
	\includegraphics[width=0.1\linewidth, height=0.07\textheight]{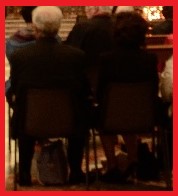}\hspace*{-0.2em}
	\includegraphics[width=0.1\linewidth, height=0.07\textheight]{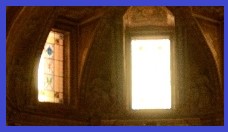}\hspace*{-0.1em}
	\includegraphics[width=0.1\linewidth, height=0.07\textheight]{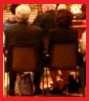}\hspace*{-0.2em}
	\includegraphics[width=0.1\linewidth, height=0.07\textheight]{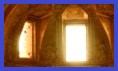}\hspace*{-0.1em}
	\includegraphics[width=0.1\linewidth, height=0.07\textheight]{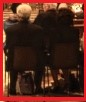}\hspace*{-0.2em}
	\includegraphics[width=0.1\linewidth, height=0.07\textheight]{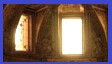}\hspace*{-0.1em}
	\includegraphics[width=0.1\linewidth, height=0.07\textheight]{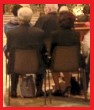}\hspace*{-0.2em}
	\includegraphics[width=0.1\linewidth, height=0.07\textheight]{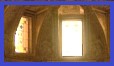}\hspace*{-0.1em}
	\includegraphics[width=0.1\linewidth, height=0.07\textheight]{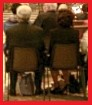}\hspace*{-0.2em}
	\includegraphics[width=0.1\linewidth, height=0.07\textheight]{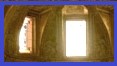}\hspace*{-0.1em}\\
	\setlength{\lineskip}{1pt}
	\includegraphics[width=0.2\linewidth, height=0.14\textheight]{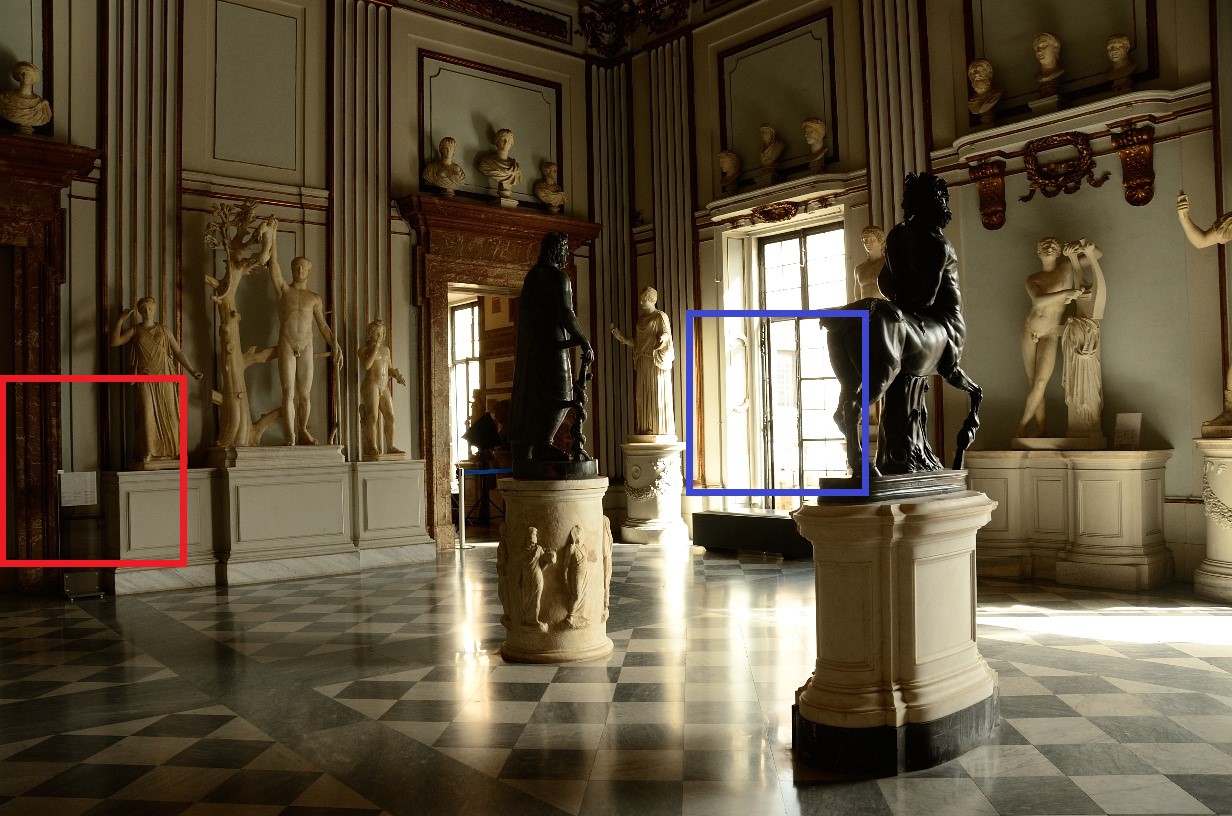}\hspace*{-0.1em}
	\includegraphics[width=0.2\linewidth, height=0.14\textheight]{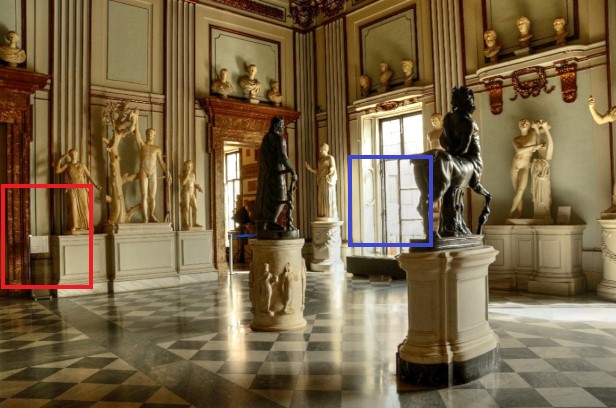}\hspace*{-0.1em}
	\includegraphics[width=0.2\linewidth, height=0.14\textheight]{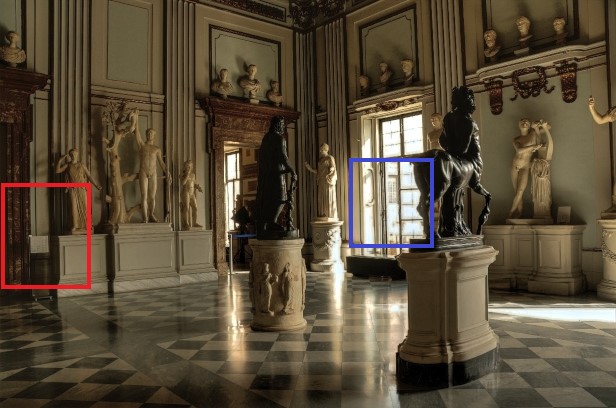}\hspace*{-0.1em}
	\includegraphics[width=0.2\linewidth, height=0.14\textheight]{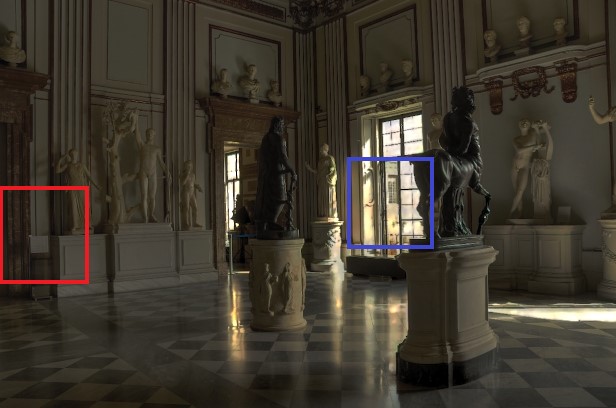}\hspace*{-0.1em}
	\includegraphics[width=0.2\linewidth, height=0.14\textheight]{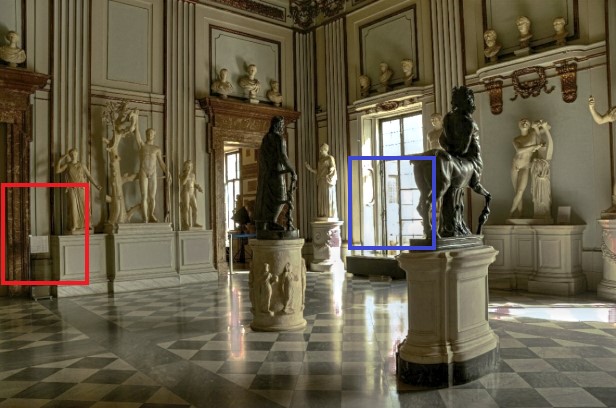}\hspace*{-0.1em}\\
	\includegraphics[width=0.1\linewidth, height=0.07\textheight]{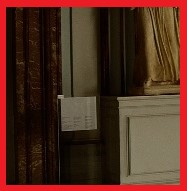}\hspace*{-0.2em}
	\includegraphics[width=0.1\linewidth, height=0.07\textheight]{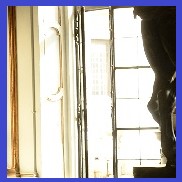}\hspace*{-0.1em}
	\includegraphics[width=0.1\linewidth, height=0.07\textheight]{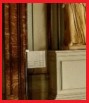}\hspace*{-0.2em}
	\includegraphics[width=0.1\linewidth, height=0.07\textheight]{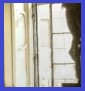}\hspace*{-0.1em}
	\includegraphics[width=0.1\linewidth, height=0.07\textheight]{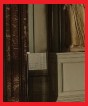}\hspace*{-0.2em}
	\includegraphics[width=0.1\linewidth, height=0.07\textheight]{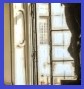}\hspace*{-0.1em}
	\includegraphics[width=0.1\linewidth, height=0.07\textheight]{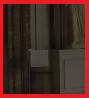}\hspace*{-0.2em}
	\includegraphics[width=0.1\linewidth, height=0.07\textheight]{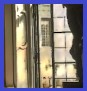}\hspace*{-0.1em}
	\includegraphics[width=0.1\linewidth, height=0.07\textheight]{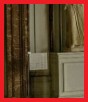}\hspace*{-0.2em}
	\includegraphics[width=0.1\linewidth, height=0.07\textheight]{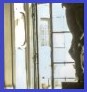}\hspace*{-0.1em}
	
	\caption{Visual comparison of our method and state-of-the-art ITM methods. From left to right: input LDR, HDRCNN \cite{eilertsen_hdr_2017}, ExpandNet \cite{marnerides_expandnet_2018}, SingleHDR \cite{liu_single-image_2020} and our MQN. }
	\label{fig:extended_comparison}
\end{figure}

We also evaluate our method and competing state-of-the-art methods using  PSNR and SSIM. To provide HDR images for these metrics, we tone map the images with the use of Photomatix suite and four tone mapping operators: Detailed, Balanced, Realistic and Photographic. Results are shown in Table \ref{tab:finaltone}. The results show that our method performs better than ExpandNet and HDRCNN in HDR Eye and Raise, and always after Single HDR. However, note that our method is 100x faster than SingleHDR both on mobile and GPU, and almost 10x faster than ExpandNet on mobile.

\begin{table}[!]
	\caption{Comparison with state-of-the-art single image HDR reconstruction methods for PSNR and SSIM performance metrics. Values reproduced with the same evaluation criteria and original code. Blue indicates the best and red indicates the second best accuracy.\label{tab:finaltone}}
	\centering
	\small
	\resizebox{\textwidth}{!}{%
		\begin{tabular}{|c|c|c|c|c|c|c|}
			\hline
			 & \multicolumn{2}{|c|}{HDR Eye} & \multicolumn{2}{|c|}{HDR Real} & \multicolumn{2}{|c|}{Raise 1K	}\\
			\hline
			\textbf{Model}  &   \textbf{PSNR$-T$} & \textbf{SSIM$-T$} & \textbf{PSNR$-T$} & \textbf{SSIM$-T$} &\textbf{PSNR$-T$} & \textbf{SSIM$-T$} \\
			\hline 
			HDRCNN \cite{eilertsen_hdr_2017} &  18.82 $\pm$ 3.49 & 0.7754 $\pm$ 0.1044 & \textcolor{red}{16.53 $\pm$ 5.65} & \textcolor{red}{0.6378 $\pm$ 0.2303} & 17.25 $\pm$ 2.81 &  \textcolor{red}{0.5950 $\pm$ 0.1213}\\
			\hline
			FHDR \cite{khan_fhdr_2019} & \textcolor{red}{20.30 $\pm$ 5.40} & 0.7794 $\pm$ 0.1897 & 16.47 $\pm$ 5.83 & 0.6436 $\pm$ 0.2305 & 17.68 $\pm$ 3.67 & 0.5653 $\pm$ 0.1315\\
			\hline
			ExpandNet \cite{marnerides_expandnet_2018}    & 19.85 $\pm$ 2.96 & 0.7854 $\pm$ 0.0954  & 16.24 $\pm$ 5.99 & 0.6175 $\pm$ 0.2564 & 16.58 $\pm$ 2.77 & 0.5264 $\pm$ 0.1306\\
			\hline
			SingleHDR \cite{liu_single-image_2020} & \textcolor{blue}{21.70 $\pm$ 4.50} & \textcolor{blue}{0.8259 $\pm$ 0.1244} & \textcolor{blue}{21.16 $\pm$ 5.33} & \textcolor{blue}{0.7409 $\pm$ 0.2150} & 15.12 $\pm$ 3.44 & 0.5688 $\pm$ 0.1253 \\
			\hline 
			DeepHDR \cite{santos_single_2020} & 19.38 $\pm$ 3.38 & 0.7723 $\pm$ 0.1131 & 16.49 $\pm$ 6.15 & 0.6252 $\pm$ 0.2591 & 17.22 $\pm$ 2.86 & 0.5861 $\pm$ 0.1243 \\
			\hline
			TwoStage \cite{sharif_two-stage_2021} & 17.97 $\pm$ 4.16 & 0.7519 $\pm$ 0.1057 &  14.57 $\pm$ 4.52& 0.5660 $\pm$ 0.2235 & \textcolor{blue}{19.12 $\pm$ 2.81} & \textcolor{blue}{0.6162 $\pm$ 0.1257}\\
			\hline
			HDRUnet \cite{chen_hdrunet_2021} & 18.58 $\pm$ 4.11 & 0.7724 $\pm$ 0.1139 & 14.59 $\pm$ 4.73 & 0.5778 $\pm$ 0.2478 & 17.70 $\pm$ 3.07 & 0.5989 $\pm$ 0.1263 \\
			\specialrule{.15em}{.15em}{.15em}
			Ours Best  & 19.97 $\pm$ 4.21 & \textcolor{red}{0.7990 $\pm$ 0.1160} & 15.95 $\pm$ 5.76 & 0.6162 $\pm$ 0.2436 & \textcolor{red}{17.71 $\pm$ 2.73} & 0.5776 $\pm$ 0.1150\\
			\hline
	\end{tabular}}
\end{table}

\subsubsection{Comparative Analysis of Latency and Accuracy}

\begin{figure}
    \centering
    \includegraphics[width=0.95\textwidth]{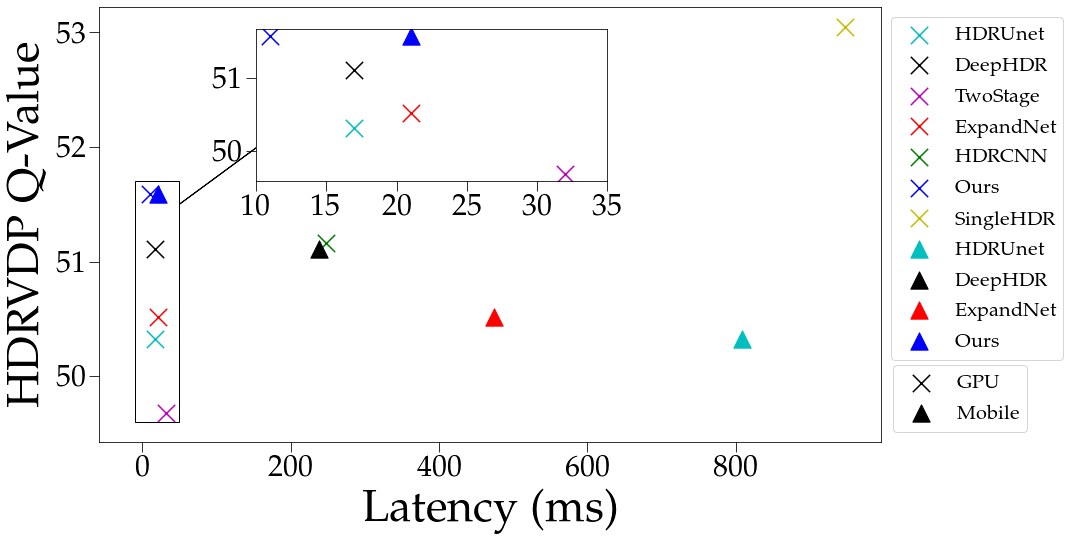}
    \caption{Latency and accuracy trade-off comparison with the HDR Eye dataset between competing methods and our proposed solution. Accuracy is measured by HDR-VDP Q value score. Latency is computed both in the mobile (CPU) and GPU platforms. }
    \label{fig:latency}
\end{figure}

Our method is intended for fast inference, while all competing methods focus on accuracy. We compare the latency and accuracy trade-off in Figure \ref{fig:latency}. We can see that our model is almost 100x faster than SingleHDR, while providing only $3\%$ accuracy loss, and while being 10x faster than ExpandNet it achieves a 1 Q point more on accuracy.

\subsubsection{Extended Qualitative Comparison}

In Figure \ref{fig:extended_comparison}, we show additional examples of predictions of our model as well as a comparison with state-of-the-art models. We see that our model is able to reconstruct well both under- and over-exposed regions. Our model shows similar accuracy to ExpandNet and shows slightly less reconstruction capacity on over-exposed regions compared to SingleHDR. Further, our model is 10x and 100x faster than ExpandNet and SingleHDR respectively.

\subsubsection{Analyses for Extreme Over-exposed Cases}

\begin{figure}[!t]		
	\setlength{\lineskip}{0pt}
	\includegraphics[width=0.32\linewidth, height=0.2\linewidth]{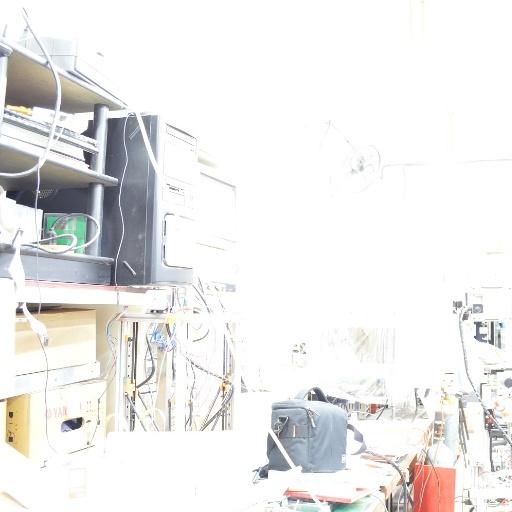}\hspace*{-0.1em}
	\includegraphics[width=0.32\linewidth, height=0.2\linewidth]{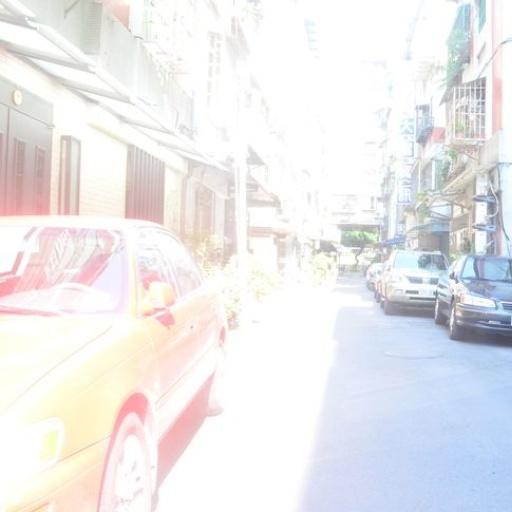}\hspace*{-0.1em}
	\includegraphics[width=0.32\linewidth, height=0.2\linewidth]{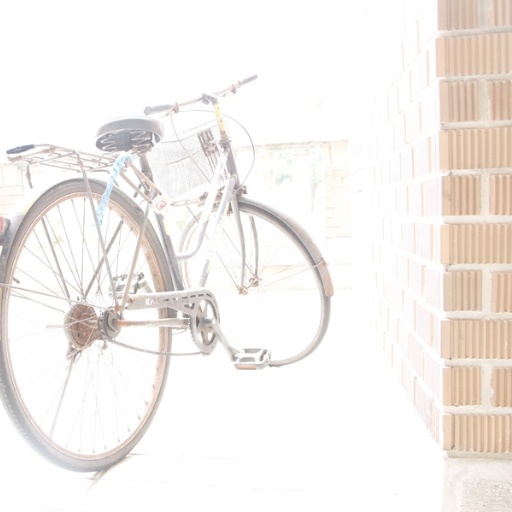}\hspace*{-0.1em}\\
	\setlength{\lineskip}{0pt}
	\includegraphics[width=0.32\linewidth, height=0.2\linewidth]{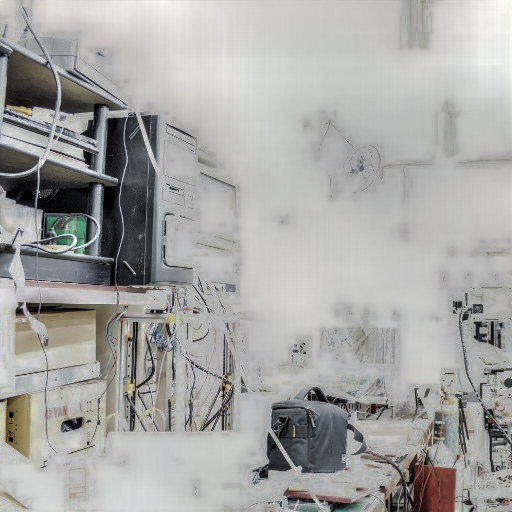}\hspace*{-0.1em}
	\includegraphics[width=0.32\linewidth, height=0.2\linewidth]{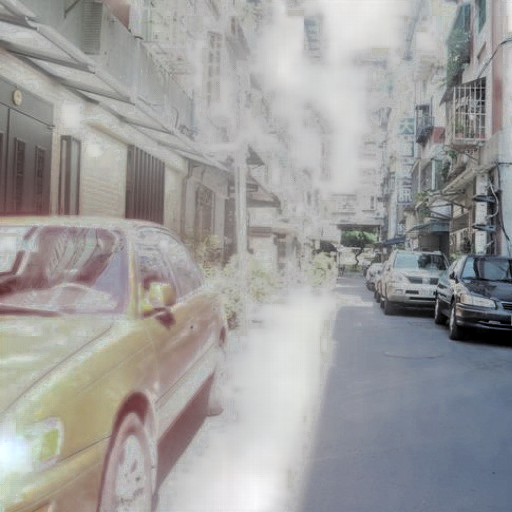}\hspace*{-0.1em}
	\includegraphics[width=0.32\linewidth, height=0.2\linewidth]{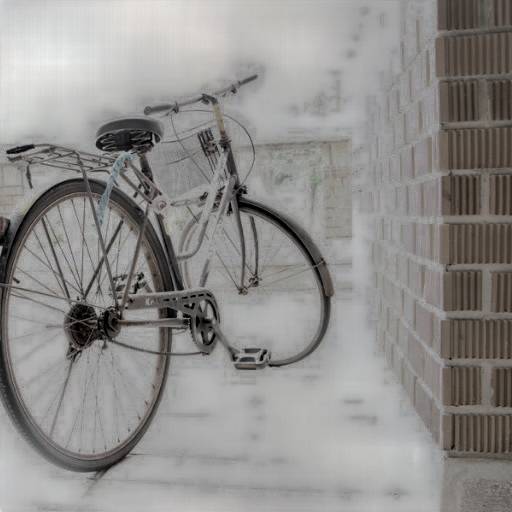}\hspace*{-0.1em}\\
	\setlength{\lineskip}{0pt}
	\includegraphics[width=0.32\linewidth, height=0.2\linewidth]{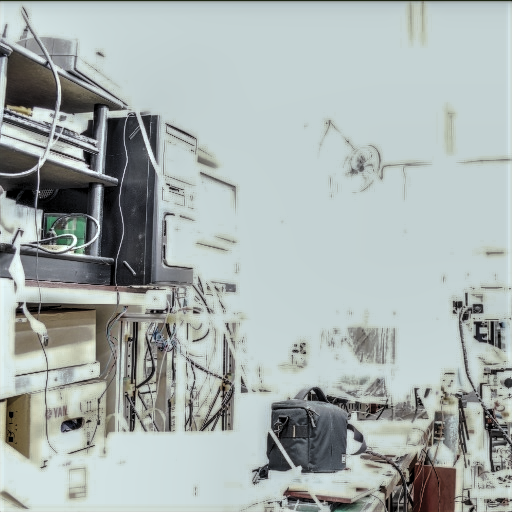}\hspace*{-0.1em}
	\includegraphics[width=0.32\linewidth, height=0.2\linewidth]{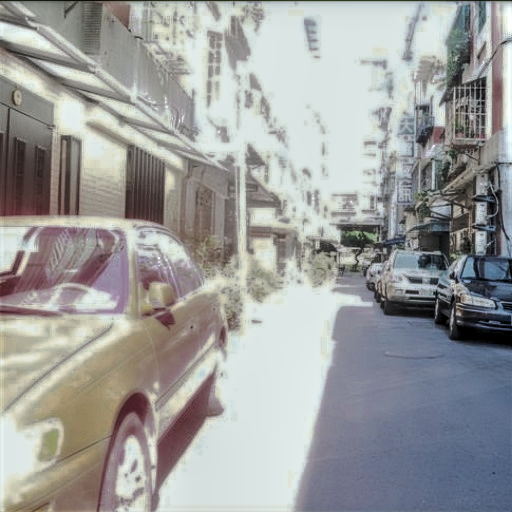}\hspace*{-0.1em}
	\includegraphics[width=0.32\linewidth, height=0.2\linewidth]{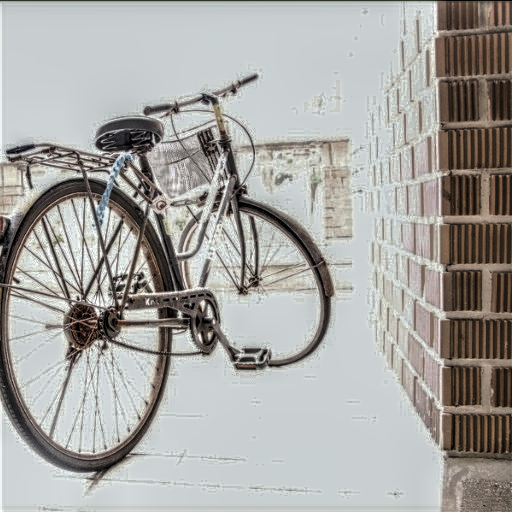}\hspace*{-0.1em}\\
	\setlength{\lineskip}{0pt}
	\includegraphics[width=0.32\linewidth, height=0.2\linewidth]{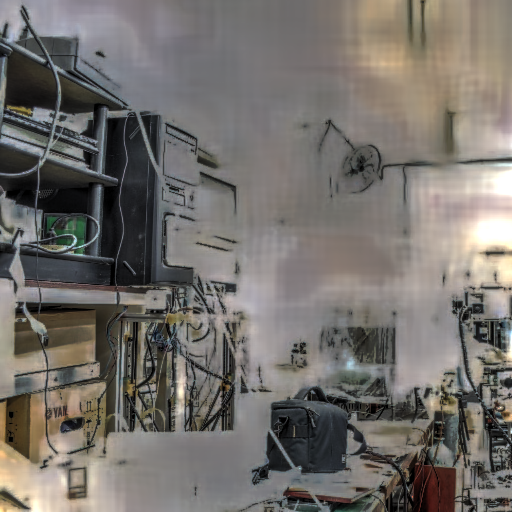}\hspace*{-0.1em}
	\includegraphics[width=0.32\linewidth, height=0.2\linewidth]{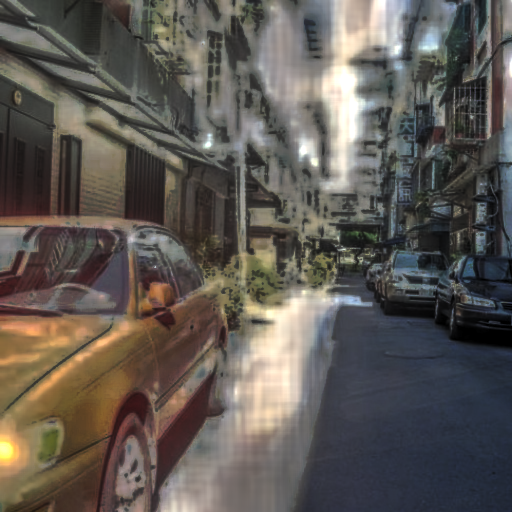}\hspace*{-0.1em}
	\includegraphics[width=0.32\linewidth, height=0.2\linewidth]{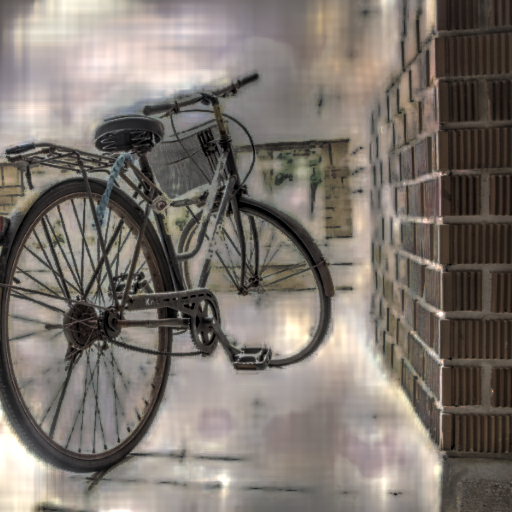}\hspace*{-0.1em}\\
	\setlength{\lineskip}{0pt}
	\includegraphics[width=0.32\linewidth, height=0.2\linewidth]{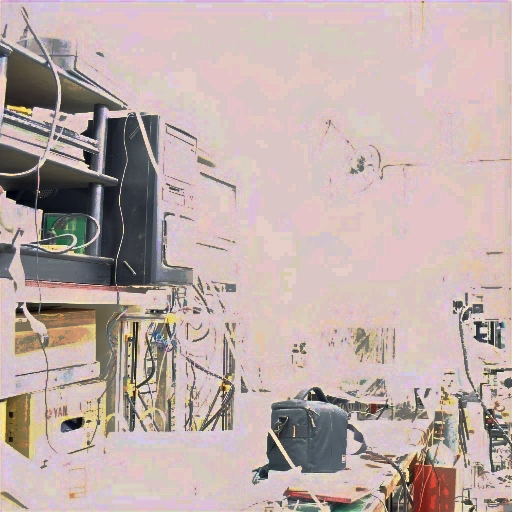}\hspace*{-0.1em}
	\includegraphics[width=0.32\linewidth, height=0.2\linewidth]{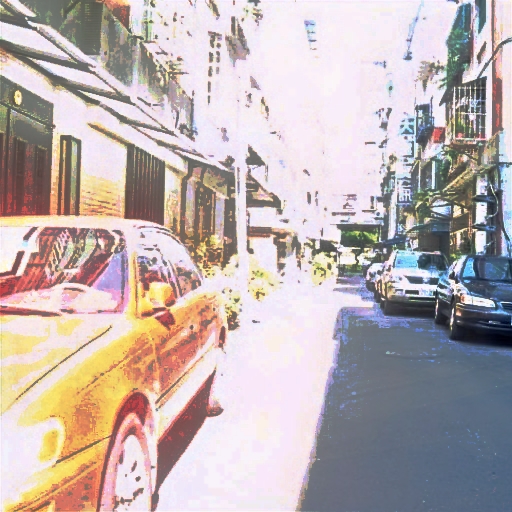}\hspace*{-0.1em}
	\includegraphics[width=0.32\linewidth, height=0.2\linewidth]{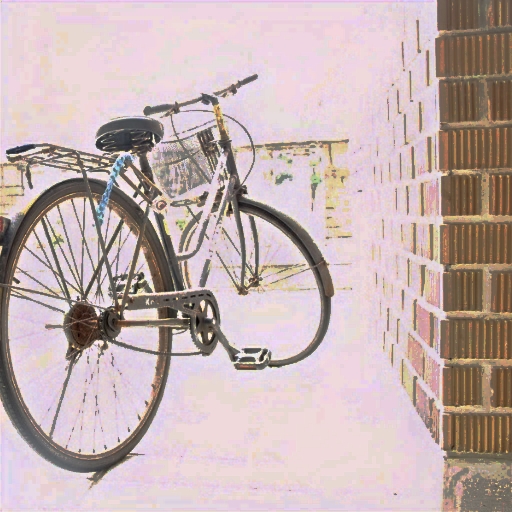}\hspace*{-0.1em}\\
	\setlength{\lineskip}{0pt}
	\includegraphics[width=0.32\linewidth, height=0.2\linewidth]{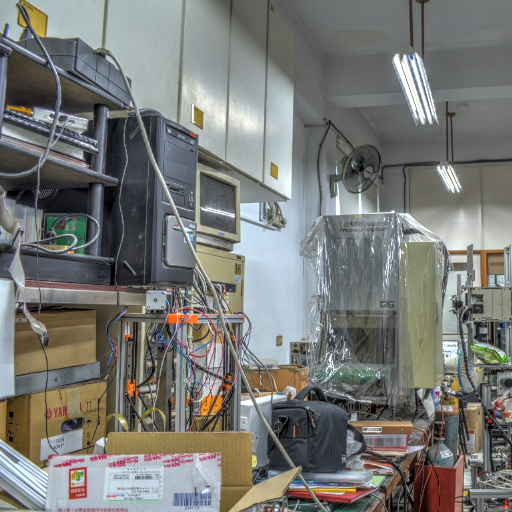}\hspace*{-0.1em}
	\includegraphics[width=0.32\linewidth, height=0.2\linewidth]{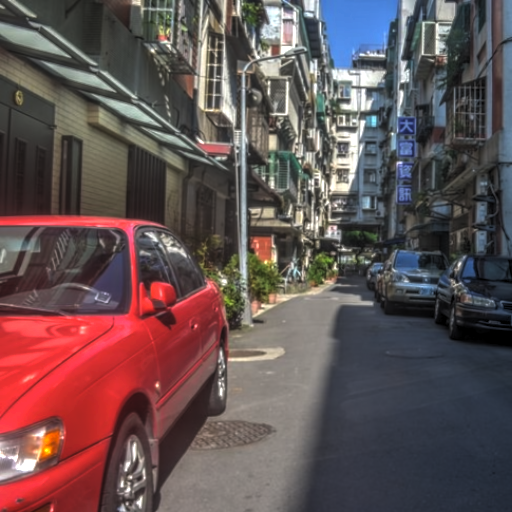}\hspace*{-0.1em}
	\includegraphics[width=0.32\linewidth, height=0.2\linewidth]{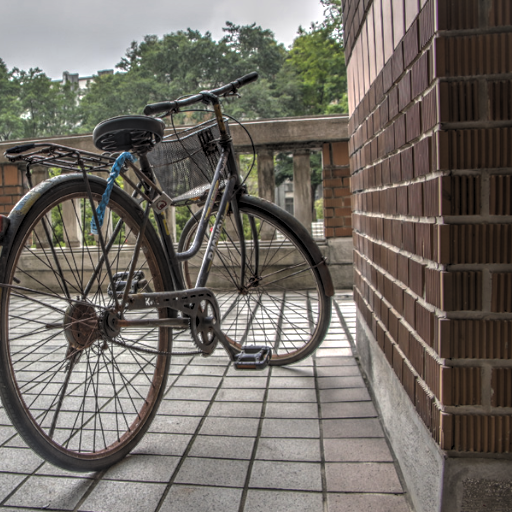}\hspace*{-0.1em}\\
	\caption{Three examples of extreme cases of over exposure in HDR Real dataset. Order from top to bottom is: input, HDRCNN \cite{eilertsen_hdr_2017}, ExpandNet \cite{marnerides_expandnet_2018}, SingleHDR \cite{liu_single-image_2020}, our method and ground truth.}
	\label{fig:overexposure}
\end{figure}

In Figure \ref{fig:overexposure}, we illustrate the difficulty of ITM for some of the samples belonging to the HDR Real with three extremely over-exposed inputs. As seen, all networks struggle to recreate the ground truth, failing in most of the content.

\subsection{Extended Ablation Studies}

\subsubsection{Int16 Scheme Comparison}

\begin{table}[!t]
	
	\centering
	\caption{Results for the different quantization schemes. Latency is measured in the deployment platform (SN20E990) with 4 CPU threads and averaged over 300 rounds.\label{tab:sr}}
	\footnotesize
	\begin{tabular}{|l|c|c|c|c|}
		\hline
		\textbf{Q. Scheme} &\textbf{ L. BA} (ms) & \textbf{L. HP} (ms) &  \textbf{\textit{PSNR}-$T$} & \textbf{\textit{SSIM}-$T$} \\
		\specialrule{.2em}{.2em}{.2em}
		MQ & 11.52 $\pm$ 0.82 & 9.63 $\pm$ 0.12 & 21.25 $\pm$ 3.11 & 0.8782 $\pm$ 0.052\\
		\hline
		Int 16 & 857.67 $\pm$ 18.2  & -   & 17.23 $\pm$ 3.87 & 0.7612 $\pm$ 0.1016 \\
		\hline
	\end{tabular}
\end{table}

An alternative solution to the MQ scheme, already available in some network optimization suits \cite{noauthor_post-training_nodate}, is quantizing network parameters to 8-bit integers and the activations to 16-bit integers. We compare both schemes and present the results in Table \ref{tab:sr} using the same network with the exception of the IN blocks which have been substituted by Batch Normalization blocks due to incompatibilities in the inference framework. The results show that there is a substantial loss in accuracy when the \textit{int}-16 scheme is used. In the case of latency, we observe a substantial increase in latency, probably due to an issue of a lack of suitability of such quantization schemes with the deployment platform or the lack of suitable implementations in the deployment suite. All in all, aside from latency measurements, analyses on accuracy of models show the benefits of our MQN in contrast to the aforementioned scheme which has 8-bit integer parameters and 16-bit integer activations.

\subsubsection{Analyses of Loss Functions}

\begin{figure}[!t]

    \setlength{\lineskip}{0pt}
	\includegraphics[width=0.22\linewidth, height=0.2\linewidth]{images/losses/inputs/8.jpg}\hspace*{-0.1em}	
    \includegraphics[width=0.22\linewidth, height=0.2\linewidth]{images/losses/3/8_zoom.jpg}\hspace*{-0.1em}	
    \includegraphics[width=0.06\linewidth, height=0.2\linewidth]{images/losses/3/8_zoom_1.jpg}\hspace*{-0.1em}	
	\includegraphics[width=0.22\linewidth, height=0.2\linewidth]{images/losses/4/8_zoom}\hspace*{-0.1em}
	\includegraphics[width=0.06\linewidth, height=0.2\linewidth]{images/losses/4/8_zoom_1}\hspace*{-0.1em}
	\includegraphics[width=0.22\linewidth, height=0.2\linewidth]{images/losses/gts/8.jpg}\hspace*{-0.1em}\\
    \setlength{\lineskip}{0pt}
	\includegraphics[width=0.22\linewidth, height=0.2\linewidth]{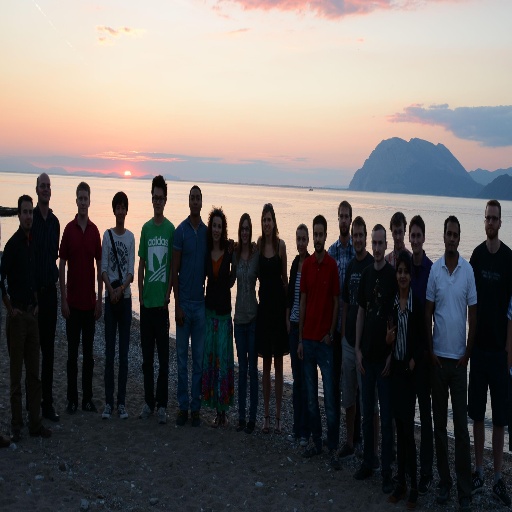}\hspace*{-0.1em}	
    \includegraphics[width=0.22\linewidth, height=0.2\linewidth]{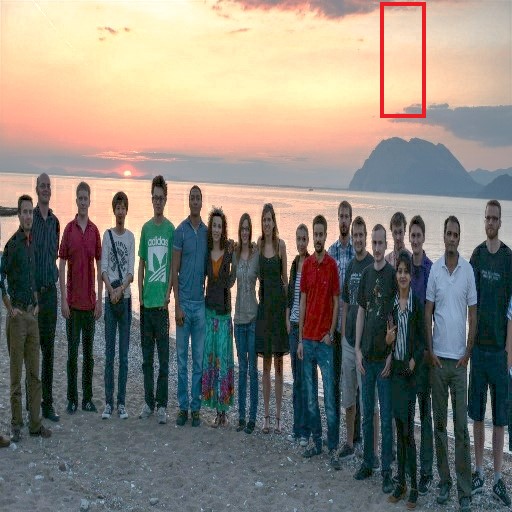}\hspace*{-0.1em}	
    \includegraphics[width=0.06\linewidth, height=0.2\linewidth]{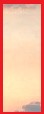}\hspace*{-0.1em}	
	\includegraphics[width=0.22\linewidth, height=0.2\linewidth]{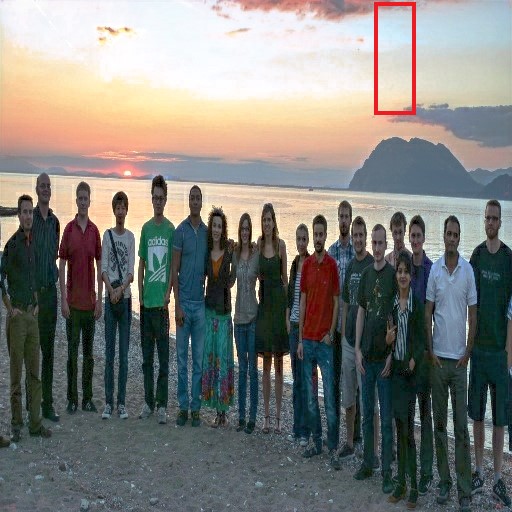}\hspace*{-0.1em}
	\includegraphics[width=0.06\linewidth, height=0.2\linewidth]{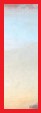}\hspace*{-0.1em}
	\includegraphics[width=0.22\linewidth, height=0.2\linewidth]{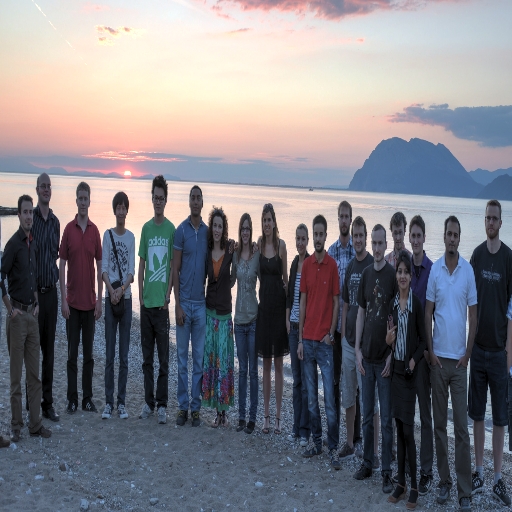}\hspace*{-0.1em}\\
    \setlength{\lineskip}{0pt}
	\includegraphics[width=0.22\linewidth, height=0.2\linewidth]{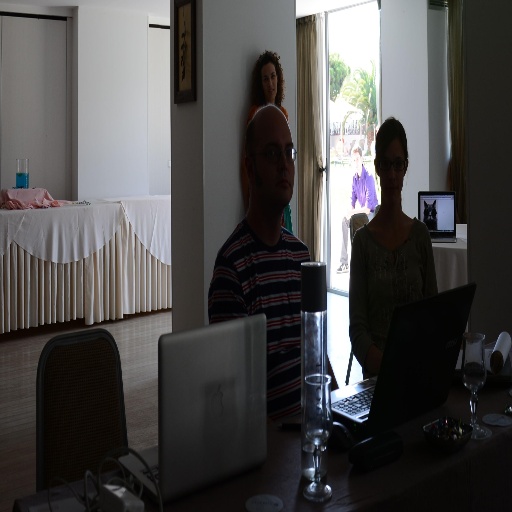}\hspace*{-0.1em}	
    \includegraphics[width=0.22\linewidth, height=0.2\linewidth]{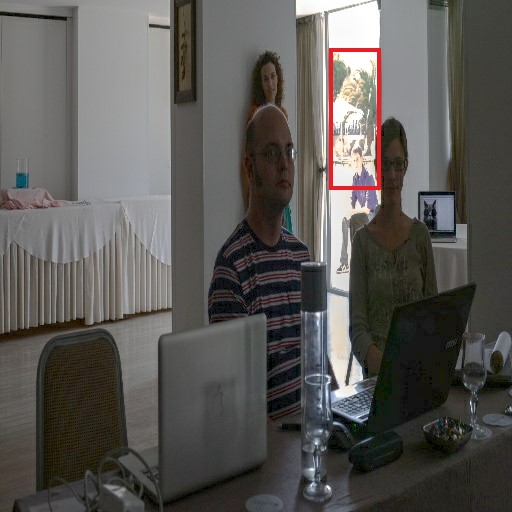}\hspace*{-0.1em}	
    \includegraphics[width=0.06\linewidth, height=0.2\linewidth]{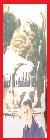}\hspace*{-0.1em}	
	\includegraphics[width=0.22\linewidth, height=0.2\linewidth]{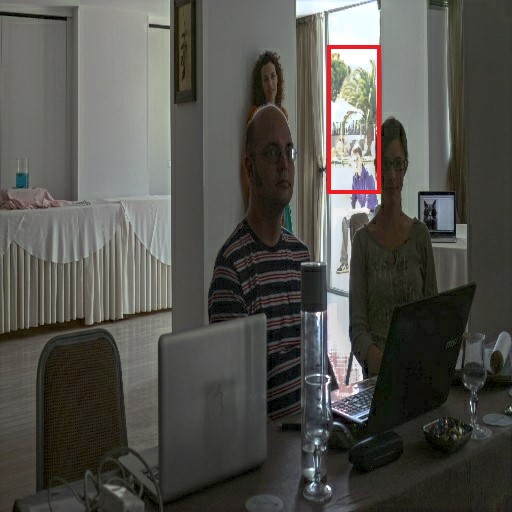}\hspace*{-0.1em}
	\includegraphics[width=0.06\linewidth, height=0.2\linewidth]{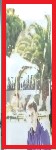}\hspace*{-0.1em}
	\includegraphics[width=0.22\linewidth, height=0.2\linewidth]{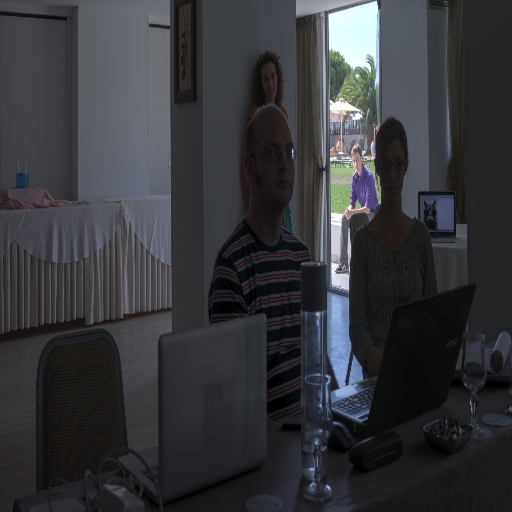}\hspace*{-0.1em}\\     \setlength{\lineskip}{0pt}
	\includegraphics[width=0.22\linewidth, height=0.2\linewidth]{images/losses/inputs/28.jpg}\hspace*{-0.1em}	
    \includegraphics[width=0.22\linewidth, height=0.2\linewidth]{images/losses/3/28_zoom.jpg}\hspace*{-0.1em}	
    \includegraphics[width=0.06\linewidth, height=0.2\linewidth]{images/losses/3/28_zoom_1.jpg}\hspace*{-0.1em}	
	\includegraphics[width=0.22\linewidth, height=0.2\linewidth]{images/losses/4/28_zoom.jpg}\hspace*{-0.1em}
	\includegraphics[width=0.06\linewidth, height=0.2\linewidth]{images/losses/4/28_zoom_1.jpg}\hspace*{-0.1em}
	\includegraphics[width=0.22\linewidth, height=0.2\linewidth]{images/losses/gts/28.jpg}\hspace*{-0.1em}
	
	    \hspace{0.06\linewidth} Input \hspace{0.1\linewidth} wo/ Perceptual Loss \hspace{0.06\linewidth} w/ Perceptual Loss \hspace{0.06\linewidth} Ground Truth
	\caption{Visual comparison of the effect on predictions by including perceptual loss as a training component. It helps models learn representations of structural color coherency and improve color details.}
	\label{fig:losses}
\end{figure}

\begin{figure}[!t]
\includegraphics[width=0.5\textwidth, height=0.4\textwidth]{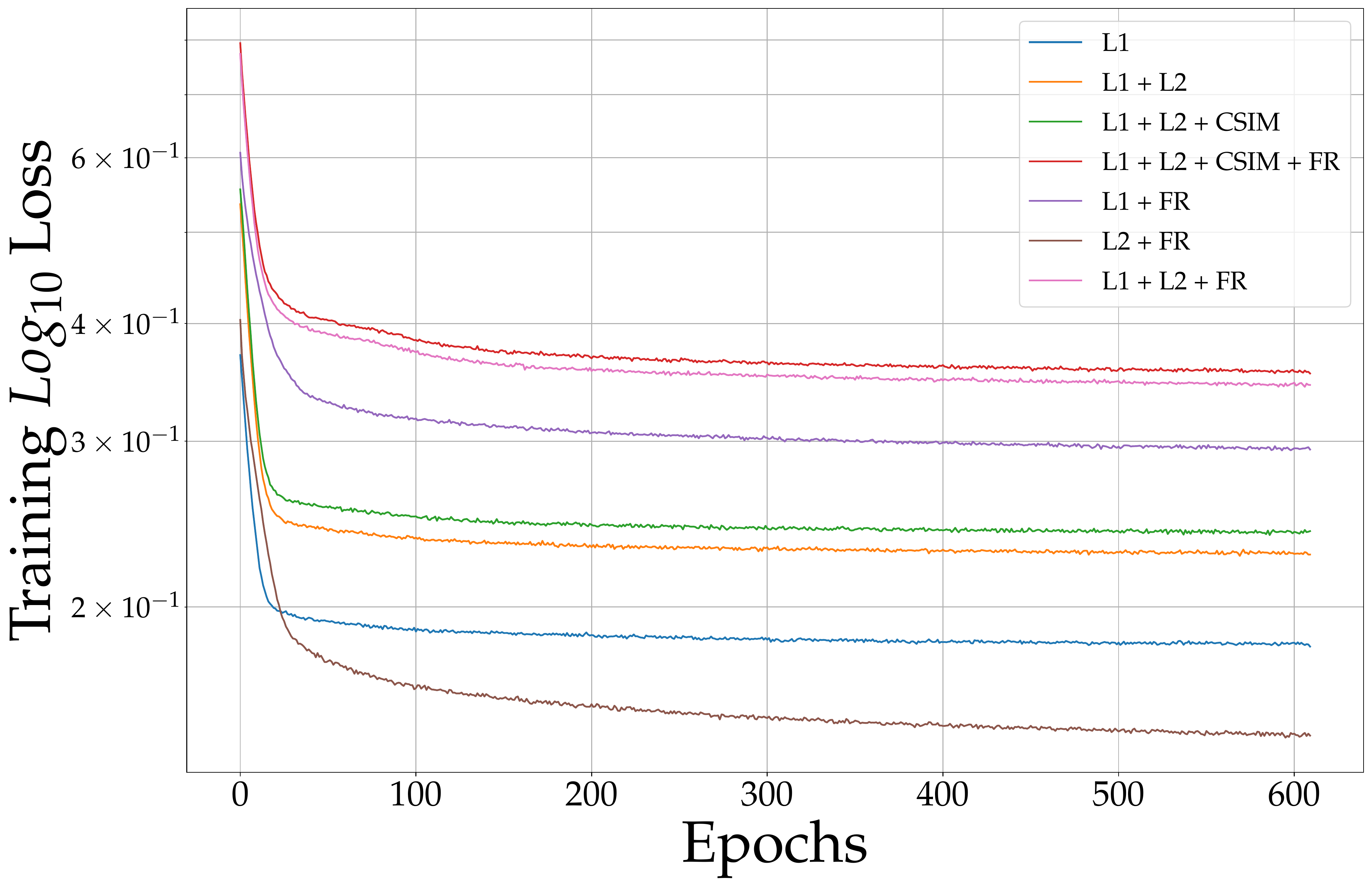}
\includegraphics[width=0.5\textwidth, height=0.4\textwidth]{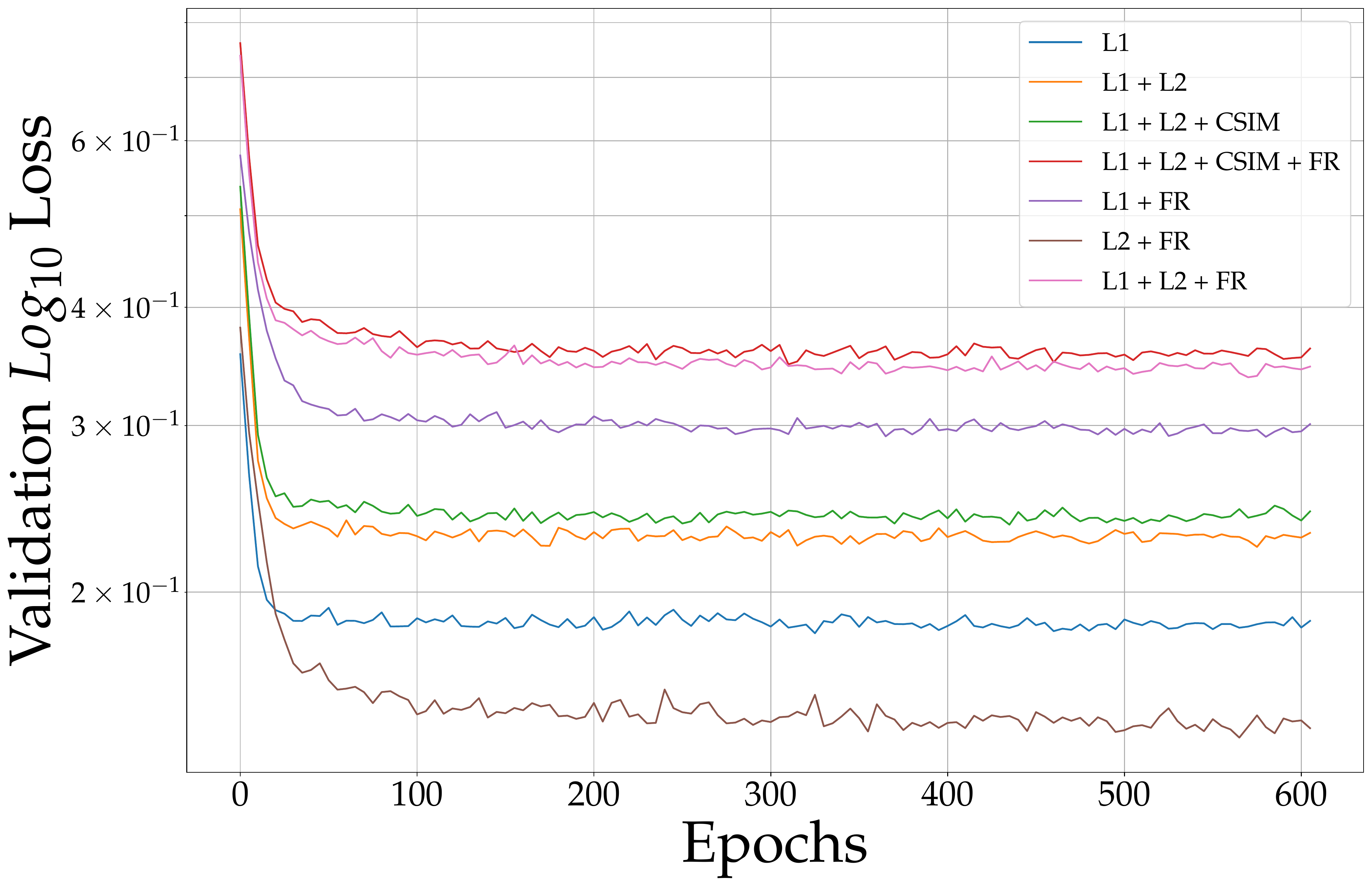}\\
\caption{Training (left) and validation (right) loss plots for each loss combination established in Table  1 (Right) in Main Paper.}
\label{fig:lossplots}
\end{figure}

In addition, the effect of training models  using the perceptual loss on predictions can be observed in Figure \ref{fig:losses}. As the results show, training models using perceptual loss helps learning feature representations of color coherency as well as color details, further enhancing the quality of the image. Examples are the structural coherency in the color-checker (row 1) or the sky color coherency (row 4). 

To provide additional information about the training process, in Figure \ref{fig:lossplots} we show the training and validation loss evolution of all combination of losses found in Table 1 in Main paper.

\section{Additional Implementation Details}

\subsection{Attention Mechanisms}

As stated in Section 3.1 of the main paper, we analyzed accuracy of the proposed MQN using three types of gated attention mechanisms. First, at the end of the first IRLB block in the decoder, we add Spatial Attention (SA) \cite{woo_cbam_2018} gated blocks. We define SA blocks by
\begin{equation}
	\mathbf{O}_{f} = ((\sigma_{1}\circ C_{1})(I_{f}))\odot I_{f}
\end{equation}
where $O_{f}$ and $I_{f}$ are the input and output respectively  with $f$ channels. $C_{1}$ denotes a convolution with a filter and kernel of size 1, $\sigma_{1}$ is a sigmoid activation, and  $\circ$ indicates function composition. 

\begin{figure*}[!ht]
	\centering
	\subfigure[]{\includegraphics[width=\linewidth]{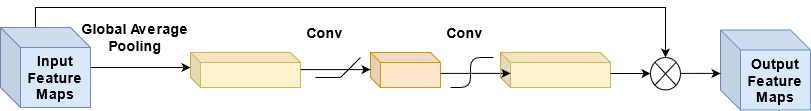}}
	\subfigure[]{\includegraphics[width=0.45\linewidth]{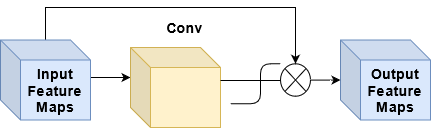}}
	\subfigure[]{\includegraphics[width=0.45\linewidth]{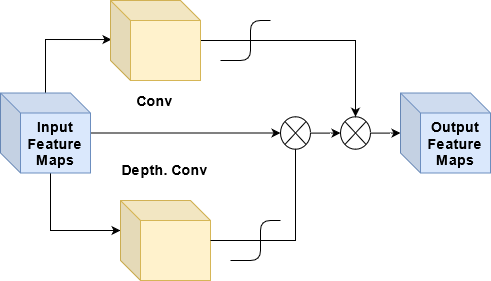}}

	\caption{Depiction of  (a) Channel Attention (CA) block, (b) Spatial Attention (SA) and Channel Spatial Attention (SCA) block. \textit{Conv} refers to a standard convolution, a soft sigmoid form denotes the sigmoid activation, the rectilinear symbol denotes ReLU activation and the $\otimes$ denotes element wise product.}
	\label{fig:attention}
\end{figure*}

To improve the previous attention mechanism, we add channel information through a depthwise convolution in parallel to the SA mechanism, which results in the channel spatial attention (CSA) mechanism defined by
\begin{equation}
	\mathbf{O}_{f} = ((\sigma_{1}\circ D_{f})(\mathbf{I}_{f}))\odot (((\sigma_{1}\circ C_{1})(\mathbf{I}_{f}))\otimes \mathbf{I}_{f})
\end{equation} 
where $D_{f}$ is the depthwise convolution with $f$ filters. Finally, although with a higher computational cost due to pooling mechanisms, we also test channel attention (CA) blocks \cite{zhang_image_2018}. This operation, inspired by both residual layers and gated attention is defined by
\begin{equation}
	\mathbf{O}_{f} = ((\sigma_1\circ C_{f}\circ\sigma_{2}\circ C_{f'}\circ GP)(\mathbf{I}_{f}))\odot \mathbf{I}_{f}
\end{equation}
where $C_{f}$ denotes convolution with $f$ filters and kernels of size 1, where $f'= f\cdot r$ and $r$ is the reduction ratio of the attention mechanism. Finally, $\sigma_{2}$ is the ReLU function and $GP$ denotes global pooling. All three attention mechanisms are depicted in Figure \ref{fig:attention}.

\subsection{Mixed Precision vs Mixed Quantization}

Mixed precision\footnote{\href{https://www.tensorflow.org/guide/mixed_precision}{Tensorflow Mixed Precision API}} \cite{micikevicius_mixed_2017} is a methodology that uses floating point data types with different bit width (16 and 32, among other specifications) to train faster and with better convergence. Each activation function and operation has a different range, and each weight takes values form different ranges. Hence, it is enough to use 16-bit for some of them, while others need higher resolution.  

However, an MQN model does not use a mixed precision training scheme. Instead, MQN uses two different types of quantization schemes: full post-training static 8-bit integer quantization\footnote{\href{https://www.tensorflow.org/lite/performance/post_training_integer_quant}{Tensorflow Post Training 8-bit Integer Quantization}} \cite{jacob_quantization_2018} and dynamic quantization\footnote{\href{https://www.tensorflow.org/lite/performance/post_training_quant}{Tensorflow Post Training Dynamic Range Quantization}}. First, we train the model in full float precision. Once we have trained the model, we quantize the Base Architecture (BA) to 8-bit integer post-training quantization. Hence, all the BA model has 8-bit weights and 8-bit activations. 

Since we aim to recover images with a higher precision for ITM with  fast inference, we use dynamic quantization for the high precision (HP) head: it quantizes the weights to 8 bits but activations have 32-bit floating point precision. For this purpose, we employ two different types of post-training quantization to be able to obtain fast inference but still have high precision output.


\subsection{Description of the Supplementary Video}

As an addition to the complementary material, we apply our method to a video sequence from Unvanquished \cite{noauthor_unvanquishedunvanquished_2021} gameplay and a cityscape at night. In the first video inference is presented in the form of a moving split screen with the original input LDR on the left side and the tonemapped HDR prediction provided by MQN on the right side. In the second video, the original LDR image is in the upper part while the HDR prediction is in the lower part. As applied to all images presented in the article, the  frames (images) of the video have been tone mapped to adapt them to commodity non-HDR-capable displays.

\subsection{Scaled Images}

In this section, we show the images that appear in the main paper, but in a higher scale for a more comfortable visualization. 

\begin{figure}[!htbp]
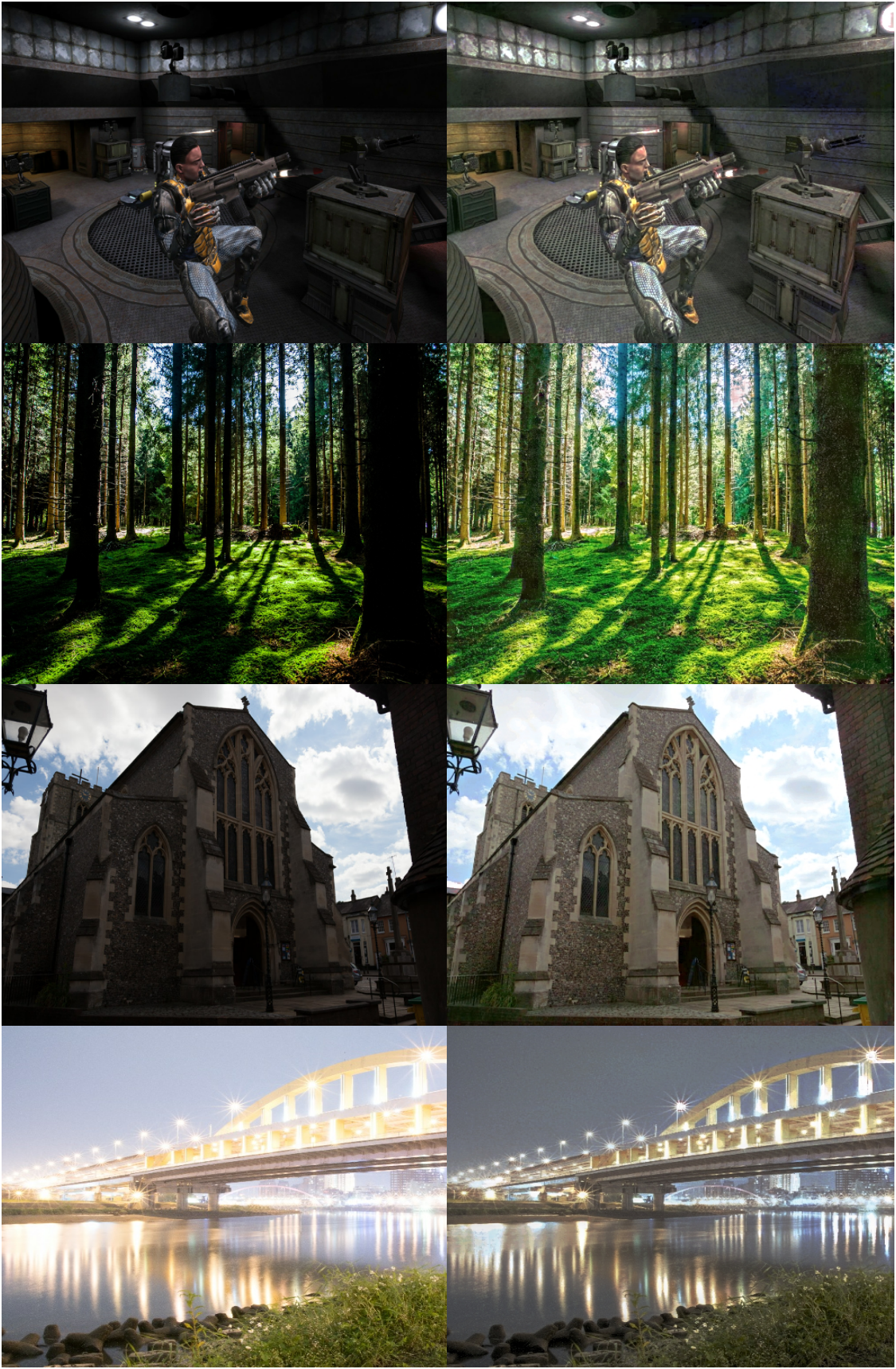

	\setlength{\lineskip}{0pt}
	\centering
	\includegraphics[width=0.45\linewidth, height=0.22\textheight]{images/Alpha_35_screenshot_of_unvanquished_resized}\hspace*{-0.3em}
	\includegraphics[width=0.45\linewidth, height=0.22\textheight]{images/Alpha_35_screenshot_of_unvanquished_Balanced}\hspace*{-0.3em}\\
	\includegraphics[width=0.45\linewidth, height=0.22\textheight]{images/00000_resized}\hspace*{-0.3em}
	\includegraphics[width=0.45\linewidth, height=0.22\textheight]{images/00000_Realistic}\hspace*{-0.3em}\\
	\includegraphics[width=0.45\linewidth, height=0.22\textheight]{images/Common_errors_-_underexposed_resized}\hspace*{-0.3em}
	\includegraphics[width=0.45\linewidth, height=0.22\textheight]{images/Common_errors_-_underexposed_Photographic}\hspace*{-0.3em}\\
	\includegraphics[width=0.45\linewidth, height=0.22\textheight]{images/00216_input_resized}\hspace*{-0.3em}
	\includegraphics[width=0.45\linewidth, height=0.22\textheight]{images/00216_input_Balanced}
	\caption{\textbf{Mobile  HDR reconstruction from a single LDR image.} Mixed quantization and efficient blocks help reduce the computational requirements of HDR reconstruction and enable its deployment to mobile platforms, achieving a latency of $\approx$ 21 ms on a Samsung Note 20 Exynos 990.}
	\label{fig:example_supplemental}
\end{figure}

\begin{figure}[!hbtp]
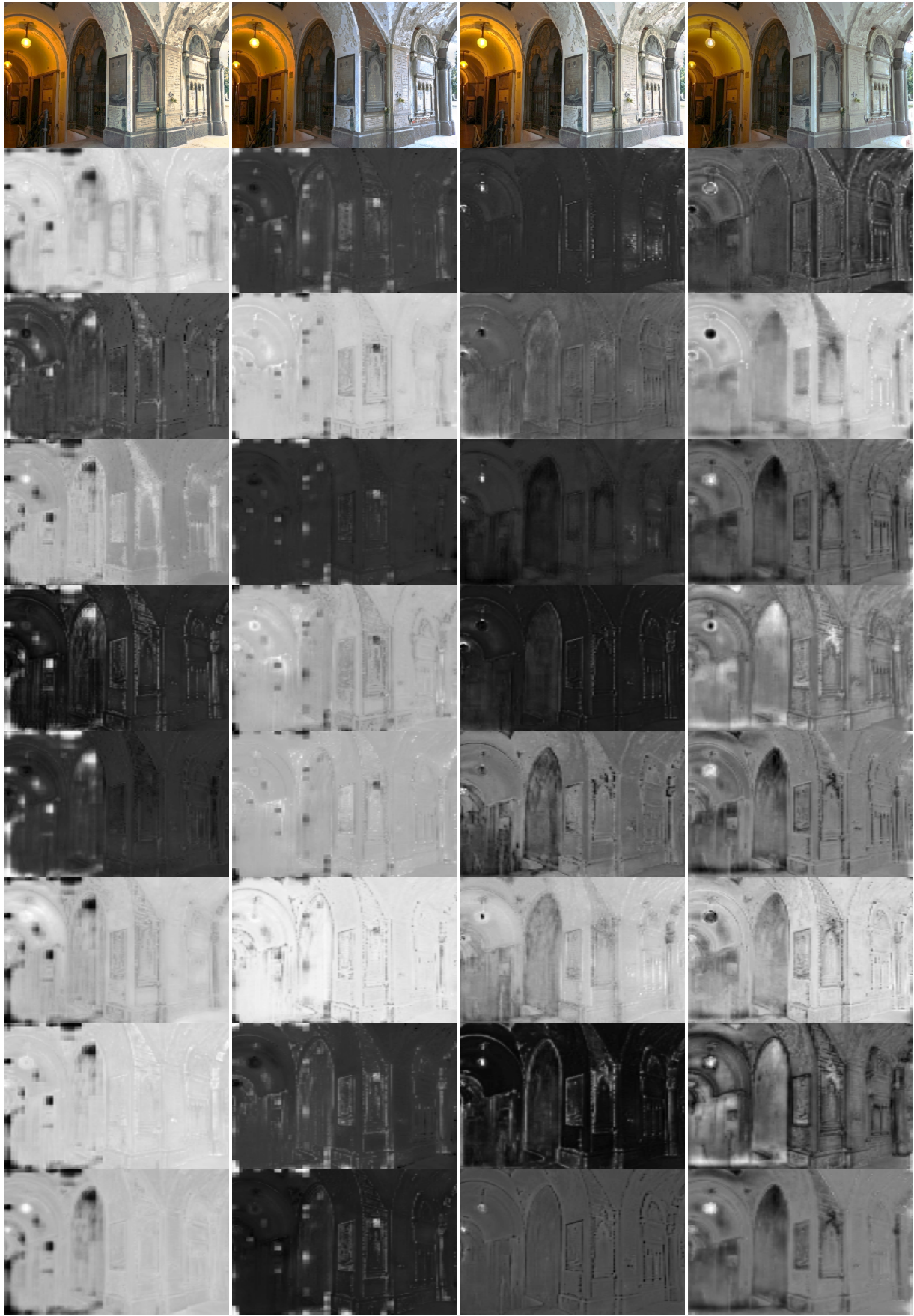

	\setlength{\lineskip}{0pt}
	\includegraphics[width=0.24\linewidth, height=0.1\textheight]{images/attention/00035_input_Balanced_a3.jpg}\hspace*{-0.1em}
	\includegraphics[width=0.24\linewidth, height=0.1\textheight]{images/attention/00035_input_Balanced_a1.jpg}\hspace*{-0.1em}
	\includegraphics[width=0.24\linewidth, height=0.1\textheight]{images/attention/00035_input_Balanced_a2.jpg}\hspace*{-0.1em}
	\includegraphics[width=0.24\linewidth, height=0.1\textheight]{images/attention/00035_input_Balanced.jpg}\hspace*{-0.1em}\\
	\includegraphics[width=0.24\linewidth, height=0.1\textheight]{images/attention/i3_0.png}\hspace*{-0.1em}
	\includegraphics[width=0.24\linewidth, height=0.1\textheight]{images/attention/i1_0.png}\hspace*{-0.1em}
	\includegraphics[width=0.24\linewidth, height=0.1\textheight]{images/attention/i2_0.png}\hspace*{-0.1em}
	\includegraphics[width=0.24\linewidth, height=0.1\textheight]{images/attention/i0_0.png}\hspace*{-0.1em}\\
	\includegraphics[width=0.24\linewidth, height=0.1\textheight]{images/attention/i3_1.png}\hspace*{-0.1em}
	\includegraphics[width=0.24\linewidth, height=0.1\textheight]{images/attention/i1_1.png}\hspace*{-0.1em}
	\includegraphics[width=0.24\linewidth, height=0.1\textheight]{images/attention/i2_1.png}\hspace*{-0.1em}
	\includegraphics[width=0.24\linewidth, height=0.1\textheight]{images/attention/i0_1.png}\hspace*{-0.1em}\\
	\includegraphics[width=0.24\linewidth, height=0.1\textheight]{images/attention/i3_2.png}\hspace*{-0.1em}
	\includegraphics[width=0.24\linewidth, height=0.1\textheight]{images/attention/i1_2.png}\hspace*{-0.1em}
	\includegraphics[width=0.24\linewidth, height=0.1\textheight]{images/attention/i2_2.png}\hspace*{-0.1em}
	\includegraphics[width=0.24\linewidth, height=0.1\textheight]{images/attention/i0_2.png}\hspace*{-0.1em}\\
	\includegraphics[width=0.24\linewidth, height=0.1\textheight]{images/attention/i3_3.png}\hspace*{-0.1em}
	\includegraphics[width=0.24\linewidth, height=0.1\textheight]{images/attention/i1_3.png}\hspace*{-0.1em}
	\includegraphics[width=0.24\linewidth, height=0.1\textheight]{images/attention/i2_3.png}\hspace*{-0.1em}
	\includegraphics[width=0.24\linewidth, height=0.1\textheight]{images/attention/i0_3.png}\hspace*{-0.1em}\\	\includegraphics[width=0.24\linewidth, height=0.1\textheight]{images/attention/i3_4.png}\hspace*{-0.1em}
	\includegraphics[width=0.24\linewidth, height=0.1\textheight]{images/attention/i1_4.png}\hspace*{-0.1em}
	\includegraphics[width=0.24\linewidth, height=0.1\textheight]{images/attention/i2_4.png}\hspace*{-0.1em}
	\includegraphics[width=0.24\linewidth, height=0.1\textheight]{images/attention/i0_4.png}\hspace*{-0.1em}\\
	\includegraphics[width=0.24\linewidth, height=0.1\textheight]{images/attention/i3_5.png}\hspace*{-0.1em}
	\includegraphics[width=0.24\linewidth, height=0.1\textheight]{images/attention/i1_5.png}\hspace*{-0.1em}
	\includegraphics[width=0.24\linewidth, height=0.1\textheight]{images/attention/i2_5.png}\hspace*{-0.1em}
	\includegraphics[width=0.24\linewidth, height=0.1\textheight]{images/attention/i0_5.png}\hspace*{-0.1em}\\
		\includegraphics[width=0.24\linewidth, height=0.1\textheight]{images/attention/i3_6.png}\hspace*{-0.1em}
	\includegraphics[width=0.24\linewidth, height=0.1\textheight]{images/attention/i1_6.png}\hspace*{-0.1em}
	\includegraphics[width=0.24\linewidth, height=0.1\textheight]{images/attention/i2_6.png}\hspace*{-0.1em}
	\includegraphics[width=0.24\linewidth, height=0.1\textheight]{images/attention/i0_6.png}\hspace*{-0.1em}\\
		\includegraphics[width=0.24\linewidth, height=0.1\textheight]{images/attention/i3_7.png}\hspace*{-0.1em}
	\includegraphics[width=0.24\linewidth, height=0.1\textheight]{images/attention/i1_7.png}\hspace*{-0.1em}
	\includegraphics[width=0.24\linewidth, height=0.1\textheight]{images/attention/i2_7.png}\hspace*{-0.1em}
	\includegraphics[width=0.24\linewidth, height=0.1\textheight]{images/attention/i0_7.png}\hspace*{-0.1em}\\
	\caption{Illustration of the feature maps learned after the last attention mechanism. Columns in order correspond to: no attention mechanism, SA, CSA and CA. Rows correspond to prediction and feature maps respectively for the first and the rest. \label{fig:attention_scaled}}
\end{figure}

\begin{figure}[!hbtp]
	\setlength{\lineskip}{0pt}
	\includegraphics[width=0.32\linewidth, height=0.11\textheight]{images/Unvanquished_tyrant_resized}\hspace*{-0.1em}
	\includegraphics[width=0.32\linewidth, height=0.11\textheight]{images/00002_input_resized}\hspace*{-0.1em}
	\includegraphics[width=0.32\linewidth, height=0.11\textheight]{images/ours/00017_input_resized}\hspace*{-0.1em}\\
	\includegraphics[width=0.32\linewidth, height=0.11\textheight]{images/Unvanquished_tyrant_Photographic}\hspace*{-0.1em}
	\includegraphics[width=0.32\linewidth, height=0.11\textheight]{images/00002_input_Photographic}\hspace*{-0.1em}
	\includegraphics[width=0.32\linewidth, height=0.11\textheight]{images/ours/00017_input_Realistic}\hspace*{-0.1em}\\
	\includegraphics[width=0.32\linewidth, height=0.11\textheight]{images/interface/aracnid_0_q}\hspace*{-0.1em}
	\includegraphics[width=0.32\linewidth, height=0.11\textheight]{images/interface/night3_0_q}\hspace*{-0.1em}
	\includegraphics[width=0.32\linewidth, height=0.11\textheight]{images/interface/stairs_0_q}\hspace*{-0.1em}\\
	\includegraphics[width=0.32\linewidth, height=0.11\textheight]{images/interface/aracnid_7_q}\hspace*{-0.1em}
	\includegraphics[width=0.32\linewidth, height=0.11\textheight]{images/interface/night3_7_q}\hspace*{-0.1em}
	\includegraphics[width=0.32\linewidth, height=0.11\textheight]{images/interface/stairs_7_q}\hspace*{-0.1em}\\
	\includegraphics[width=0.32\linewidth, height=0.11\textheight]{images/interface/aracnid_diff}\hspace*{-0.1em}
	\includegraphics[width=0.32\linewidth, height=0.11\textheight]{images/interface/night3_diff}\hspace*{-0.1em}
	\includegraphics[width=0.32\linewidth, height=0.11\textheight]{images/interface/stairs_diff}\hspace*{-0.1em}\\
	\includegraphics[width=0.32\linewidth, height=0.11\textheight]{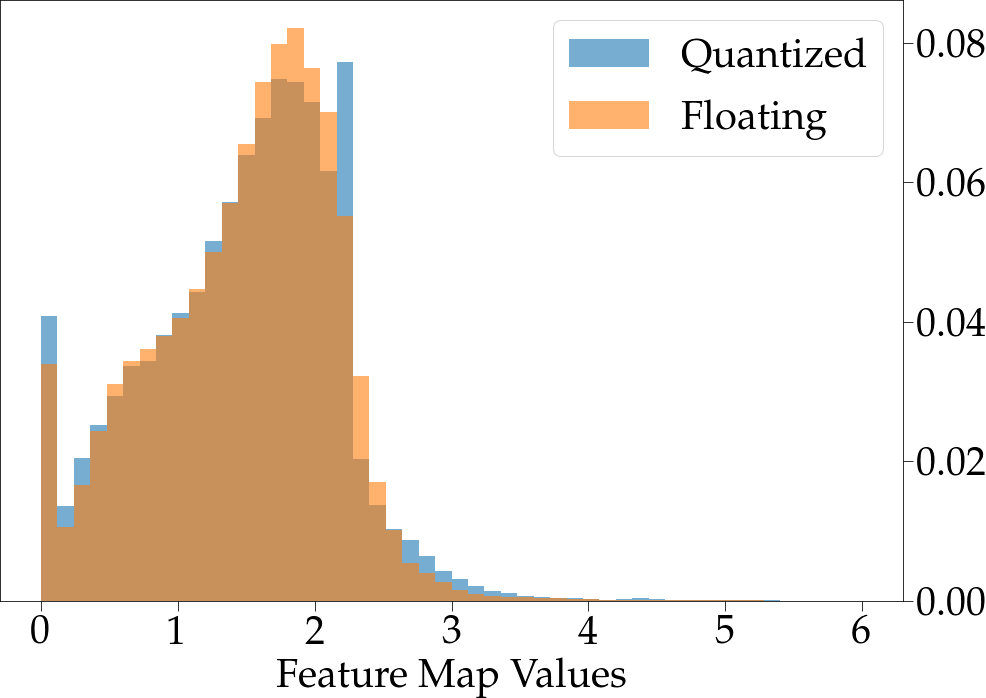}\hspace*{-0.1em}
	\includegraphics[width=0.32\linewidth, height=0.11\textheight]{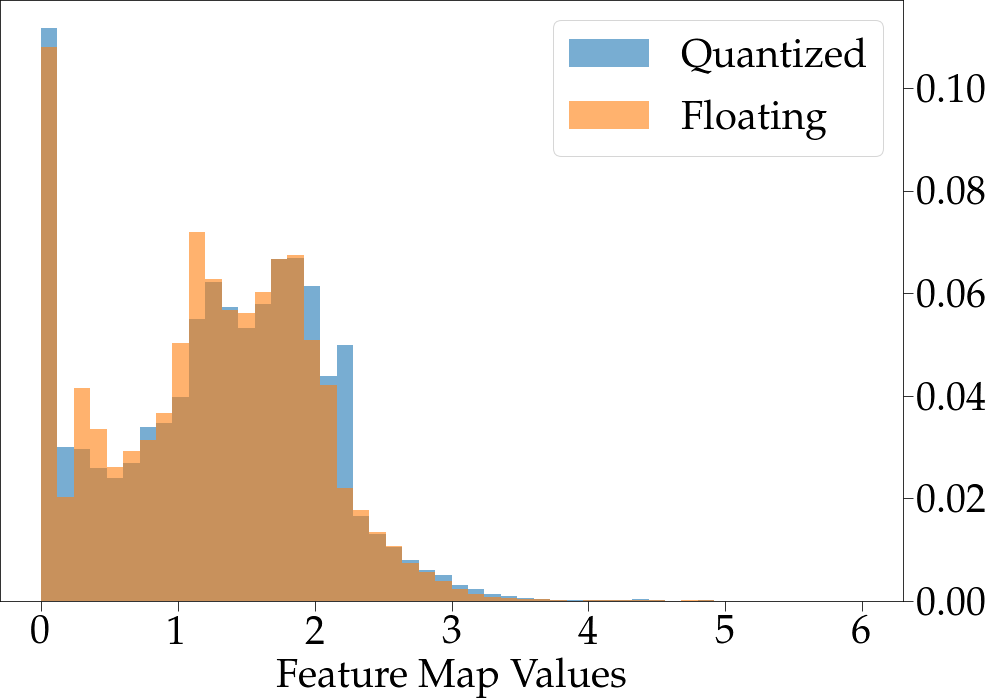}\hspace*{-0.1em}
	\includegraphics[width=0.32\linewidth, height=0.11\textheight]{images/interface/stairs_dist}\hspace*{-0.1em}\\

	\caption{Analyses of the quantized $Q(\mathbf{f})$ and floating point $F(\mathbf{f})$ features $\mathbf{f}$ learned by the MQN at the ConvBnReLU3 layer at Figure 5 using  three sample input images (first row) with predictions $\hat{H}$ (second row). Visualization of the first channel $f_1$ and the second channel $f_2$ of $\mathbf{f}$ (third and fourth rows, respectively), and the difference map $\Delta f_2 = \| Q(f_2) - F(f_2)\|_1$ (fifth row). We show the probability mass function (PMF) of $Q(f_2)$ and $F(f_2)$ (last row). \label{fig:interface_scaled}}
\end{figure}

\begin{figure}[!hbtp]		

	\setlength{\lineskip}{0pt}
	\includegraphics[width=0.2\linewidth, height=0.2\linewidth]{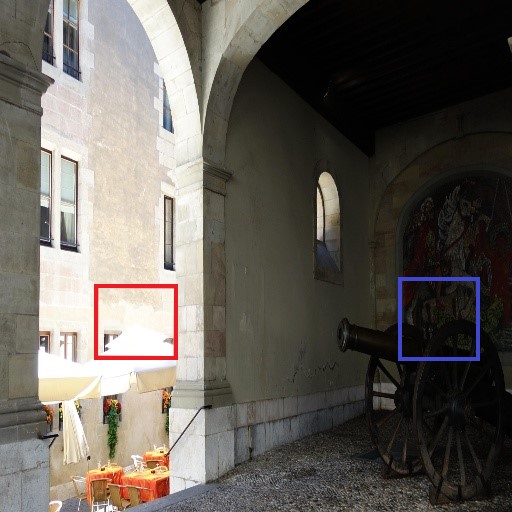}\hspace*{-0.1em}
	\includegraphics[width=0.2\linewidth, height=0.2\linewidth]{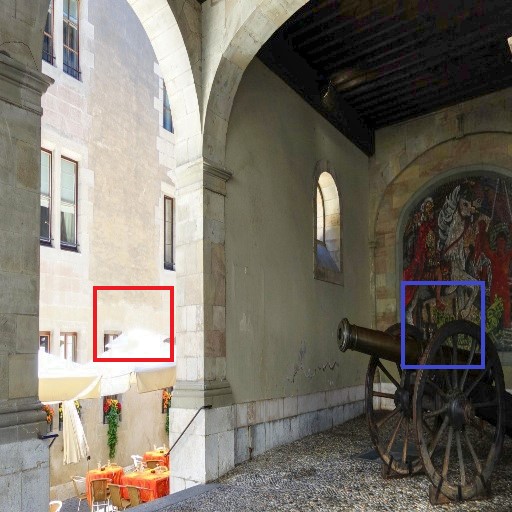}\hspace*{-0.1em}
	\includegraphics[width=0.2\linewidth, height=0.2\linewidth]{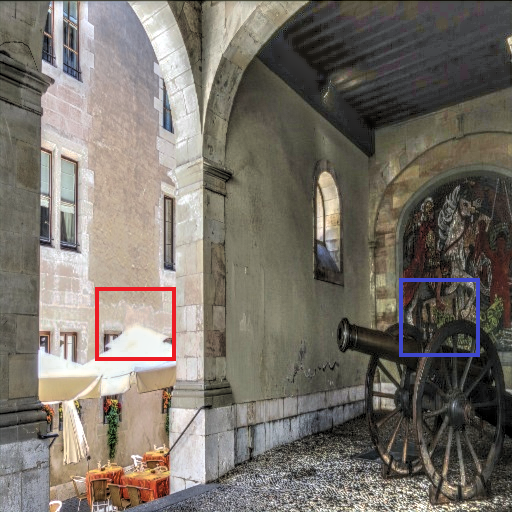}\hspace*{-0.1em}
	\includegraphics[width=0.2\linewidth, height=0.2\linewidth]{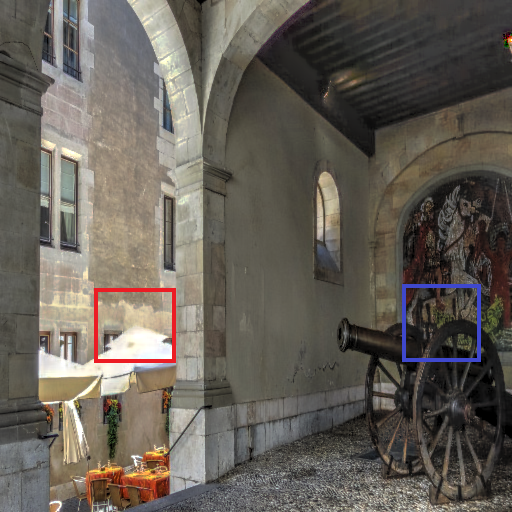}\hspace*{-0.1em}
	\includegraphics[width=0.2\linewidth, height=0.2\linewidth]{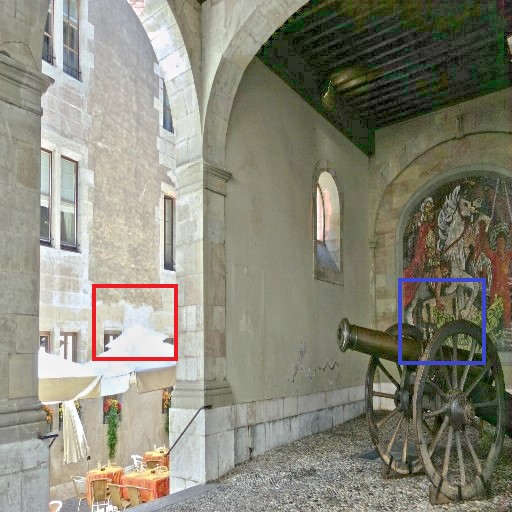}\hspace*{-0.1em}
\\
\setlength{\lineskip}{1pt}
	\includegraphics[width=0.1\linewidth, height=0.08\linewidth]{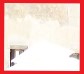}\hspace*{-0.2em}
	\includegraphics[width=0.1\linewidth, height=0.08\linewidth]{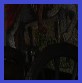}\hspace*{-0.1em}
	\includegraphics[width=0.1\linewidth, height=0.08\linewidth]{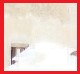}\hspace*{-0.2em}
	\includegraphics[width=0.1\linewidth, height=0.08\linewidth]{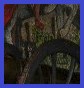}\hspace*{-0.1em}
	\includegraphics[width=0.1\linewidth, height=0.08\linewidth]{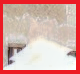}\hspace*{-0.2em}
	\includegraphics[width=0.1\linewidth, height=0.08\linewidth]{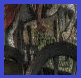}\hspace*{-0.1em}
	\includegraphics[width=0.1\linewidth, height=0.08\linewidth]{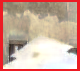}\hspace*{-0.2em}
	\includegraphics[width=0.1\linewidth, height=0.08\linewidth]{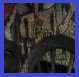}\hspace*{-0.1em}
	\includegraphics[width=0.1\linewidth, height=0.08\linewidth]{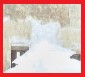}\hspace*{-0.2em}
	\includegraphics[width=0.1\linewidth, height=0.08\linewidth]{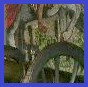}\hspace*{-0.1em}\\
	\setlength{\lineskip}{0pt}
	\includegraphics[width=0.2\linewidth, height=0.2\linewidth]{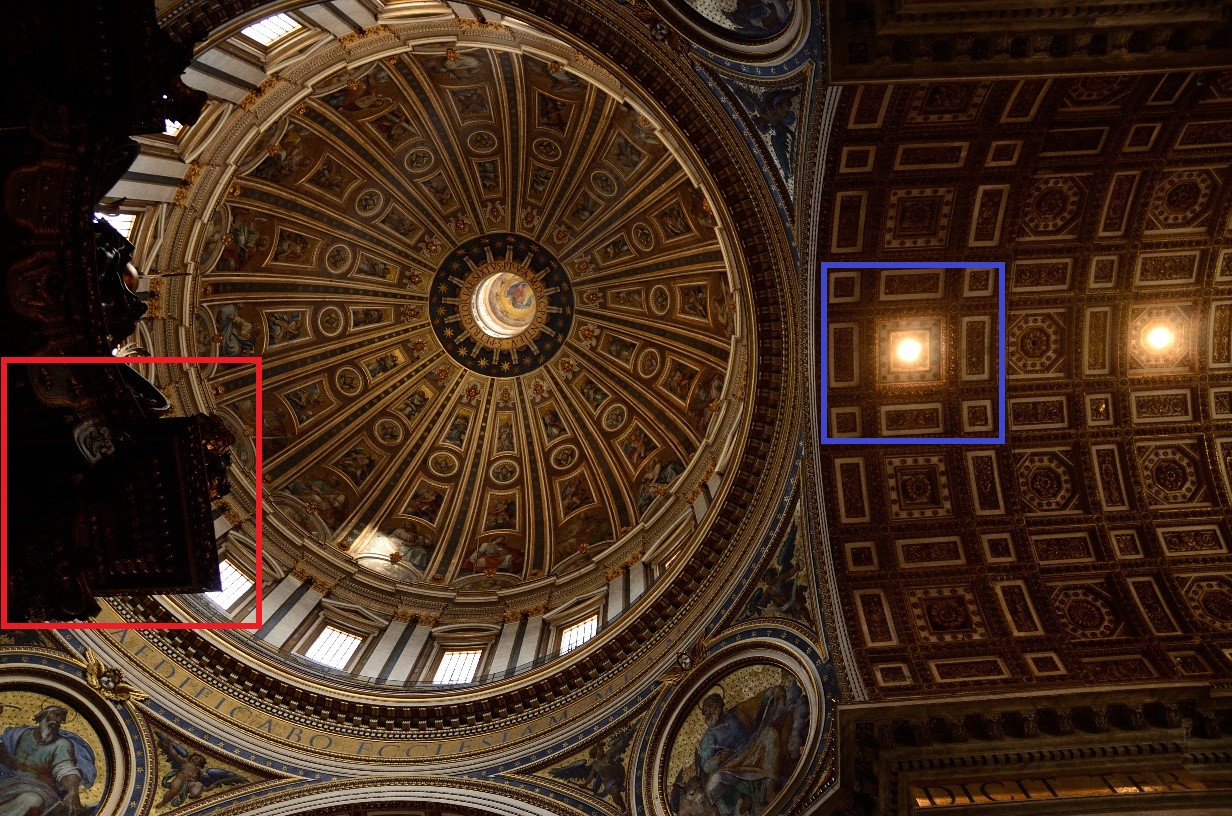}\hspace*{-0.1em}
	\includegraphics[width=0.2\linewidth, height=0.2\linewidth]{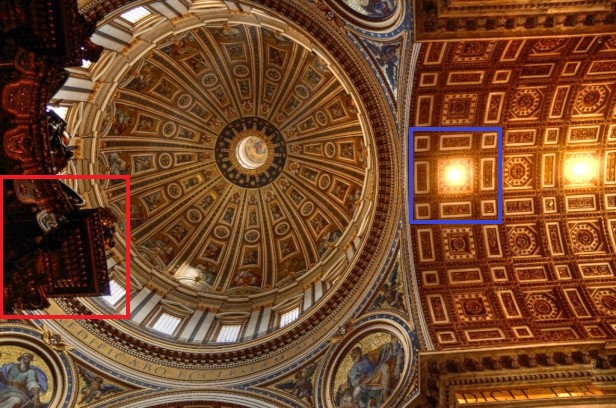}\hspace*{-0.1em}
	\includegraphics[width=0.2\linewidth, height=0.2\linewidth]{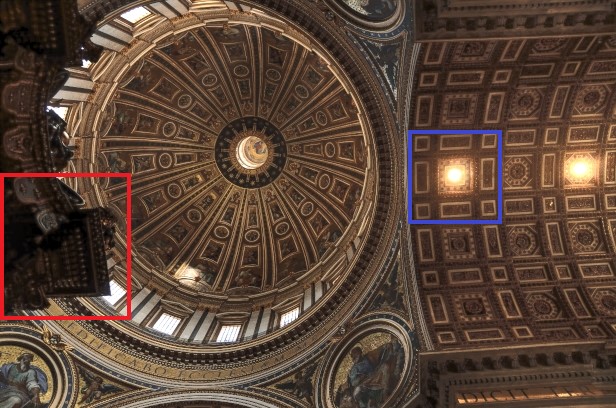}\hspace*{-0.1em}
	\includegraphics[width=0.2\linewidth, height=0.2\linewidth]{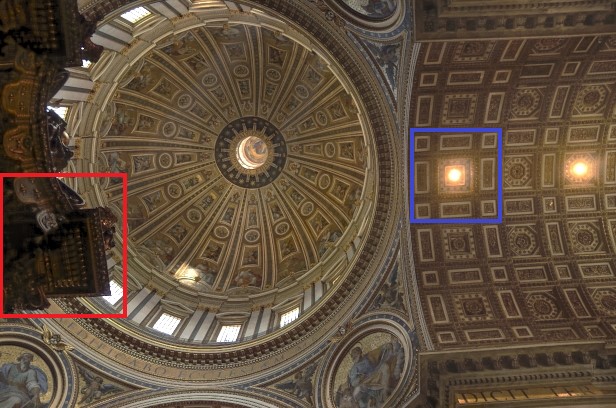}\hspace*{-0.1em}
	\includegraphics[width=0.2\linewidth, height=0.2\linewidth]{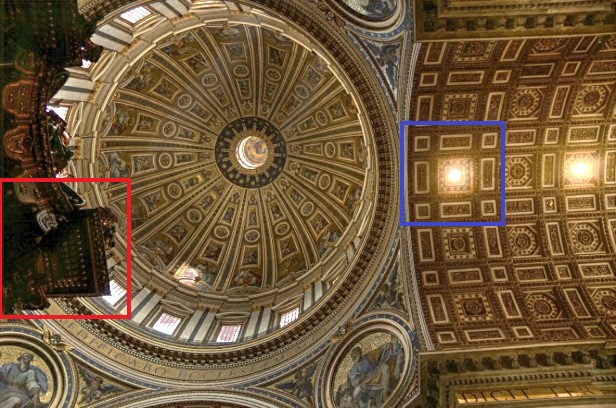}\hspace*{-0.1em}\\
    \setlength{\lineskip}{1pt}
	\includegraphics[width=0.1\linewidth, height=0.09\linewidth]{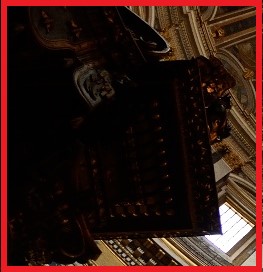}\hspace*{-0.2em}
	\includegraphics[width=0.1\linewidth, height=0.09\linewidth]{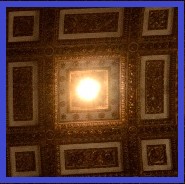}\hspace*{-0.1em}
	\includegraphics[width=0.1\linewidth, height=0.09\linewidth]{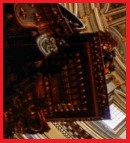}\hspace*{-0.2em}
	\includegraphics[width=0.1\linewidth, height=0.09\linewidth]{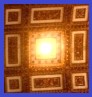}\hspace*{-0.1em}
	\includegraphics[width=0.1\linewidth, height=0.09\linewidth]{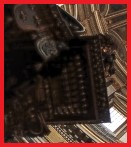}\hspace*{-0.2em}
	\includegraphics[width=0.1\linewidth, height=0.09\linewidth]{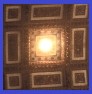}\hspace*{-0.1em}
	\includegraphics[width=0.1\linewidth, height=0.09\linewidth]{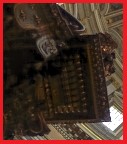}\hspace*{-0.2em}
	\includegraphics[width=0.1\linewidth, height=0.09\linewidth]{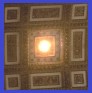}\hspace*{-0.1em}
	\includegraphics[width=0.1\linewidth, height=0.09\linewidth]{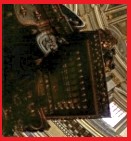}\hspace*{-0.2em}
	\includegraphics[width=0.1\linewidth, height=0.09\linewidth]{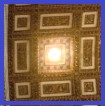}\hspace*{-0.1em}\\
	\setlength{\lineskip}{1pt}
	\includegraphics[width=0.2\linewidth, height=0.2\linewidth]{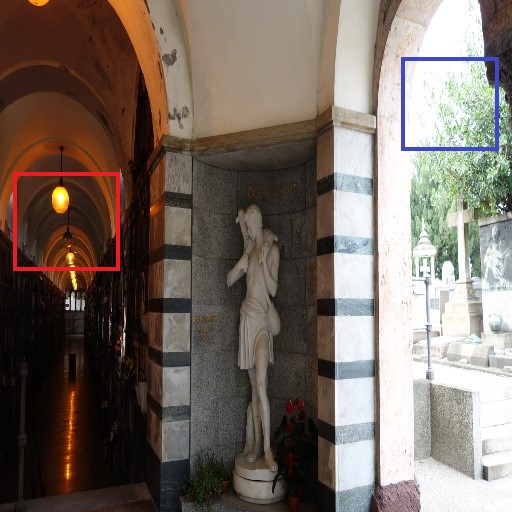}\hspace*{-0.1em}
	\includegraphics[width=0.2\linewidth, height=0.2\linewidth]{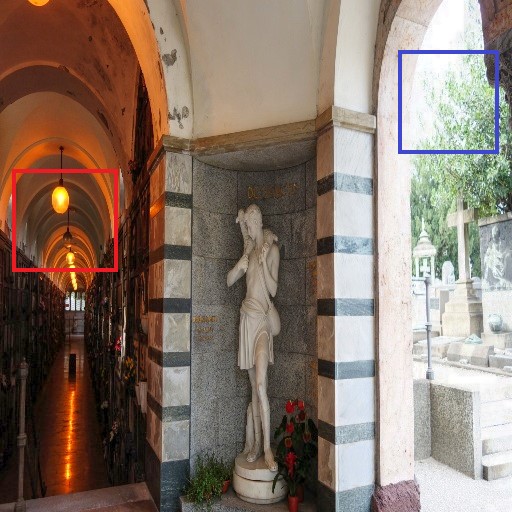}\hspace*{-0.1em}
	\includegraphics[width=0.2\linewidth, height=0.2\linewidth]{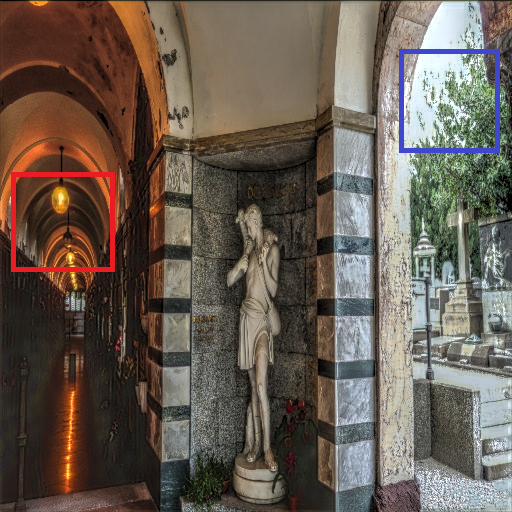}\hspace*{-0.1em}
	\includegraphics[width=0.2\linewidth, height=0.2\linewidth]{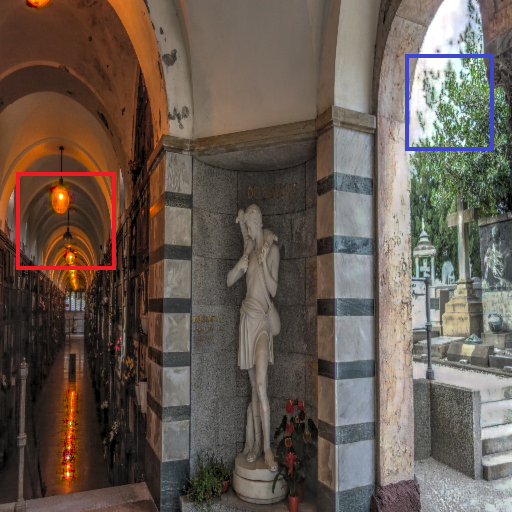}\hspace*{-0.1em}
	\includegraphics[width=0.2\linewidth, height=0.2\linewidth]{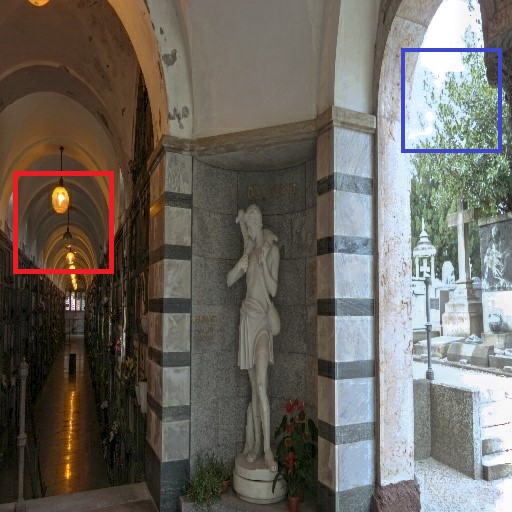}\hspace*{-0.1em}\\
	\includegraphics[width=0.1\linewidth, height=0.08\linewidth]{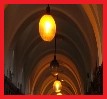}\hspace*{-0.2em}
	\includegraphics[width=0.1\linewidth, height=0.08\linewidth]{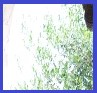}\hspace*{-0.1em}
	\includegraphics[width=0.1\linewidth, height=0.08\linewidth]{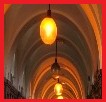}\hspace*{-0.2em}
	\includegraphics[width=0.1\linewidth, height=0.08\linewidth]{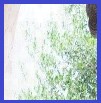}\hspace*{-0.1em}
	\includegraphics[width=0.1\linewidth, height=0.08\linewidth]{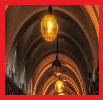}\hspace*{-0.2em}
	\includegraphics[width=0.1\linewidth, height=0.08\linewidth]{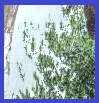}\hspace*{-0.1em}
	\includegraphics[width=0.1\linewidth, height=0.08\linewidth]{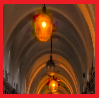}\hspace*{-0.2em}
	\includegraphics[width=0.1\linewidth, height=0.08\linewidth]{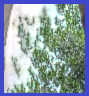}\hspace*{-0.1em}
	\includegraphics[width=0.1\linewidth, height=0.08\linewidth]{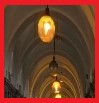}\hspace*{-0.2em}
	\includegraphics[width=0.1\linewidth, height=0.08\linewidth]{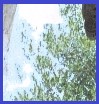}\hspace*{-0.1em}
	
	\caption{Visual comparison of our MQN and state-of-the-art ITM methods. From left to right: input LDR, HDRCNN \cite{eilertsen_hdr_2017}, ExpandNet \cite{marnerides_expandnet_2018}, SingleHDR \cite{liu_single-image_2020} and our MQN. 
	}
	\label{fig:comparison_scaled}
\end{figure}

\newpage
\bibliography{toCite}